\documentclass[twocolumn,tighten]{aastex701}
\usepackage{xspace}
\usepackage[figuresright]{rotating}
\usepackage{enumitem}

\usepackage[normalem]{ulem} % for revision

\newcommand{\lya}    {Ly$\alpha$\xspace}

\newcommand{\heii}   {\ion{He}{2}\xspace}
\newcommand{\civ}    {\ion{C}{4}\xspace}

\newcommand{\NV}     {\ion{N}{5}\xspace}

\newcommand{\heiilamb}  {\ion{He}{2}\,$\lambda$1640\xspace}

\newcommand{\unitcgssb}   {erg\,s$^{-1}$\,cm$^{-2}$\,arcsec$^{-2}$\xspace}

\newcommand{\unitcgslum}  {erg\,s$^{-1}$\xspace}

\newcommand{\unitcovolume} {cMpc$^{-3}$\xspace}
\newcommand{\kms}         {km\,s$^{-1}$\xspace}

\newcommand{\sqarcsec}    {arcsec$^{2}$\xspace}

\newcommand{\sqdeg}       {deg$^{2}$\xspace}

\newcommand{\tractor}     {\texttt{Tractor}\xspace}
\newcommand{\tractormethod}     {extended-beyond-continuum\xspace}
\newcommand{\traditional} {extended-LAE\xspace}
\newcommand{\stractor}    {{\tt SExtractor}\xspace}
\newcommand{\psfex}       {{\tt PSFEx}\xspace}

\newcommand{\frecv}       {{$f_\mathrm{recv}$}\xspace}

\shorttitle{ODIN: A New \lya Blob Selection and Analysis at $z\sim3.1$}
\shortauthors{Moon et al.}
\begin{document}

\title{ODIN: A New Lyman Alpha Blob Selection Method, Sample, and Statistical Analysis at $z\sim3.1$}

\author[0009-0008-4022-3870]{Byeongha Moon}
\affiliation{Korea Astronomy and Space Science Institute, 776 Daedeokdae-ro, Yuseong-gu, Daejeon, Korea 34055}
\affiliation{Korea National University of Science and Technology, Daejeon, Korea 34113}
\email[show]{bhmoon@kasi.re.kr}

\author[0000-0003-3078-2763]{Yujin Yang}
\affiliation{Korea Astronomy and Space Science Institute, 776 Daedeokdae-ro, Yuseong-gu, Daejeon, Korea 34055}
\affiliation{Korea National University of Science and Technology, Daejeon, Korea 34113}
\email{yyang@kasi.re.kr}

\author[0000-0003-3004-9596]{Kyoung-Soo Lee}
\affiliation{Department of Physics and Astronomy, Purdue University, 525 Northwestern Avenue, West Lafayette, IN 47907, USA}
\email{soolee@purdue.edu}
\author[0000-0003-1530-8713]{Eric Gawiser}
\affiliation{Department of Physics and Astronomy, Rutgers, the State University of New Jersey, Piscataway, NJ 08854, USA}
\email{gawiser@physics.rutgers.edu}
\author[0000-0002-4928-4003]{Arjun Dey}
\affiliation{NSF’s NOIRLab, 950 N. Cherry Ave., Tucson, AZ 85719, USA}
\email{arjun.dey@noirlab.edu}

\author[0000-0001-5567-1301]{Francisco Valdes}
\affiliation{National Optical Astronomy Observatory, 950 N. Cherry Avenue, Tucson, AZ 85719, USA}
\email{frank.valdes@noirlab.edu}

\author[0000-0002-1172-0754]{Dustin Lang}
\affiliation{Perimeter Institute for Theoretical Physics, 31 Caroline Street North, Waterloo, ON N2L 2Y5, Canada}
\affiliation{Department of Physics and Astronomy, University of Waterloo, Waterloo, ON N2L 3G1, Canada}
\email{dstndstn@gmail.com}

\author[0000-0002-1328-0211]{Robin Ciardullo}
\affiliation{Department of Astronomy \& Astrophysics, The Pennsylvania
State University, University Park, PA 16802, USA}
\affiliation{Institute for Gravitation and the Cosmos, The Pennsylvania
State University, University Park, PA 16802, USA}
\email{rbc3@psu.edu}

\author[0000-0001-6842-2371]{Caryl Gronwall}
\affiliation{Department of Astronomy \& Astrophysics, The Pennsylvania
State University, University Park, PA 16802, USA}
\affiliation{Institute for Gravitation and the Cosmos, The Pennsylvania
State University, University Park, PA 16802, USA}
\email{caryl@astro.psu.edu}

\author[0000-0001-6047-8469]{Ann Zabludoff}
\affiliation{Astronomy Department and Steward Observatory, University of Arizona, 933 North Cherry Avenue, Tucson, Arizona, 85721, USA}
\email{aiz@arizona.edu}

\author[0000-0002-9176-7252]{Vandana Ramakrishnan}
\affiliation{Department of Physics and Astronomy, Purdue University, 525 Northwestern Avenue, West Lafayette, IN 47907, USA}
\email{ramakr18@purdue.edu}

\author[0000-0002-9811-2443]{Nicole M. Firestone}
\affiliation{Department of Physics and Astronomy, Rutgers, the State University of New Jersey, Piscataway, NJ 08854, USA}
\email{nmf82@physics.rutgers.edu}

\author[0009-0003-5244-3700]{Ethan Pinarski}
\affiliation{Department of Physics and Astronomy, Purdue University, 525 Northwestern Avenue, West Lafayette, IN 47907, USA}
\email{epinarsk@purdue.edu}

\author[0000-0002-0112-5900]{Seok-jun Chang}
\affiliation{Max-Planck-Institut f\"{u}r Astrophysik, Karl-Schwarzschild-Stra$\beta$e 1, 85748 Garching b. M\"{u}nchen, Germany}
\email{sjchang@MPA-Garching.MPG.DE}

\author[0000-0002-4902-0075]{Lucia Guaita}
\affiliation{Universidad Andres Bello, Facultad de Ciencias Exactas, Departamento de Fisica y Astronomia, Instituto de Astrofisica, Fernandez Concha 700, Las Condes, Santiago RM, Chile}
\affiliation{Millennium Nucleus for Galaxies (MINGAL)}
\email{lucia.guaita@gmail.com}

\author[0000-0001-9991-8222]{Sungryong Hong}
\affiliation{Korea Astronomy and Space Science Institute, 776 Daedeokdae-ro, Yuseong-gu, Daejeon, Korea 34055}
\email{shongscience@kasi.re.kr}

\author[0000-0003-3428-7612]{Ho Seong Hwang}
\affiliation{Astronomy Program, Department of Physics and Astronomy, Seoul National University, 1 Gwanak-ro, Gwanak-gu, Seoul 08826, Republic of Korea}
\affiliation{SNU Astronomy Research Center, Seoul National University, 1 Gwanak-ro, Gwanak-gu, Seoul 08826, Republic of Korea}
\email{hwang.ho.seong@gmail.com}

\author[0009-0003-9748-4194]{Sang Hyeok Im}
\affiliation{Astronomy Program, Department of Physics and Astronomy, Seoul National University, 1 Gwanak-ro, Gwanak-gu, Seoul 08826, Republic of Korea}
\affiliation{Korea Institute for Advanced Study, 85 Hoegi-ro, Dongdaemun-gu, Seoul 02455, Republic of Korea}
\email{sanghyeok.im97@gmail.com}

\author[0000-0002-2770-808X]{Woong-Seob Jeong}
\affiliation{Korea Astronomy and Space Science Institute, 776 Daedeokdae-ro, Yuseong-gu, Daejeon, Korea 34055}
\affiliation{Korea National University of Science and Technology, Daejeon, Korea 34113}
\email{jeongws@kasi.re.kr}

\author[0009-0007-9489-8278]{Eunsoo Jun}
\affiliation{Korea Astronomy and Space Science Institute, 776 Daedeokdae-ro, Yuseong-gu, Daejeon, Korea 34055}
\affiliation{Korea National University of Science and Technology, Daejeon, Korea 34113}
\email{esjun@kasi.re.kr}

\author[0009-0002-3931-6697]{Seongjae Kim}
\affiliation{Korea Astronomy and Space Science Institute, 776 Daedeokdae-ro, Yuseong-gu, Daejeon, Korea 34055}
\affiliation{Korea National University of Science and Technology, Daejeon, Korea 34113}
\email{seongjkim@kasi.re.kr}

\author[0000-0002-6810-1778]{Jaehyun Lee}
\affiliation{Korea Astronomy and Space Science Institute, 776 Daedeokdae-ro, Yuseong-gu, Daejeon, Korea 34055}
\email{jaehyun@kasi.re.kr}

\author[0000-0001-5342-8906]{Seong-Kook Lee}
\affiliation{Astronomy Program, Department of Physics and Astronomy, Seoul National University, 1 Gwanak-ro, Gwanak-gu, Seoul 08826, Republic of Korea}
\affiliation{SNU Astronomy Research Center, Seoul National University, 1 Gwanak-ro, Gwanak-gu, Seoul 08826, Republic of Korea}
\email{s.joshualee@gmail.com}

\author[0000-0002-0905-342X]{Gautam Nagaraj}
\affiliation{Laboratoire d’Astrophysique, EPFL, 1015 Lausanne, Switzerland}
\email{gautam.nagaraj@epfl.ch}

\author[0000-0002-7356-0629]{Julie B. Nantais}
\affiliation{Facultad de Ciencias Exactas, Departamento de Física y Astronomía, Instituto de Astrofísica, Universidad Andrés Bello, Fernández Concha 700, Edificio C-1, Piso 3, Las Condes, Santiago, Chile}
\email{julie.nantais@unab.cl}

\author[0000-0001-9850-9419]{Nelson Padilla}
\affiliation{Instituto de Astronomía Teórica y Experimental, CONICET-Universidad Nacional de Córdoba, Laprida 854, X5000BGR, Córdoba, Argentina}
\email{n.d.padilla@gmail.com}

\author[0000-0001-9521-6397]{Changbom Park}
\affiliation{Korea Institute for Advanced Study, 85 Hoegi-ro, Dongdaemun-gu, Seoul 02455, Republic of Korea}
\email{cbp@kias.re.kr}

\author[0000-0002-4362-4070]{Hyunmi Song}
\affiliation{Department of Astronomy and Space Science, Chungnam National University, 99 Daehak-ro, Yuseong-gu, Daejeon, 34134, Republic of Korea}
\email{hmsong@cnu.ac.kr}

\author[0000-0001-6162-3023]{Paulina Troncoso}
\affiliation{Escuela de Ingeniería, Universidad Central de Chile, Avenida Francisco de Aguirre 0405, 171-0614 La Serena, Coquimbo, Chile}
\email{p.troncoso.iribarren@gmail.com}

% AUTHORS HERE



\begin{abstract}
\lya blobs (LABs) are large, spatially extended \lya-emitting objects whose nature remains unclear. Their statistical properties such as number densities and luminosity functions are still uncertain because of small sample sizes and large cosmic variance. The One-hundred-deg$^2$ DECam Imaging in Narrowbands (ODIN) survey, with its large volume, offers an opportunity to overcome these limitations. We describe our LAB selection method and present 112 new LABs in the 9 \sqdeg E-COSMOS field. We begin with the conventional LAB selection approach, cross-matching LAEs with extended \lya sources, yielding 89 LAB candidates. To obtain a more complete LAB sample, we introduce a new selection pipeline that models all galaxies detected in deep broadband imaging, subtracts them from the narrowband image, and then directly detects extended \lya emission. This method successfully identifies 23 additional low-surface-brightness LABs which could otherwise be missed by the conventional method. The number density of ODIN LABs near an ODIN protocluster ($n=7.5\times10^{-5}$ \unitcovolume) is comparable to that found in the SSA22 proto-cluster and is four times higher than the average across the field. The cumulative \lya luminosity function within the protocluster regions is similar to that measured for the LABs in the SSA22 proto-cluster, suggesting a large excess of luminous LABs relative to the average field. These findings suggest the \lya luminosities and number densities of LABs are environment-dependent. ODIN will provide an expansive LAB and protocluster samples across six additional fields and two more redshifts, allowing us to investigate the nature of LABs in relation to their environments.
\end{abstract}

\keywords{\uat{Galaxies}{573} --- \uat{Intergalactic medium}{813} --- \uat{Lyman-alpha galaxies}{978} --- \uat{Circumgalactic medium}{1879} --- \uat{High-redshift galaxy clusters}{2007}}
 


\section{Introduction} \label{sec:intro}
\lya blobs (LABs or ``nebulae'') are extended halos of \lya emitting gas at $z$ $\sim$ 2--6 that are typically $\gtrsim$\,5\arcsec\ ($\gtrsim$\,50\,pkpc) with line luminosities of $L_{\rm{Ly\alpha}}\gtrsim10^{43}$\,\unitcgslum \citep{Keel99, Steidel00, Francis01, Palunas04, Matsuda04, Dey05, Saito06, Yang09, Yang10, Erb11, Mingyu24}.
Originally introduced by \cite{Steidel00} to describe extended \lya-emitting objects lacking obvious AGN or UV continuum sources, the term ``LAB'' is now widely used to refer to spatially extended \lya sources with characteristic luminosity and sizes described above. In this work, we adopt the broader definition. 

More generally, extended \lya emitting objects have a wide range of sizes and luminosities \citep[][for a review]{Ouchi20}.
\lya halos (LAHs) surrounding identifiable galaxies have \lya luminosities of $L_{\rm{Ly\alpha}} \lesssim 10^{43}$ \unitcgslum and smaller sizes ($<$30 pkpc). LAHs are thought to be powered not only by star formation, but also by mechanisms such as gas cooling and fluorescence \citep{Momose16, Leclercq17, Kusakabe22}. 
On the brighter end, enormous \lya nebulae reach $L_{\rm{Ly\alpha}}\sim10^{45}$ \unitcgslum and are illuminated by QSOs \citep{Cantalupo14, Hennawi15, Cai17, Fabrizio18a}.

The nature of LABs, which lie between these two extremes, is not well understood. Several mechanisms have been proposed to explain their formation, potentially acting together: star-formation within embedded galaxies \citep{Geach16, Kato18} photoionization by AGNs \citep{Dey05, Geach09, Yang14b}, and gravitational cooling \citep{Rosdahl12, Ao20, Daddi20, Fabrizio22}. Resonant scattering may play an important role in the extent of \lya emission \citep{Hayes11, You2017, Eunchong20}.

LABs appear to be closely tied to their environments, e.g., overdensities of compact \lya emitting galaxies (LAEs) \citep{Matsuda04, Matsuda11, Yang10, Prescott08, Prescott12a, Kikuta19, Zhang25}. 
For example, \cite{Badescu17} has shown that LABs often reside on the outskirts of LAE overdensities, which suggests that LABs trace infalling substructures or protogroups and that the extended \lya emitting gas represents a proto-intragroup medium and/or gas stripped through galaxy interactions.
\citet{Erb11} report an potential alignment of LABs along large-scale filaments in the HS1700 protocluster at $z$ = 2.3. More recently,
\citet{Umehata19} and \citet{Ramakrishnan23} demonstrate that LABs live in and near cosmic filaments.
Because of their link to large-scale structure, their complex nature, and their extensive gaseous halos, LABs are laboratories for studying hierarchical structure formation and the interplay between galaxies and the circumgalactic medium (CGM) and intergalactic medium (IGM).

To date, many LABs have been discovered through narrowband surveys with a variety of filters ($2<z<7$), survey fields, surface brightness limits and survey areas \citep{Francis01, Palunas04, Matsuda04, Matsuda11, Saito06, Yang09, Yang10, Erb11, Shibuya18, Kikuta19, Zhang20, Mingyu24}. 
Because LABs are rare and exhibit strong field-to-field variation \citep{Yang10}, their statistical properties---such as number densities, luminosity functions,  clustering properties---remain poorly constrained \citep[e.g.,][]{Saito06, Shibuya18, Zhang25}.
Yet these properties are essential for understanding the dark matter halos of LABs, including their masses and occupation fractions, as shown for LAEs \citep[e.g.,][]{Hong19, Umeda25, Herrera25}.
Analyzing these properties across all known LABs has been challenging because of inconsistencies in selection techniques and surface brightness. To robustly characterize LAB statistics, a larger sample from a uniform, well-defined survey is essential.

The One-hundred-deg$^2$ DECam Imaging in Narrowbands (ODIN) survey, a NSF Optical-Infrared Laboratory (NOIRLab) large survey program, is currently the most extensive survey in both area and volume using three narrowband filters corresponding to the three cosmic slices ($z\sim2.4$, 3.1 and 4.5).
Thus, ODIN overcomes past limitations and provides an opportunity to explore the nature of LABs within the context of large-scale structure, discovering more than 65,000 LAEs, 800 LABs, and about 42 Coma-like clusters over three redshifts \citep{Soo23}. 

The extensive coverage of the ODIN survey ($\sim$100 \sqdeg) necessitates efficient methods for searching for rare objects like \lya blobs. In this paper, we employ a hybrid approach to identify \lya blobs that addresses the challenge of managing numerous spurious detections while simultaneously pushing the detection threshold to include fainter surface brightness objects.

This paper is organized as follows. 
In Section~\ref{sec:data}, we review our ODIN data and describe the Gemini/GMOS observations used for spectroscopic confirmation.
In Section~\ref{sec:LAB selection}, we present our two LAB selection methods that are applied to all ODIN data. 
In Section~\ref{sec:result}, we report the ODIN LAB samples and their spectroscopic confirmation. We present statistical analyses for the LABs, including the luminosity function and number density evolution. 
Throughout the paper, we assume a flat $\Lambda$CDM cosmology, with $\Omega_{\rm M}=0.3$, $\Omega_{\Lambda}=0.7$, and $H_0$ = 70\,\kms\,Mpc$^{-1}$.

%----------------------------------------------------------------------
% Table
\begin{deluxetable*}{lccccl}[t]
\tablenum{1}
\tablecaption{Summary of Gemini/GMOS observation \label{tab:GMOS_obs}}
\tablewidth{0pt}
\tabletypesize{\scriptsize}
\tablehead{
\colhead{LAB ID}  & 
\colhead{Slit width [\arcsec]}  & 
\colhead{$R$\tablenotemark{\footnotesize 1}}  & 
\colhead{$T_{\mathrm{exp}}$ [s]} & %Exposure Time [s]
\colhead{N(LAEs)\tablenotemark{\footnotesize 2}} &
\colhead{Note} 
}
\startdata
ODIN-COSMOS-z3p1-LAB004 & 1.0 & 844 & 7000 & 8 & MOS mode; protocluster\tablenotemark{\footnotesize 3} \\ 
ODIN-COSMOS-z3p1-LAB012 & 1.5 & 562 & 6000 & 4 & MOS mode \\ 
ODIN-COSMOS-z3p1-LAB014 & 1.5 & 562 & 6600 & \nodata & Longslit mode; \lya on the bad column \\ 
ODIN-COSMOS-z3p1-LAB018 & 1.5 & 562 & 5100 & 4 & MOS mode \\ 
ODIN-COSMOS-z3p1-LAB031 & 2.0 & 422 & 7200 & 8 & MOS mode; protocluster\tablenotemark{\footnotesize 3} \\ 
ODIN-COSMOS-z3p1-LAB050 & 2.0 & 422 & 7200 & 8 & MOS mode; protocluster\tablenotemark{\footnotesize 3} \\ 
\enddata
\tablecomments{Spectral redshifts of LABs are included in Table~\ref{tab:lab_cat}.}
\tablenotetext{1}{Spectral resolution.}
\tablenotetext{2}{Number of LAEs in the same MOS mask with each ODIN LAB.}
\tablenotetext{3}{COSMOS-z3.1-C in COSMOS field from \citet{Ramakrishnan23} and \citet{Ramakrishnan25b}.}
\end{deluxetable*}
%----------------------------------------------------------------------

\section{Data and Observation} \label{sec:data}
In this work, we use ODIN's $N501$ narrowband data obtained with the Dark Energy Camera \citep[DECam;][]{Flaugher15}, along with $g$ and $r$ broadband images from the Hyper Suprime-Cam Subaru Strategic Program (HSC SSP) second public data release \citep{Aihara19}.
The narrowband image covers $\sim$12 \sqdeg of the extended COSMOS field with a seeing of 0\farcs9 on a pixel scale of 0\farcs27. The image has a nearly uniform 5$\sigma$ depth of 25.6 mag in the central 10 \sqdeg \citep{Ramakrishnan23}.
The narrowband filter has a central wavelength of $\lambda_c=5014$ \AA~and bandwidth of $\Delta\lambda=76$ \AA, designed to be sensitive to redshifted \lya emission at $z\sim3.1$ (3.09 $<z<$ 3.15, corresponding to 57 cMpc).
We refer readers to \cite{Soo23} for an overview of the ODIN program (e.g., narrowband filter specifications, survey strategy, field definitions, and data products), and \citet{Ramakrishnan23} and \citet{Firestone24} for details on the dataset used for LAB and LAE selection at $z\sim3.1$.

Although the background subtraction procedure is described in \cite{Soo23}, we elaborate on it here because of its importance for the detection of low–surface-brightness objects. The large scale background across the field of view is found by fitting a template for a camera reflection pattern and sky gradients by a surface fit. Smaller scale variations within a CCD are determined by detecting and masking sources and measuring the background in blocks of $256\times256$ pixels ($70\times70$ arcsec). The blocks are linearly interpolated to every pixel. The chosen block size is large enough to sample the surrounding background of each LAB while avoiding over-subtraction at the blob position and preserving its detection.

We report spectroscopic observations conducted to confirm the LABs.
We observed six LAB candidates using Gemini/GMOS \citep{GMOS} in long-slit mode during the 2021A semester as a Fast Turnaround (FT) program, and in Multi-Object Spectroscopy (MOS) mode during the 2021B, 2023A and 2025A semesters. The primary objective of these programs is to confirm the presence of \lya emission lines from ODIN LAB candidates by identifying characteristic \lya profiles such as double-peaked or asymmetric profile shapes. We used the B600 grating to obtain moderate sensitivity over a wide spectral coverage (between 4500\,\AA\ and 6800\,\AA) including \lya, \civ and \heii at $z\sim3.1$.

The FT long-slit program targeted ODIN-COSMOS-z3p1-LAB014, the first LAB discovered through our visual inspection. 
However, the peak of \lya emission coincided with a bad column, which enabled redshift confirmation of the LAB but prevented any analysis of the line profile.
The MOS mode programs employed five masks: two targeted bright LAB samples, and three focused on LAB candidates in one of the ODIN protoclusters (COSMOS-z3.1-C; \citealt{Ramakrishnan23}). Each MOS mask includes at least one LAB and 4 -- 8 LAE candidates. Fifteen out of 24 LAEs were confirmed through narrow \lya profiles and no emission lines were detected in the spectra of the remaining sources that are mostly fainter. Details of the LAE and proto-cluster confirmation are described in \cite{Ramakrishnan25b}. Table~\ref{tab:GMOS_obs} summarizes our targets, slit widths, spectral resolution, exposure times, and the number of LAEs per mask.

We reduced the data using the {\tt Gemini IRAF} package and successfully confirmed the \lya emission from all the targeted LAB candidates. We report the spectroscopic confirmation in Section~\ref{sec:spec}.

\section{Selection of \lya blob candidates} \label{sec:LAB selection}
The precise definition of LAB varies greatly depending on the depth of the data and the selection strategy.
In their pioneering work, \cite{Matsuda04} identifies \lya blobs based on their size being larger than point sources in \lya images, regardless of the presence of embedded compact LAEs. Thus, due to the combination of this selection method and the depth provided by Subaru telescope, their sample includes LABs with exceptionally diffuse \lya emission. 
Alternatively, \cite{Yang09, Yang10}, \cite{Badescu17}, \cite{Shibuya18},  \cite{Zhang20} and \cite{Mingyu24} define \lya blobs as LAEs with \lya sizes larger than those of point sources at given \lya luminosities. Similarly, \cite{Saito06} and \cite{Ouchi09} identify \lya blobs as sources with sizes in narrow- or intermediate-band images larger than their point spread functions (PSF).

We adopt two complementary methods for selecting \lya blobs using the ODIN {\sl N501} dataset. The first is a traditional approach based on cross-matching extended \lya sources with LAEs, as demonstrated in previous studies \citep{Yang09, Yang10, Shibuya18}. To extend our search to diffuse LABs without identifiable LAE counterparts such as those found in the SSA22 field \citep{Steidel00, Matsuda04}, we develop a new selection technique that does not rely on associated LAEs. This method utilizes forced photometry with \tractor \citep{Lang16} to identify spatially extended \lya emission in the absence of compact sources (i.e., galaxies). In the following sections, we describe both the traditional and new LAB selection strategies. Throughout this paper, we refer to the traditional method as the ``extended-LAE'' method and to the new \tractor-based method as the ``\tractormethod'' method.

\subsection{Extended-LAE Method}
\label{sec:traditional_LAB_selection}
We select LABs candidates in two steps, following the approach outlined in \cite{Yang10}:
(1) identifying bright cores of LABs (i.e., LAEs), and 
(2) detecting extended \lya emission around these galaxies in the continuum-subtracted \lya images.

We first construct an LAE catalog to cross-match with extended \lya emission using 
2\arcsec\ diameter aperture, comparable to the seeing of the image, photometry from \stractor \citep{Bertin96}. 
We adopt following criteria for LAE selection as \citet{Yang10} and \citet{Badescu17}:
(1) $N501 < 25.6$, 
(2) $gr - N501 > 0.8$, corresponding to the rest-frame equivalent width cut, EW$_\mathrm{rest} > 20$\,\AA, and
(3) $gr - N501>3\sigma_{gr - N501}$, where the $\sigma_{gr - N501}$ is uncertainty of the color ($gr - N501$).
The $gr$ represents the continuum only magnitude corresponding to the continuum only flux density ($f^{N\!B}_\mathrm{cont}$) in Equation~\ref{eq:continnum-subtraction},
\begin{eqnarray} \label{eq:continnum-subtraction}
f^{B1}_\mathrm{cont}    &=& \frac{F_{B1} - \epsilon_{B1} F_{N\!B}}{\Delta\lambda_{B1} - \Delta\lambda_{N\!B}}; \\
\nonumber
f^{B2}_\mathrm{cont}    &=& \frac{F_{B2} - \epsilon_{B2} F_{N\!B}}{\Delta\lambda_{B2} - \Delta\lambda_{N\!B}}; \\
\nonumber
f^{N\!B}_\mathrm{cont} &=& \frac{(\lambda_{B2} - \lambda_{N\!B}) f^{B1}_\mathrm{cont} +
                                 (\lambda_{N\!B} - \lambda_{B1}) f^{B2}_\mathrm{cont}}
                                {\lambda_{B2}-\lambda_{B1}}; \\
\nonumber
F_\mathrm{line}        &=& F_{N\!B}  - f^{N\!B}_\mathrm{cont} ~ \Delta\lambda_{N\!B}.
\end{eqnarray}
In the Equation, $\Delta\lambda$ and $\lambda$ denote the FWHM and the central wavelengths of the filters, respectively. We define the observed flux ($F$) as the product of $\Delta\lambda$ and flux density ($f$) to calculate continuum-only flux and line-only map. The $f^{N\!B}_\mathrm{cont}$ is estimated through linear interpolation between the two broadband fluxes. Here, $B1$ and $B2$ correspond to broadband filters whose central wavelengths are shorter and longer than those of narrowband filters, respectively. For {\sl N501}, $B1$ and $B2$ correspond to the $g$ and $r$ bands. 
To correct for the differences in central wavelength between the narrow- and broadband filters, as well as the wavelength-dependent transmission, we adopt correction factors $\epsilon$. These are defined as the ratio of the broadband mean throughput within the narrowband FWHM to that within the broadband FWHM. We adopt $\epsilon_g = 1.13$ and $\epsilon_r = 0$.
Although this LAE selection method is cruder than the improved hybrid-weighted method \citep{Firestone24}, we employ it to ensure a comparable EW in both the \lya image and the LAEs, consistent with previous works \citep{Yang09, Yang10, Badescu17}.

We create a \lya image by subtracting the continuum-only flux image from the narrow-band image as described in Equation~\ref{eq:continnum-subtraction}. Prior to continuum subtraction, the broadband images are smoothed using Gaussian kernels to match the seeing conditions of the narrowband image.

From the \lya image, we search for extended \lya emission as follows: after applying a 7-pixel 2D Gaussian filter with an FWHM of 3 pixels ($\sim$0.8\arcsec), we identify all sources with an isophotal area (A$_{\rm ISO}$), the area of pixels with SB $>$ 1.5$\sigma_{\rm SB_1}$, exceeding $\sim$3 \sqarcsec ({\tt DETECT\_MINAREA} = 42 pixels) above a surface brightness threshold of $3.3 \times 10^{-18}$~\unitcgssb. 
This SB threshold is selected based on tests using visually identified ODIN LABs. It corresponds to $1.5\sigma_{\rm SB_1}$, where $\rm \sigma_{\rm SB_1}$ is the surface brightness limit {\it per} 1 \sqarcsec aperture ($\sigma_{\rm SB_1}=2.2\times10^{-18}$ \unitcgssb).
We test the radial variation of surface brightness depth following \citet{Nagaraj25}. The depth remains nearly constant within $\pm$5\% over 1.5$^\circ$ radius, and varies within $\pm$10\% across 90\% of the survey area. Therefore, we adopt a constant surface brightness depth and detection threshold, even though the depth becomes shallower toward the edges of the survey area.

The \lya image contains numerous artifacts originating from both the narrow- and broadband images, including detector artifacts, internal reflections, and background over-subtraction  near bright stars, and galactic cirrus. 
To mitigate these artifacts, which can mimic diffuse emission in the \lya map, we apply a masking procedure to exclude these large and bright artifacts.
We first run \stractor on the $g$ and $r$ band image with {\tt DETECT\_THRESH} $=$ 3$\sigma_\mathrm{pix}$, where the $\sigma_\mathrm{pix}$ is a pixel-wise root-mean-square (RMS), and with {\tt DETECT\_MINAREA} $=$ 10,000 pixels ($\sim$700 \sqarcsec). The resulting segmentation maps from each band are then combined to create a mask for bright and extended sources.

To account for over-subtracted background regions in broadbands that can mimic \lya sources, we run \stractor on the inverted $g$ and $r$ band images with the same {\tt DETECT\_MINAREA} as above.
In this case, we calculate {\tt DETECT\_THRESH} for the negative broadband images using Equation~\ref{eq:continnum-subtraction} by fixing the $F_{\rm line}$ as \lya detection threshold and assuming that only one broadband (e.g., $g$) contains negative flux while the other has zero flux.
This allows us to derive a detection threshold in the negative broadband image that corresponds to the \lya detection level of 1.5$\sigma_{\rm SB_1}$. We repeat this for the other broadband image and all resulting masks are combined and applied during the selection process.
This eliminates spurious sources and ensure reliable identification of diffuse \lya emission.

Even after applying the mask, many artifacts persist because the masking procedure was optimized for clearly identifiable artifacts with large sizes and high detection thresholds.
In previous studies with smaller survey areas, such features could be manually rejected through visual inspection. However, this approach is not feasible for the $\sim$100 deg$^2$ area covered by ODIN. Cross-matching the identified extended sources with compact LAEs solves this problem, as it provides an effective way to eliminate artifacts. A caveat, however, is that this method may miss \lya blobs without LAE counterparts, particularly those that are diffuse and have low surface brightness. The next section describes our approach to recovering these diffuse sources.

%--------------------------------------------------------
\begin{figure}
\includegraphics[width=0.47\textwidth]{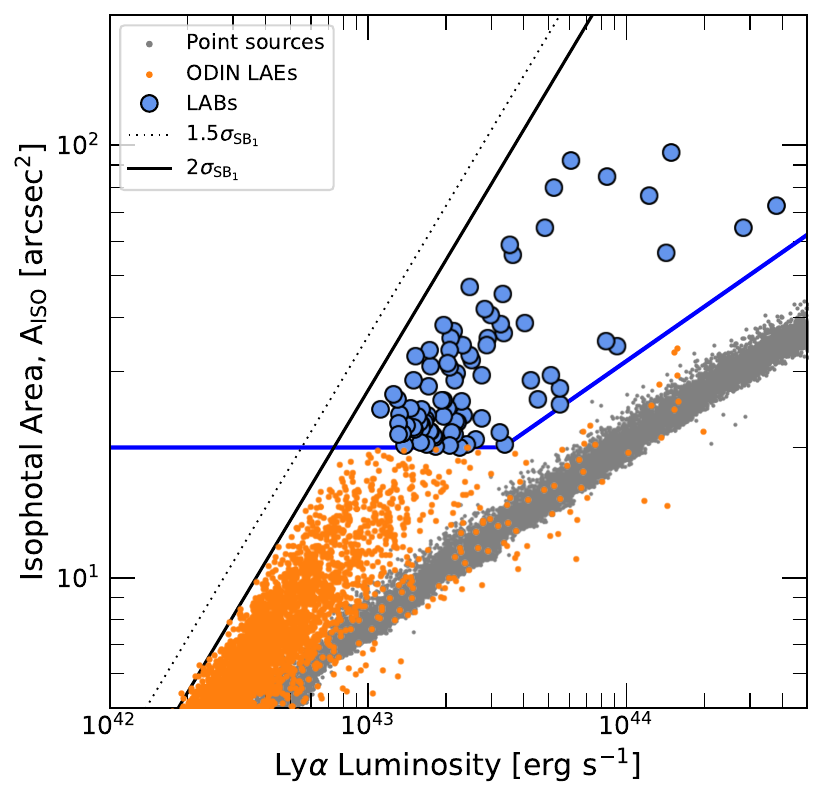}
\caption{
Size-luminosity relation for LAB candidates selected using the extended-LAE selection method. Blue circles represent the LAB candidates. 
The ODIN LAE sample \citep[orange dots;][]{Firestone24} shows two sequences. One extends to LABs implying that they could be LAH candidates, while the other follows the simulated point sources, indicating that our LAB candidates are distinct from bright and compact point sources.
Black dotted and solid lines represent 1.5$\sigma_{\rm SB_1}$ and 2$\sigma_{\rm SB_1}$, respectively, where $\sigma_{\rm  SB_1}$ represents the detection threshold per 1 arcsec$^2$. The blue solid lines outline the LAB selection criteria: size over 20\;\sqarcsec and 3$\sigma$ above the size-luminosity relation defined by the simulated point sources.
\label{fig:size_lum}}
\end{figure}
%--------------------------------------------------------

Among the LAB candidates with LAE counterparts, bright point-like sources can appear as extended \lya emitters, as their apparent sizes increase above certain surface brightness thresholds. 
To ensure that we distinguish true extended sources from bright point sources, we compare the size-luminosity distribution of LAB candidates with that of simulated point sources following \cite{Yang10}. 
To establish the size-luminosity relation for point sources, we inject approximately 10,000 simulated point sources into empty regions of the \lya image. 
These simulated sources span luminosities from $10^{42}$ to $10^{45}$~\unitcgslum\ and are placed to avoid masked areas and existing sources. We measure their sizes and luminosities using the same photometry parameters applied to the LAB candidates.

Figure~\ref{fig:size_lum} shows the size-luminosity distribution of the cross-matched LAB candidates, ODIN LAEs and simulated point sources. 
We isolate 89 LAB candidates (blue circles) with isophotal areas larger than 20 \sqarcsec\ that lie more than 3$\sigma$ above the size–luminosity relation defined by the simulated point sources, where $\sigma$ is the RMS scatter of the simulated point-sources. 
The independently identified ODIN LAEs \citep[orange dots;][]{Firestone24} shows two sequences. One extends to the LAB size-luminosity relation indicating that these sources could be LAH candidates \citep{Ouchi20}, while the other follows the simulated point sources (grey dots), confirming the reliability of our point source simulation.
Some ODIN LAEs are not selected as LABs even though they satisfy the selection criteria. We note that most of these are identified as LABs using the alternative LAB selection method (Section~\ref{sec:discuss_method}).

In Appendix~\ref{app:SB_profile}, we further test whether LAB candidates near the size-luminosity relation of point sources truly exhibit more extended \lya emission than LAEs with similar \lya luminosities. We compare the surface brightness profiles of LAB candidates located just above the point source relation with those of LAEs lying within the relation. The LAB candidates show extended surface brightness profiles than LAEs, lending support to the validity of our LAB selection method.

\subsection{A New Selection Technique for Diffuse LABs: \tractormethod Method} 
\label{sec:new_labselcection}

\begin{figure}
\includegraphics[width=0.47\textwidth]{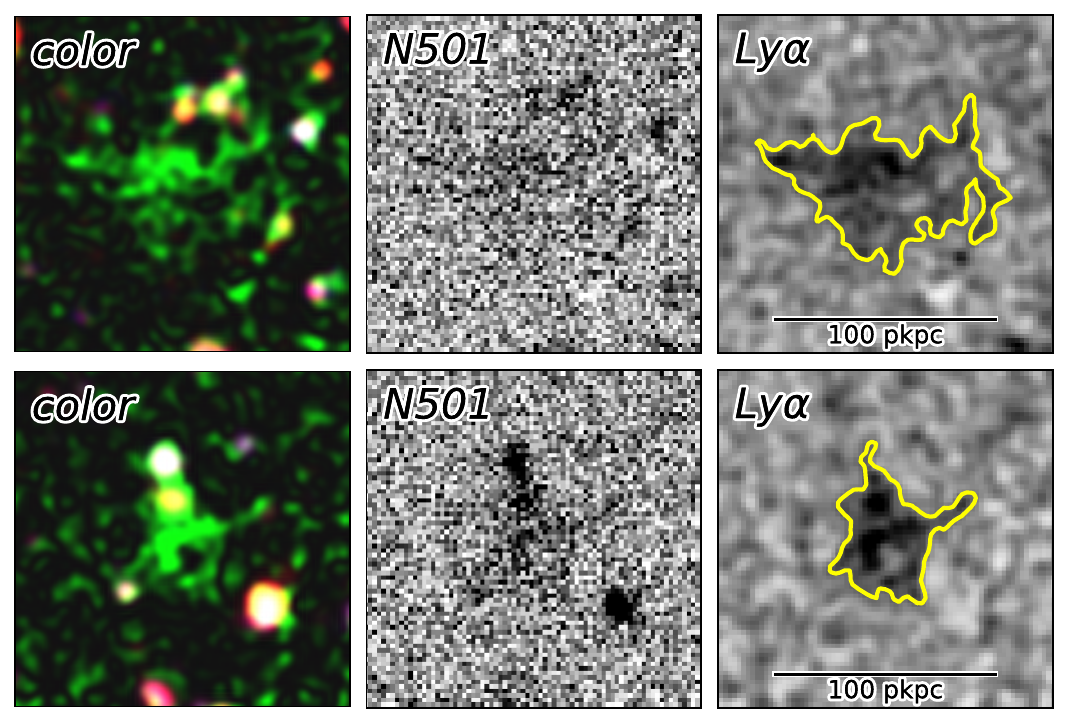}
\caption{Postage stamp ($20\times20$ \sqarcsec) images for examples of diffuse and extended \lya halos identified by our visual inspection.  These objects are not selected as LAB in extended-LAE method, requiring additional LAB selection is necessary for assembling a complete LAB samples.
{\bf Left:} Color image constructed with $N501$, $g$, and $r$ bands. Green shows the diffuse and extended signal in the $N501$ image.
{\bf Middle:} High contrast $N501$ image revealing the diffuse and extended source.
{\bf Right:} Filtered \lya image. Yellow contours ($1.5\sigma_{\rm SB1}=3.3\times10^{-18}$ \unitcgssb) highlighting the diffuse and extended \lya emission.
\label{fig:tractor_need}}
\end{figure}

While the \traditional method effectively identifies a wide range of \lya blobs, our extensive visual inspection of the ODIN data revealed diffuse and extended \lya emission (Figure~\ref{fig:tractor_need}) that were not recovered by the selection method described in Section~\ref{sec:traditional_LAB_selection}. 
These diffuse sources resemble SSA22-LAB18 \citep{Matsuda04}, which shows low surface brightness without any compact LAE counterparts.
When attempting to identify these sources and construct a complete LAB sample, it is necessary to adopt a relatively aggressive detection threshold for both LAEs and LABs. However, this inevitably increases the number of spurious detections, thereby requiring extensive visual inspection due to various artifacts arising from the construction of the continuum-subtracted image (e.g., negative backgrounds from broadband images) over a large survey area. To mitigate this issue and enable source detection without excessive manual inspection, we introduce a new LAB selection method.

The key feature of the new pipeline is the subtraction of all galaxies detected in the broadband image from the narrowband image, enabling direct detection of the remaining extended emission. 
This approach is conceptually similar to the wavelet decomposition technique \citep{Prescott12b},
but we extend it by incorporating multi-band information.
To efficiently and accurately remove known galaxies from the narrowband image, we adopt a forced-photometry tool, \tractor. This tool includes modules for optimizing galaxy shapes and brightness, constructing model images, and performing forced photometry across multiple bands \citep{Lang16, Lang16b, Dey19}. Using \tractor, we extract galaxy model parameters from a broadband image and subtract the corresponding galaxy models from the narrowband image. 
By using this NB residual image---which is largely free from artifacts arising from broadband subtraction---as the detection image, we can minimize the number of spurious detections, as described below.
In the following, we describe the workflow of this \tractor-based LAB selection pipeline in detail.

\paragraph{(1) Broadband source detection} 
Because \tractor requires source positions and initial shape parameters, we begin by detecting sources in the HSC SSP $g$-band image using \stractor. Detecting objects near bright sources is particularly challenging, as \stractor tends to overestimate the local background in such regions, reducing sensitivity to nearby faint sources. These undetected objects may persist in both the broadband and narrowband images and can be mistakenly identified as LABs.
To mitigate this issue, we adopt a lower detection threshold ({\tt DETET\_THRESH} $=$ 0.9 and {\tt DETECT\_MINAREA} $=$ 4 pixels), which improves completeness while introducing more false detections. However, these false detections are not a critical concern because the \tractor will assign them near-zero flux in the narrowband images, effectively excluding them from the final residual image.

\paragraph{(2) Cutout images}

\begin{figure}
\includegraphics[width=0.47\textwidth]{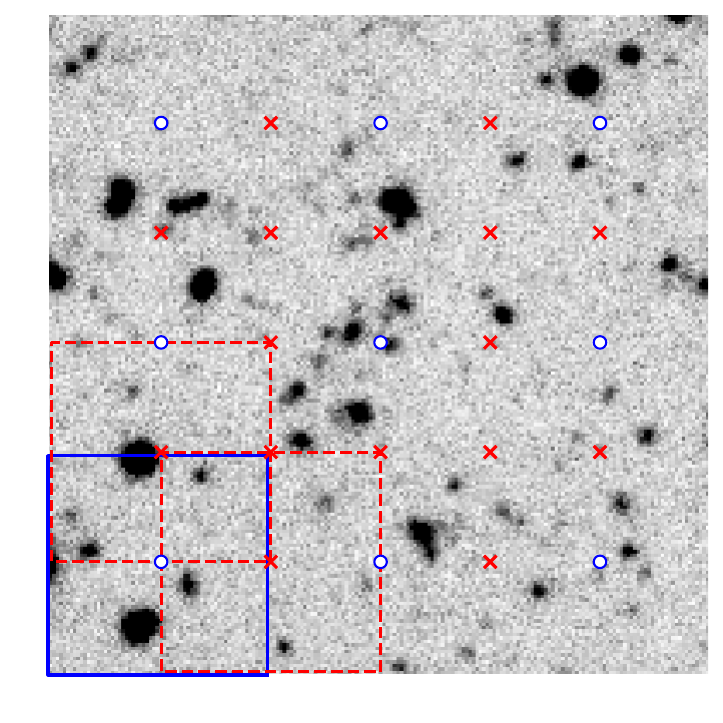}
\caption{
    Example $192\times192$ pixel$^2$ (52\arcsec$\times$52\arcsec) HSC SSP $g$ band image that illustrates our tiling strategy. We use $64\times64$ pixel$^2$ sub-tile images to perform model fitting for $g$-band galaxies. To ensure accurate model fitting and avoid fragmentation near the sub-tile edges, 
    We offset each tile by half a tile-size from its nearest adjoining tile. This pattern is illustrated by the solid blue and red dashed squares, whose centers are defined by the blue circles and red crosses, respectively. 
    \label{fig:bordering}}
\end{figure}

We cut out sky-subtracted $g$-band and $N501$ images into $64 \times 64$ pixel sub-tiles. This small tile size reduces computation time by limiting the number of galaxies modeled in each sub-tile. 
However, when a galaxy falls near the edge of a sub-tile, the \tractor modeling procedure may fail. To avoid this issue, we introduce overlapping sub-tiles with a half-tile offset in both the X and Y directions (Figure~\ref{fig:bordering}). This overlap minimizes galaxy fragmentation at tile boundaries and improves model accuracy. 

If the source size is significantly larger than 64 pixels on a side, \tractor tends to underestimate its size, often modeling the object as a point source. However, this does not critically affect our analysis, as such sources are low-redshift galaxies and will be excluded by the EW$_{\rm rest}$ criterion applied in the final step of this method.

\paragraph{(3) Extract galaxy shape}
Galaxy shapes are modeled from the broadband ($g$) image using \tractor. To optimize galaxy fits, \tractor convolves each source model with the image PSF (constructed with \psfex) and determines the best-fit model and integrated flux density by $\chi^2$ minimization. 
We adopt six different \tractor source models: {\tt PointSource}, {\tt GaussianGalaxy}, {\tt  ExpGalaxy}, {\tt DevGalaxy}, {\tt SersicGalaxy}, and {\tt FixedCompositeGalaxy} corresponding to a point source, a gaussian profile, an exponential profile, a de Vaucouleurs profile \citep{deVaucouleurs}, a Sersic profile \citep{Sersic} and the combination of ExpGalaxy and DevGalaxy models.
They are defined by position, brightness and elliptical shape. The elliptical shape is parameterized by {\tt EllipseESoft}, the sigmoid-softened ellipticities provided by \tractor.
Initial input parameters for the model fitting are taken from the \stractor output (e.g., {\tt FLUX\_AUTO}, {\tt A\_IMAGE}, {\tt B\_IMAGE}, and {\tt THETA\_IMAGE}). To avoid over-estimation of galaxy sizes, we impose a size limit of $r_e$ $<$ 10\arcsec\ (50 pixels) based on repeated tests. For the same reason noted above, this upper limit does not affect results because it will be removed in subsequent steps. If a galaxy appears in multiple overlapping sub-tiles, we keep only the model from the sub-tile in which the galaxy is closest to the tile center. 

\begin{figure}
\includegraphics[width=0.47\textwidth]{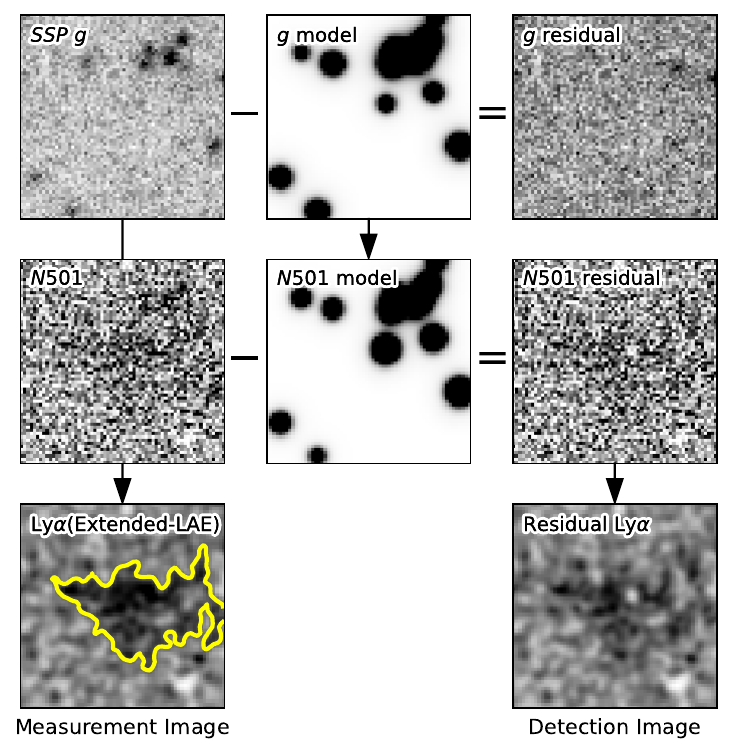}
\caption{Postage stamp images ($64\times64$ pixel$^2$) demonstrating \tractormethod method. 
\textbf{Top row:} HSC SSP $g$-band image (left), the \tractor model image of galaxies (center), and the residuals from the fits (right). The Tractor models do an excellent job reproducing the continuum emission from the galaxies. \textbf{Middle row:} the same set of images for the $N501$ frame. The galaxy models have the exact same shape parameters as for the $g$-band data. Note the excess \lya flux on the residual image. \textbf{Bottom row:} The residual \lya image which is used for the detection of LABs in the \tractormethod method (right). The image is constructed by only the residual $N501$ image. The flux and sizes are measured using the \lya image from \traditional method (left).}
\label{fig:tractor_example}
\end{figure}

\paragraph{(4) Construct $N501$ Model and Residual images}
We perform {\it forced} photometry on the $N501$ image using the source models obtained in the step \textit{(3)}. In this step, only the amplitudes of the model components are fitted, while the shapes and positions remain fixed. 
The philosophy of our technique is to remove all features attributable to galaxies, thereby minimizing stellar components while preserving extended emission from the CGM.
\tractor generates a best-fit image by convolving the source models with the  $N501$ PSF\null.
Figure~\ref{fig:tractor_example} shows an example of a \tractor model and the corresponding residual images, in which compact sources have been subtracted, leaving only the extended \lya emission. By subtracting the model image (``N501 model'') from the observed $N501$ image (``N501''), we obtain a residual map (``N501 residual'') that highlights extended emission without continuum. 
To avoid missing sources larger than the cutout size, we construct the model image to match the full coverage of the observed N501 field, and detection is performed on this image in the subsequent step.

\begin{figure}
\includegraphics[width=0.47\textwidth]{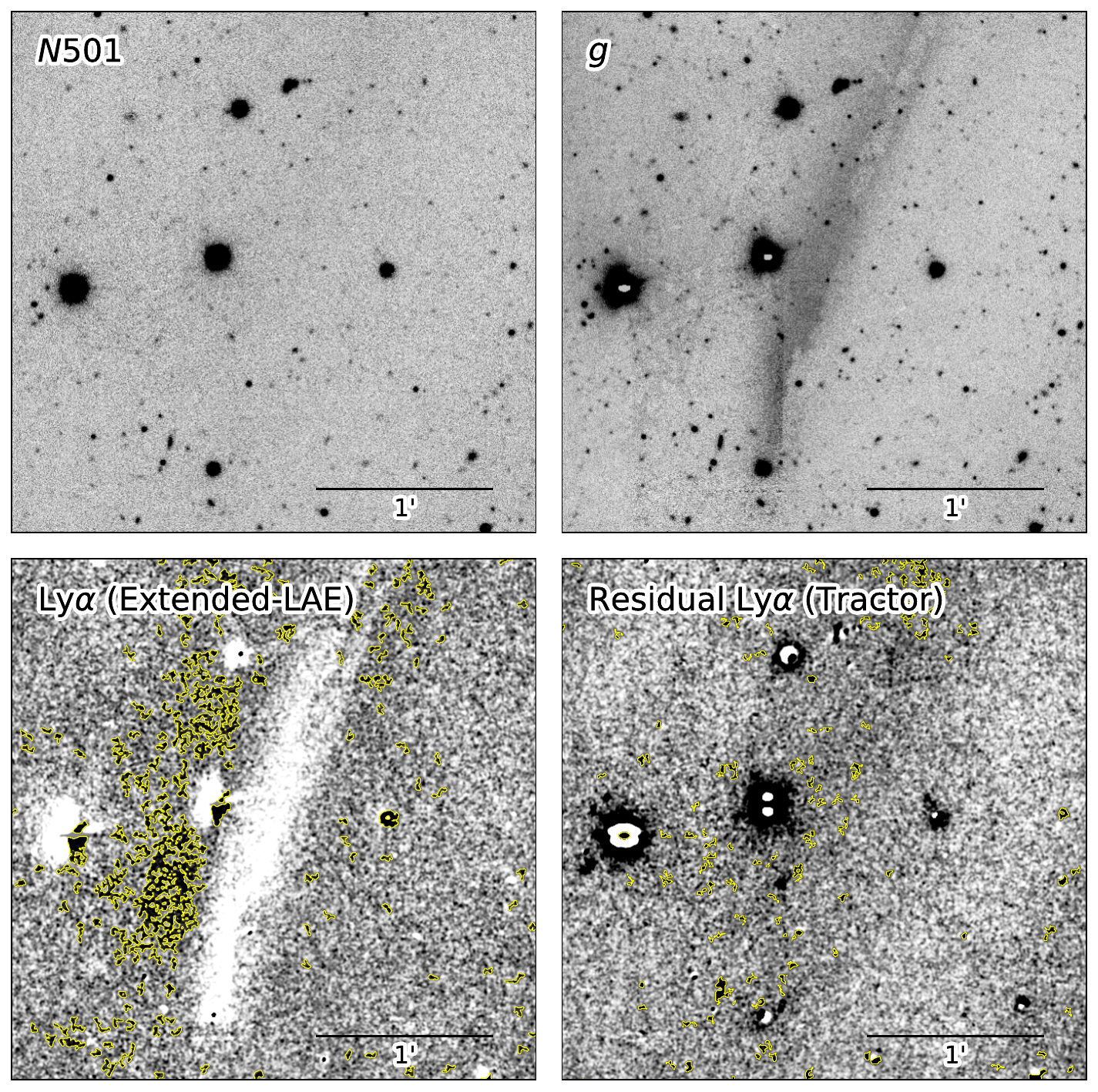}
\caption{
Cutout images ($3\arcmin\times3\arcmin$) illustrating the robustness of the \tractor-based residual \lya detection image.
{\bf Top}: $N501$ and $g$ band images. A bright internal reflection (dark stripe) is clearly visible across the $g$-band image, introducing large-scale artifacts.
{\bf Bottom}: \lya detection images from the \traditional (left) and the \tractormethod methods, 
respectively. Yellow contours indicate regions exceeding the detection threshold for \lya blob candidates identified by \texttt{SExtractor}.
In the \traditional method (Section~\ref{sec:traditional_LAB_selection}), the artifact is propagated into the \lya detection image, and the over-subtracted background near this artifact leads to numerous false detections (yellow contours). In contrast, the residual \lya\ image from \tractor is effectively free of this artifact, significantly reducing spurious detections.
\label{fig:tractor_detection_better}}
\end{figure}

\paragraph{(5) Detection and Measurement}
We convert the residual $N501$ map to the residual \lya map assuming that the continuum emission has been effectively modeled and removed.
We detect LAB candidates from this residual \lya map after filtering the image with the same Gaussian kernel and applying the same detection criteria used in the \traditional pipeline: a SB threshold of $3.3 \times 10^{-18}\,$\unitcgssb and a minimum isophotal size of 3\,\sqarcsec (see Section~\ref{sec:traditional_LAB_selection}).
We note that this model-subtracted and smoothed \lya image is used solely for source detection.
For measuring isophotal sizes and fluxes, we use the continuum-subtracted \lya-only image from Section~\ref{sec:traditional_LAB_selection} to ensure fair \lya luminosity and size measurements. 
We employ the {\tt ASSOCIATION} feature of \stractor to link detections from the residual image (from \tractor) with their corresponding measurements in the continuum-subtracted image (from \traditional method).

Figure~\ref{fig:tractor_detection_better} illustrates the advantages of using the residual \lya image from the \tractormethod method for selecting diffuse sources. 
If we use the \lya image from the \traditional method (i.e., {\sl NB} $-$ $gr$) as detection image, the over-subtracted background near the artifact (a reflected stripe in Figure~\ref{fig:tractor_detection_better}) leads to numerous false detections (yellow contours) as \stractor splits them into clumps of various sizes.
The residual \lya image from \tractor avoids this issue by relying solely on the narrowband image and proper modeling of continuum sources. As a result, the \tractormethod method significantly reduces the number of spurious detections and alleviates the need for extensive visual inspection, which inevitably introduces subjective bias.

\paragraph{(6) LAB Candidate Selection}

\begin{figure}
\includegraphics[width=0.47\textwidth]{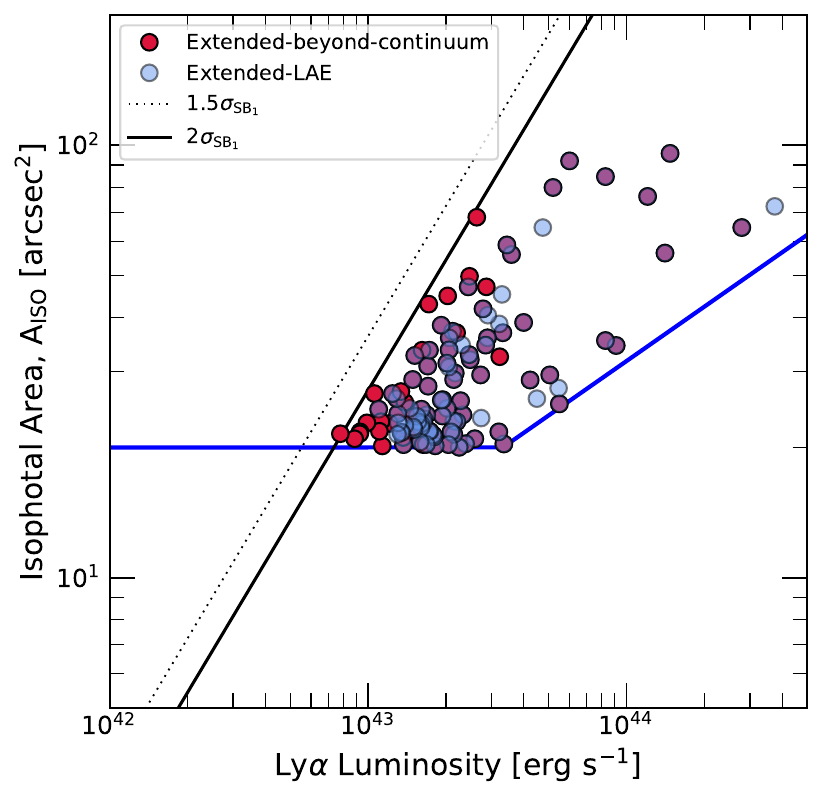}
\caption{
Same as Figure~\ref{fig:size_lum}, but for both of the selection methods.  Blue circles represent LABs identified using the \traditional pipeline described in Section \ref{sec:traditional_LAB_selection}, while red circles denote LAB candidates from the \tractormethod method. 
Of the 89 LABs from the \traditional method, 72 are also recovered by the \tractormethod pipeline, indicating the robustness of our LAB selection and the reliability of the sample. LABs identified only by the \tractormethod method tend to exhibit lower surface brightness and lie closer to the 2$\sigma_{\rm SB_1}$ limit (black solid line), suggesting that the method effectively identifies diffuse LABs missed by traditional methods.
}
\label{fig:tractor_size_lum}
\end{figure}

We select LAB candidates using the same criteria described in Section \ref{sec:traditional_LAB_selection}, with a EW$_\mathrm{rest}$ cut based on isophotal fluxes:
(1) an isophotal area larger than 20~\sqarcsec, 
(2) a $3\sigma$ measurement above that predicted from the point-source size-luminosity relation, and 
(3) EW$_\mathrm{rest}$ $>$ 20\AA, matching the equivalent width threshold used for LAEs in the \traditional pipeline.
Figure~\ref{fig:tractor_size_lum} shows the LAB candidates selected by the \tractormethod pipeline (red circles) on the size–luminosity diagram. 

The new method identifies 95 LABs, including 23 newly discovered LABs. Out of the 89 LABs identified by the \traditional method, 72 are also recovered by the \tractormethod pipeline, demonstrating the consistency and reliability of both LAB selection methods. 
On the other hand, there are 17 LABs that are not recovered by the \tractormethod method because their measured sizes fall below the minimum detection area ($< 3~\mathrm{arcsec}^2$). This sometimes occurs when forced photometry removes the galaxy component: the bright core of the Ly$\alpha$ emission in the N501 image can be over-subtracted, creating a hole in the NB–continuum image and leaving only a smaller Ly$\alpha$ halo in the residual image that does not meet the minimum area threshold. Consequently, the LAB catalogs produced by the \traditional and the \tractormethod approaches are complementary, and together they provide a more complete LAB catalog.

\subsection{Comparison of the Two Selection Methods} \label{sec:discuss_method}
%-------------
\begin{figure}
\includegraphics[width=0.47\textwidth]{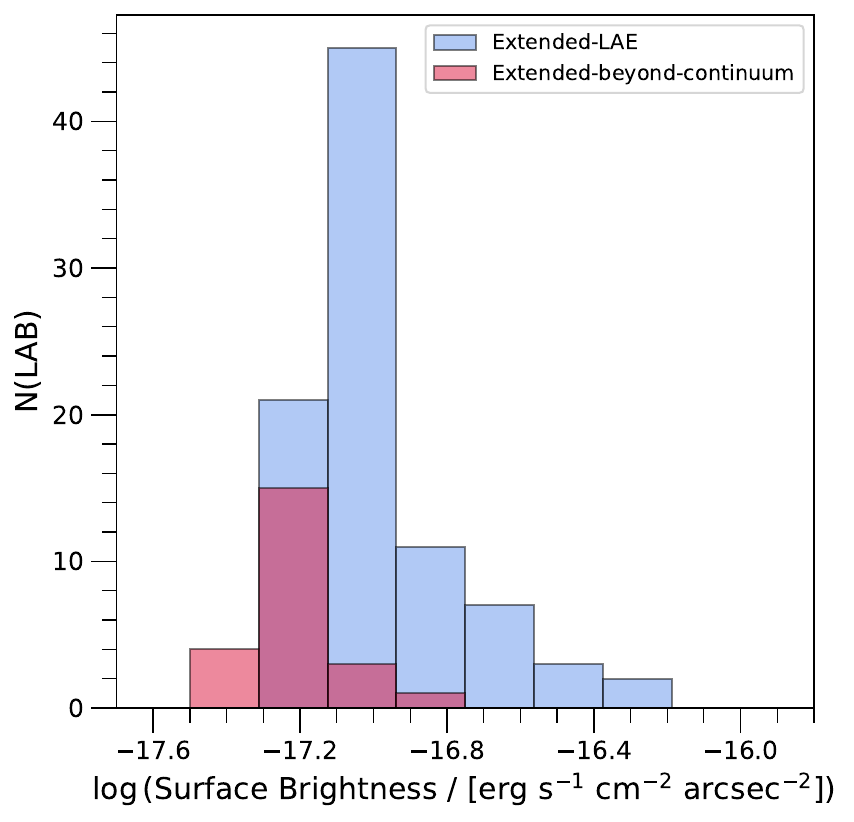}
\caption{Average surface brightness within the isophotes for the \traditional pipeline and the \tractormethod pipeline. The bin size (0.1875 in a log) is defined following \cite{Hogg08}.
The LABs recovered only with the \tractormethod pipeline (red) have lower surface brightnesses.
}
\label{fig:SB_hist}                                                                 
\end{figure}
%-------------

To compare the LAB samples selected by the \traditional\ and \tractormethod methods, we compute the average surface brightness of each sample. As shown in Figure~\ref{fig:SB_hist}, the \tractormethod pipeline is effective at identifying LABs with low surface brightness. Notably, out of 112 total LABs, 23 (20\%) are uniquely identified by the \tractormethod pipeline.

The \tractormethod method is also capable of detecting dual-LAE systems (e.g., Figure~\ref{fig:RGB_LAB}, LAB024), which are typically missed by the \traditional pipeline. In these cases, \texttt{SExtractor} deblends the dual-LAE into two separate compact LAEs, artificially reducing the apparent size of the extended Ly$\alpha$ emission. 
In contrast, in the \tractormethod pipeline, the continuum sources (i.e., the primary source peaks) are effectively removed from the detection image by modeling. Therefore the de-blending is not triggered. The \tractormethod pipeline detects the diffuse, extended Ly$\alpha$ halo directly, thus preserving the full extent of the emission and enabling the identification of this interesting class of LABs.

In terms of pipeline complexity and required human effort, the visual inspection of low-surface-brightness features is substantially less demanding in the \tractormethod approach. For example, in the \traditional method, the initial size–luminosity cut yields $\sim$10,000 extended \textit{features}, which must then be cross-matched with LAEs to reduce the sample to $\sim$280 candidates, followed by visual inspection to obtain the final sample of 89 LABs. In contrast, the \tractormethod method reduces the initial number of extended features to only $\sim$250, smaller by a factor of $\sim$40 compared to the \traditional method, thereby greatly simplifying the visual inspection process.

On the other hand, LABs identified by the \traditional\ method have been spectroscopically confirmed over the past two decades \citep[e.g.,][]{Yamada12, Yang11, Yang14b, Shibuya18, Zhang20, Mingyu24}, making them a reliable benchmark for validating the performance of the \traditional\ pipeline. However, the method is less sensitive to low surface brightness LABs and tends to miss dual-LAE systems.

To construct a more complete and robust LAB sample while minimizing the need for intensive visual inspection and mitigating artifacts in large survey areas, we adopt both selection methods in parallel.

\subsection{Recovery Test and Uncertainties \label{sec:recovery}}

To evaluate the reliability of detections and quantify uncertainties in the sizes and luminosities of LAB candidates, we conduct a recovery test following the method of \cite{Yang09}. We extract a smoothed \lya postage stamp image ($25 \times 25$\,\sqarcsec) for each LAB candidate and inject it into $\sim$4,000 randomly selected empty sky regions within the \lya image. 
The empty sky regions are defined as the survey areas that avoid detected sources and masked regions, based on the \stractor segmentation map and the masks described in Section~\ref{sec:traditional_LAB_selection}. 
We then perform the same isophotal photometry and apply the LAB selection criteria described in Section \ref{sec:traditional_LAB_selection} to the injected sources.

We evaluate how many of the injected mocks are successfully recovered based on the size--luminosity selection criteria (blue line in Figure~\ref{fig:tractor_size_lum}). The recovery fraction (\frecv) for each LAB is defined as the ratio of recovered mocks to the total number of injected mocks.
To eliminate outliers caused by unmasked faint halos of bright sources or over-subtracted sky regions, we apply a 3$\sigma$ clipping algorithm to the distribution of recovered luminosities prior to computing \frecv.
Variations in surface brightness depth across the image lead to fluctuations in the apparent sizes of LABs, which in turn affect their likelihood of being selected as LABs. We adopt the standard deviations of the recovered size and luminosity distributions as the uncertainties associated with each source.

The majority of LABs (99\%) are recovered with a recovery fraction \frecv\ $>$ 0.5. A substantial fraction of LABs (71\%) exhibit \frecv\ $>$ 0.95. The recovery fractions, along with the uncertainties in size and luminosity, are included in the final \lya blob catalog (Table~\ref{tab:lab_cat}).

\section{Results and Discussion} \label{sec:result}
\subsection{LABs in E-COSMOS at $z\sim3.1$ \label{sec:LAB_samples}}

\begin{figure*}
\centering
\includegraphics[width=\textwidth]{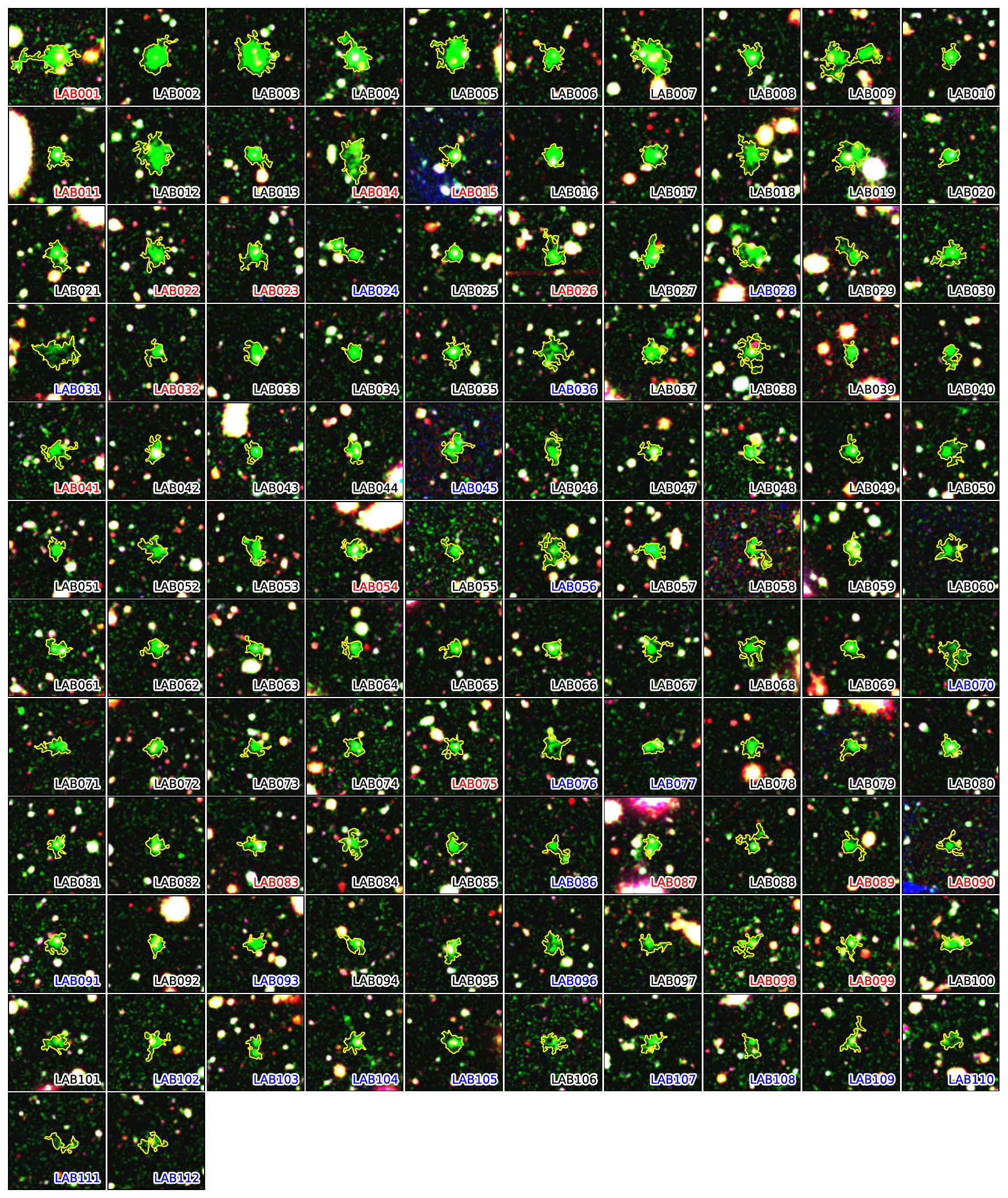}
\caption{
$30 \times 30$ \sqarcsec postage stamp RGB images of 112 LABs. The color represented are composed of Subaru/HSC $r$ (red), ODIN $N501$ (green), and Subaru/HSC $g$ (blue) images. In each case, the yellow contours indicate the surface brightness detection threshold ($1.5\sigma_{\rm SB_1}=3.3 \times 10^{-18} $\unitcgssb). 
The LAB IDs are colored red, blue, and black, corresponding to LABs selected only by the extended-LAE method, only by the extended-beyond-continuum method, and by both methods, respectively.
}
\label{fig:RGB_LAB}
\end{figure*}

We discover a total of 112 \lya blobs using both the \traditional and the \tractormethod selection methods.
This constitutes one of the largest LAB samples to date, covering a wide survey area of $\sim$9~\sqdeg with a uniform surface brightness threshold. We expect the LAB sample to grow substantially as the ODIN survey progresses to its goal of $\sim$100\,\sqdeg coverage across three redshift windows.
Figure~\ref{fig:RGB_LAB} presents RGB postage stamp images of the LABs, constructed using HSC SSP $r$, ODIN $N501$, and HSC SSP $g$ bands as the red, green, and blue channels, respectively.

Table~\ref{tab:lab_cat} lists the key properties of LABs: source ID, coordinates, \lya luminosity, isophotal areas, end-to-end extents of the \lya contour, and recovery fractions. Within the catalog, the LAB IDs are assigned in order of decreasing luminosity \lya, with lower ID numbers corresponding to brighter sources. We also indicate which selection method was used to identify each LAB.
We highlight a few notable examples from the LAB sample.

ODIN-COSMOS-z3p1-LAB001 is the most luminous LAB in our sample, exhibiting a narrow, linear feature extending toward the east. To access the robustness of this diffuse linear structure, we perform a bootstrap test by combining subsets of the exposures.
We randomly choose five exposures from ten exposures and construct a stacked image. We repeat this 100 times and confirm that the narrow and elongated \lya structure persists, indicating that the observed morphology is not an artifact from certain exposure combinations.

ODIN-COSMOS-z3p1-LAB009 appears to be a pair of LAB or an interacting system with $\sim$150 kpc end-to-end size. 
Recently, deep VLT/MUSE observations discovered much larger ($\sim$700 pkpc) \lya nebulae pair at $z\sim3$. This system consists of a QSO pair and a \lya filament connecting them \citep{Lusso19, Tornotti25}. 
Measurement of systematic redshifts of two components and deep IFU observation could provide insight into the nature of this source. We redefine their positions as the mean coordinate of the two spatial peaks of the \lya emissions.

ODIN-COSMOS-z3p1-LAB024 consists of the two LAEs that are individually listed in the ODIN-LAE catalog \citep{Firestone24}. In the \traditional pipeline, the source is deblended into two components by \stractor. In contrast, the \tractormethod pipeline does not deblend the system, as it removes the compact galaxies and retains only the diffuse, connected \lya halo. Since no distinct sources remain in the residual image, deblending is not triggered, and the halo is detected as a single extended structure. We classify this type of object as a ``dual-LAE'' system.

ODIN-COSMOS-z3p1-LAB028 exhibits a tail-like feature extending toward the northeast. Because of a continuum source between the tail and the central emission, it is deblended by the current pipeline and the tail is not included as part of this LAB.

ODIN-COSMOS-z3p1-LAB031 is one of the LABs identified by initial visual inspection and later recovered by the \tractormethod pipeline. It exhibits a large, diffuse \lya halo with only a faint optical counterpart visible in the HSC SSP $g$-band image, indicating the absence of any clearly identifiable powering source. This example highlights the effectiveness of the \tractormethod pipeline in discovering low-surface-brightness LABs that may be missed by traditional methods.

\begin{figure*}[ht]
\centering
  \begin{tabular}{ll}
    \includegraphics[width=.49\textwidth]{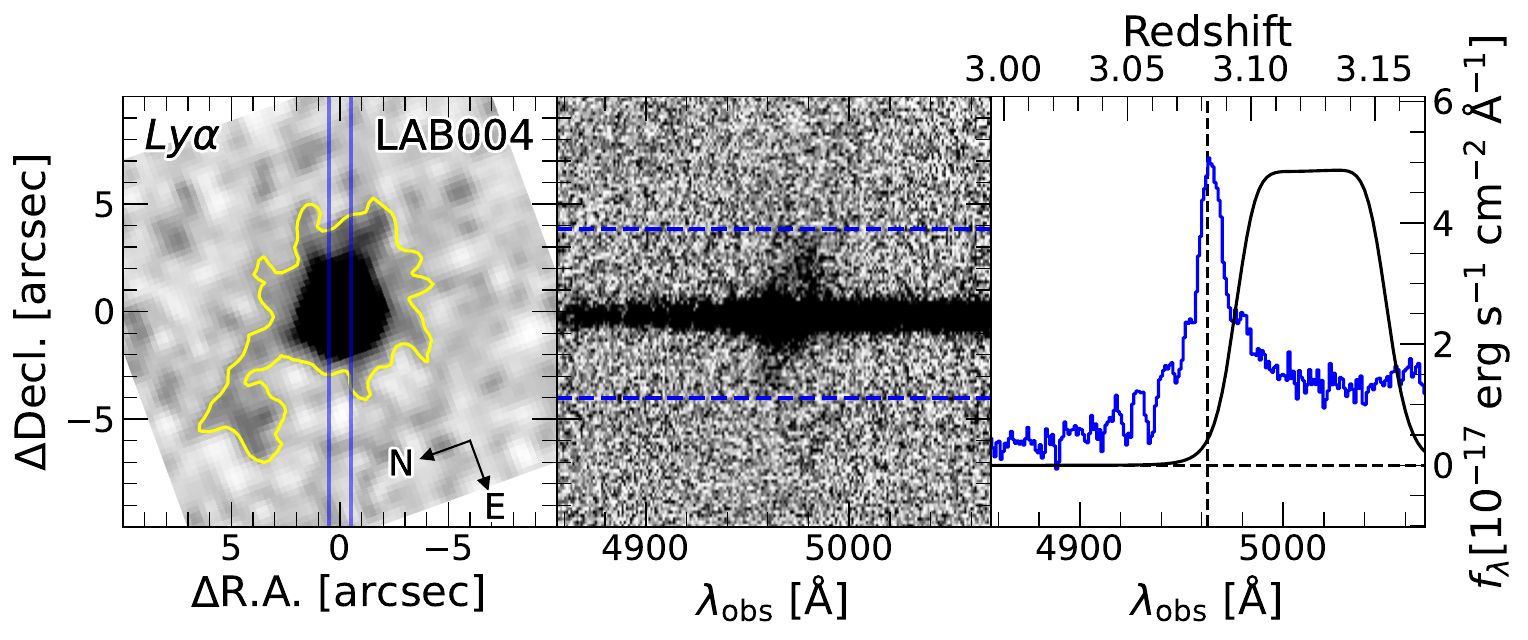} &
    \includegraphics[width=.49\textwidth]{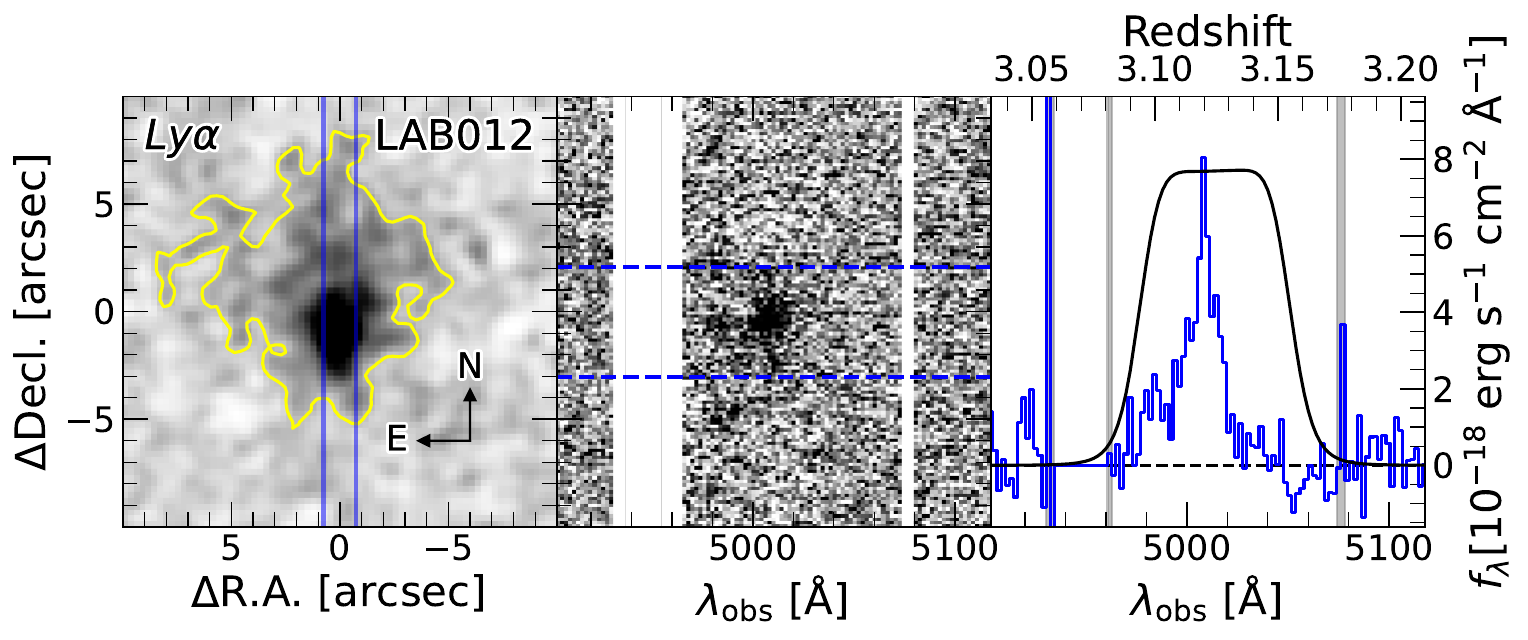} \\ 
    \includegraphics[width=.49\textwidth]{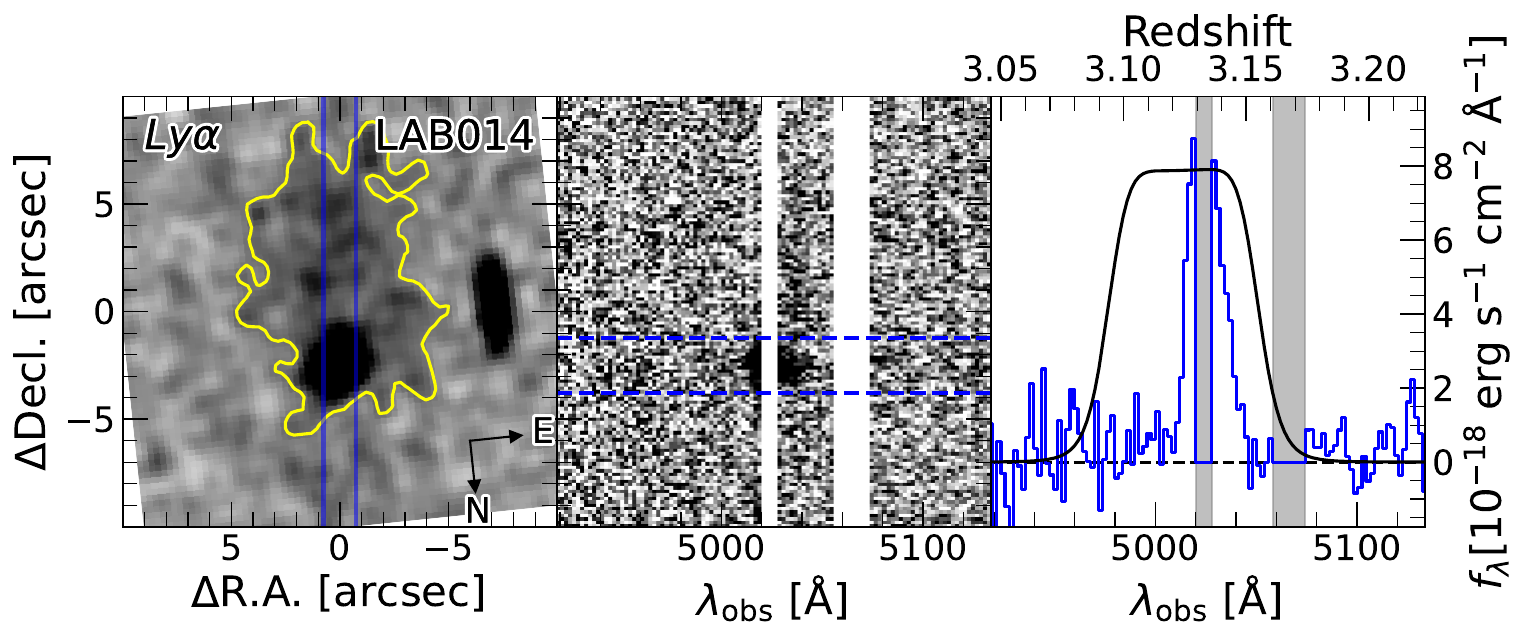} &
    \includegraphics[width=.49\textwidth]{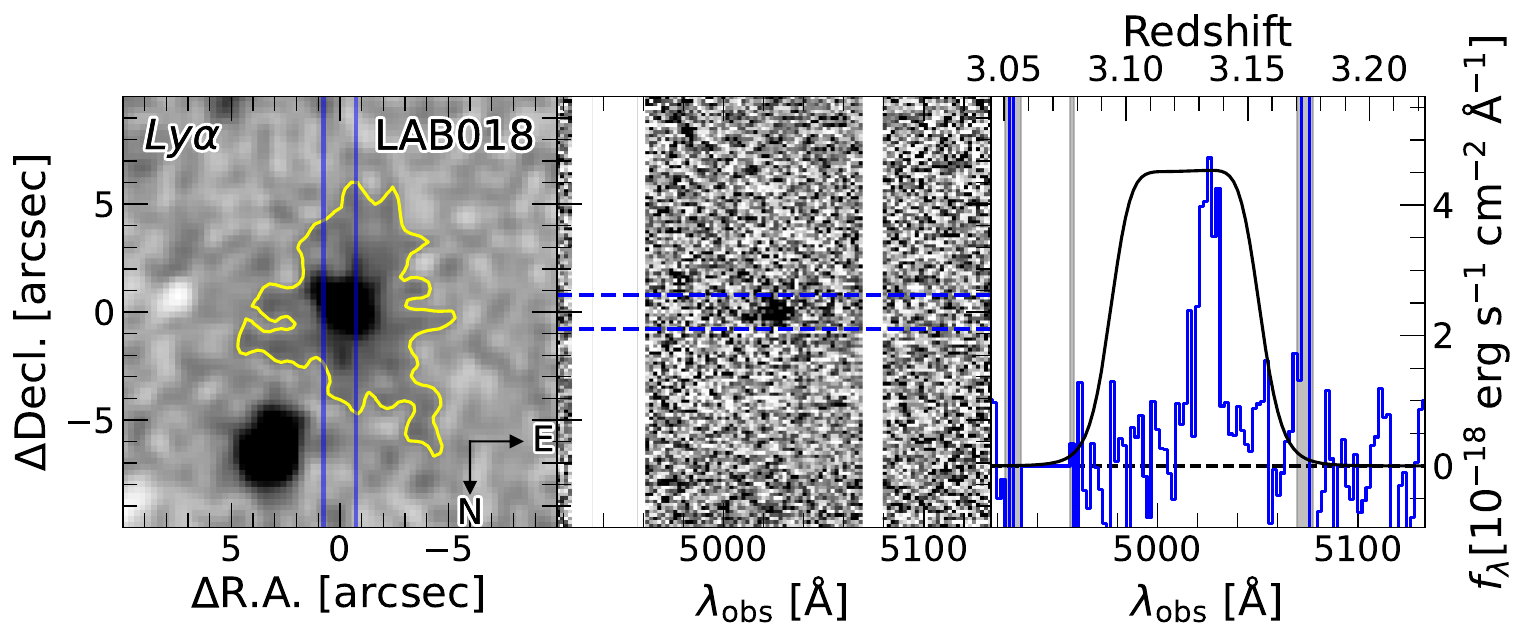} \\
    \includegraphics[width=.49\textwidth]{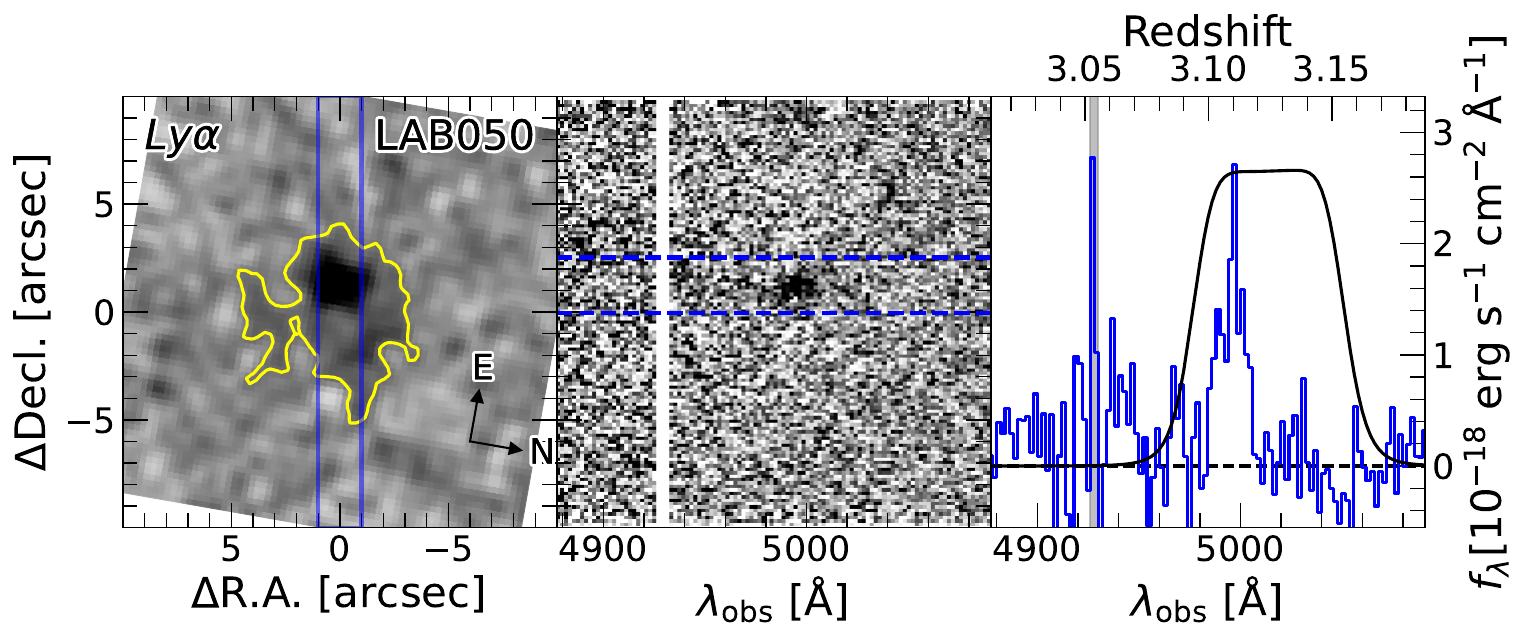}
  \end{tabular}
  \caption{
  Gemini/GMOS spectra (Section~\ref{sec:data}) alongside \lya images. 
  The first column in each panel shows the \lya image with the GMOS slit position overlaid in blue.
  The second column is the corresponding 2D spectrum; blue dashed lines indicate the apertures used for extracting the 1D spectrum. The vertical axis corresponds to the spatial direction, aligned with the orientation in the \lya images.
  The third column displays the extracted 1D spectrum in the vicinity of \lya, overlaid with the $N501$ filter transmission curve (in arbitrary units).
  The Gray-shaded regions indicate areas affected by poor sky subtraction or bad columns. 
  For LAB004, The black dashed vertical line indicates the redshift measured from the \heiilamb emission line. The detection of spatially extended \lya emission even at the $N501$ filter bandpass edge supports the reliability of our LAB selection.
  }
\label{fig:GMOS}
\end{figure*}

ODIN-COSMOS-z3p1-LAB084 is associated with a galaxy group exhibiting an extended \lya halo. Previous Keck/KCWI observations \citep[RO-0959;][]{Daddi22} yield a systemic redshift of $\sim3.096$, placing it near the blue edge of the $N501$ filter transmission curve \citep{Soo23}. Despite the reduced filter throughput at this redshift, RO-0959 is still robustly selected as a LAB, confirming the reliability of the ODIN LAB sample.

% differences of LAB catalog
We note that the LAB catalog in the E-COSMOS field at $z$ = 3.1 has been updated relative to \cite{Ramakrishnan23}, which identified 122 LABs without applying the \tractormethod\ method. In the earlier catalog, the sizes of some LABs were slightly overestimated due to imperfect deblending, allowing certain sources to exceed the size selection criterion and thus be classified as LABs, inflating the sample size. Nevertheless, approximately 70\% of the LABs are common to both catalogs. Importantly, the scientific results presented in \cite{Ramakrishnan23} and in this work remain unchanged regardless of which catalog is used.

\subsection{Spectroscopic Confirmation \label{sec:spec}}

Five out of six LABs are spectroscopically confirmed through our Gemini/GMOS program. 
We successfully detect \lya emission for the five LABs in both the 1D and 2D spectra.
\lya emission is not detected from LAB031, which has one of the lowest surface brightnesses, because the Gemini/GMOS observations are approximately a factor of two shallower than the ODIN data.
Figure~\ref{fig:GMOS} presents the \lya spectra for these sources. Among them, two (and possibly three) LABs exhibit spatially extended \lya emission in the 2D spectra, consistent with their classification as extended halos. 

\begin{figure}[t]
\centering
\includegraphics[width=0.485\textwidth]{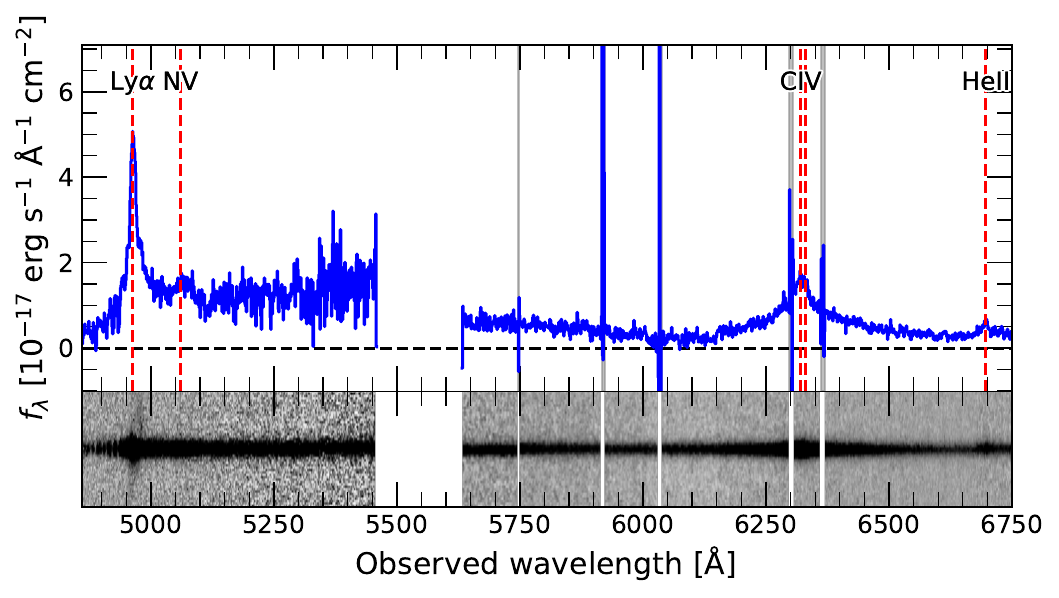}
\caption{
Gemini/GMOS 1D and 2D spectrum for LAB004 and a central QSO. 
The gray-shaded regions indicate areas affected by poor sky subtraction or bad columns. 
A large gap at $\lambda\sim5550$ \AA\ is because of the bad amplifier. 
Red dashed lines indicate observed emission lines. 
The broad hard ionization lines (\NV and \civ) implies that the QSO is the powering source of LAB004.
\label{fig:QSO_wide_spec}}
\end{figure}

\paragraph{ODIN-COSMOS-z3p1-LAB004} 
This object shows not only a clear continuum but also spatially extended \lya emission. Broad rest-frame UV emission lines (\NV and \civ) and relatively narrow \heii emission (FWHM = $549\pm32$ \kms) are also detected (Figrue~\ref{fig:QSO_wide_spec}).
While the peak of the \lya\ emission lies near the edge of the $N501$ filter bandpass, the spatially extended component extends to wavelengths where the filter transmission is $\sim$50\% ($\sim$4990\,\AA), ensuring that a significant portion of the \lya emission is still captured by the narrowband imaging.

We estimate the spectroscopic redshift to be $z = 3.0825 \pm 0.0002$ based on the \heii
emission line. Interestingly, the peak wavelength of the \lya emission closely matches the systemic redshift derived from \heii, suggesting either direct photoionization by a central AGN \citep{Dey05, Yang14b, Langen23}, or the presence of clumpy neutral hydrogen clouds within the CGM that allows for efficient \lya escape without significant resonant scattering \citep{Gronke17b, Chang22}.

\paragraph{ODIN-COSMOS-z3p1-LAB012} 
We confirm spatially extended \lya emission over $\sim$5\arcsec\ in the GMOS 2D spectrum. This extended emission coincides with the brightest region of the \lya blob.

\paragraph{ODIN-COSMOS-z3p1-LAB014} 
The peak of \lya emission coincided with a bad column, allowing us to confirm the redshift of the LAB but preventing analysis of the line profile. Spatially extended \lya emission is marginally detected on the red side of the bad column.

\paragraph{ODIN-COSMOS-z3p1-LAB018 and LAB050} 
Figure~\ref{fig:GMOS} shows a compact and narrow \lya\ emission in both the 2D and 1D spectra. 
The detection of only a single emission line within the 3600\AA\ - 6730\AA\ spectral range strongly suggests that the feature is not from a lower-redshift source. Lower redshift features such as \ion{C}{4}, \ion{He}{2}, [\ion{O}{2}], and [\ion{O}{3}] would typically be accompanied by additional bright emission lines in the GMOS spectra.
The absence of spatially extended diffuse \lya emission may be due to the relatively short exposure time and/or missing the low surface brightness part.

\bigskip

To further test the reliability of our LAB candidates, we cross-match them with known spectroscopic redshifts from the literature. 
We use datasets from SDSS-DR14Q \citep{Paris14}, PRIMUS \citep{Cool13}, zCOSMOS \citep{Lilly09}, Subaru/FMOS survey \citep{Silverman15}, 3D-HST grism survey \citep{Momcheva16}, MOSDEF survey \citep{Kriek07}, DEIMOS 10k survey \citep{Hasinger18} and other literatures. 
In addition, we visually inspect DESI DR1 spectra \citep{DESI_DR1} of 45 DESI targets that coincide with the positions of LABs. We confirm \lya emission in 40 LABs regardless of DESI redshift. 
HETDEX successfully confirm \lya emission from one LAB (\citealt{Hill21_hetdex} and Mentuch Cooper et al. in preparation).
We confirm the redshift of six LAB from Keck/DEIMOS observation \citep{Ramakrishnan25b}. 
In summary, we find that 52 LABs (out of a total of 112) with known spectroscopic redshifts are consistent with $z \sim 3.1$ without lower-redshift interlopers (Table \ref{tab:lab_cat}). This constitutes one of the largest spectroscopically confirmed LAB samples to date.

%------------------------------------------------------
\begin{figure*}
\centering
\includegraphics[width=\textwidth]{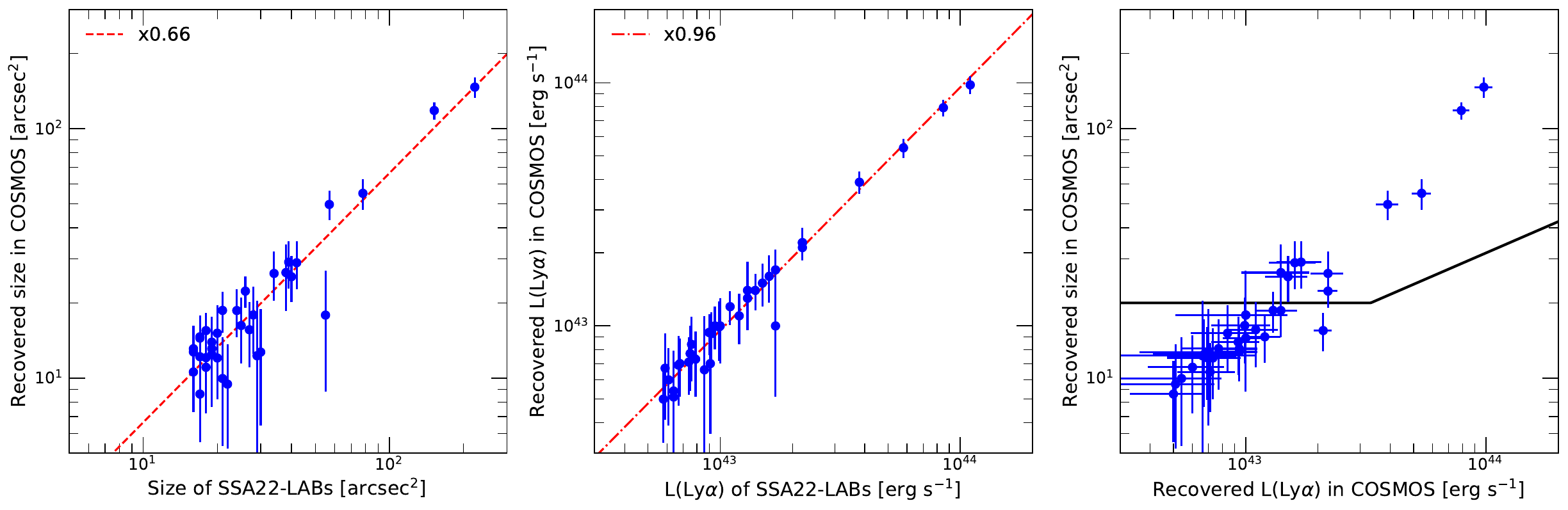}
\caption{
Recovery test results for SSA22 LABs. 
{\bf Left}: Recovered isophotal size vs. original size. 
{\bf Middle}: Recovered luminosity vs. original luminosity. 
{\bf Right}: The recovered size--luminosity relation for the SSA22 LABs overlaid with ODIN LAB selection criteria.
Because of ODIN's shallower surface brightness depth, the sizes and luminosities of the SSA22 LABs would be measured as smaller than their original values by factors of $\sim0.66$ and $\sim0.96$, respectively. Ten of the 35 SSA22 LABs would be recovered as LABs under the ODIN selection criteria in the size--luminosity diagram (black solid line). For a fair comparison, we adopt the number density derived from this recovery test for the SSA22 LABs.
}
\label{fig:ssa22_recover}
\end{figure*}
%------------------------------------------------------

\subsection{Environmental Effects on Number density and Luminosity.}
\label{sec:f2f}
The number density and luminosity function (LF) of LABs remain poorly constrained due to their rarity and strong field-to-field variation \citep[e.g.,][]{Yang10}. Furthermore, different selection criteria and survey depths across studies make statistical analyses particularly challenging.
With one of the largest LAB samples to date from the ODIN survey, we investigate the LAB luminosity function and its dependence on environment. 
ODIN enables a significant breakthrough, offering a large survey area with a uniform surface brightness threshold and environment information, all of which are crucial for obtaining reliable statistical analysis.

A detailed analysis of the redshift evolution of LABs and field-to-field variation across the disjoint ODIN survey fields will be presented in a forthcoming paper (B.\ Moon in preparation). In this work, we present initial results on the number density and luminosity functions (LFs) of LABs in the E-COSMOS field, which hosts two massive proto-clusters \citep{Ramakrishnan23, Ramakrishnan24, Ramakrishnan25}.

We measure the number density of LABs across the entire E-COSMOS survey area (9\,deg$^2$; $6.5\times10^6$\,cMpc$^3$ assuming the FWHM of the $N501$ filter as the line-of-sight thickness), as well as within the overdense regions (i.e., proto-clusters) separately.
For the entire survey area, the number density of LABs at $z \sim 3.1$ is $n = 1.7 \pm 0.2\times 10^{-5}$\,cMpc$^3$, with the uncertainty coming from the Poisson shot noise of the LAB number counts.
To enable direct comparison with previous works, we also compute the LAB number density within a survey area equivalent to that used by \cite{Matsuda04}: a $31\arcmin \times 23\arcmin$ FOV (0.2\,deg$^2$), typical of pre-HSC and pre-DECam surveys.
Moreover, the similar survey redshifts allow us a direct comparison of LAB number densities between our sample and that of the SSA22 proto-cluster field, one of the most well-studied overdense environments.
Within the sampling box that encompasses the most overdense regions \citep[Complexes A and C;][]{Ramakrishnan23} with a local 2D surface overdensity of $\delta_\Sigma \sim 5.5$ comparable to the SSA22 proto-cluster \citep{Steidel00, Matsuda11}, we identify 11 and 8 LABs in each complex, respectively. This yields a significantly elevated number density of $6.5\pm1.5 \times 10^{-5}$~\unitcovolume on average, a factor of $\sim$4 higher than that of the entire E-COSMOS LAB sample.

For a direct comparison with the SSA22 sample, which has a deeper detection threshold of $2.2\times10^{-18}$ \unitcgssb than the ODIN survey, we test how many SSA22 LAB would be recovered as LABs under the shallower depth of the ODIN survey, using the same methodology described in Section~\ref{sec:recovery}. 
For each SSA22 LAB, we inject its image into 4,000 empty sky regions in the ODIN \lya image and measure the median recovered isophotal sizes and luminosities at our detection threshold. The uncertainties are estimated as the standard deviations of these measurements.
Figure~\ref{fig:ssa22_recover} shows that the recovered SSA22 LABs, if observed by ODIN, would have isophotal sizes that are, on average, $\sim$66\% smaller than the original measurements by \citet{Matsuda04}, due to the shallower depth. Out of the 35 SSA22 LABs, ten would meet the LAB selection criteria in the ODIN survey, resulting in a number density of $n = 7.7\pm2.4 \times 10^{-5}$~\unitcovolume. This recovered number density is consistent with that of the ODIN LABs found in proto-cluster regions.

%----------------------------------------------------
\begin{figure}
\includegraphics[width=0.47\textwidth]{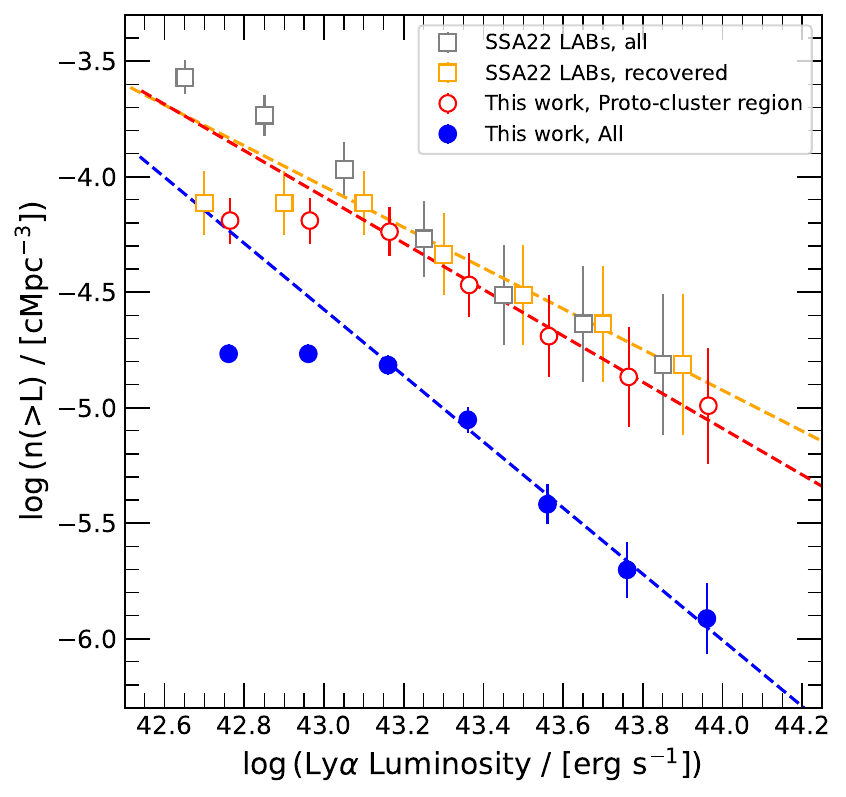}
\caption{
Cumulative luminosity functions (CLFs) of the SSA22 and E-COSMOS LABs.
Data points are slightly offset for clarity. Proto-cluster–associated LABs are shown as open symbols, while blue filled circles represent all ODIN LABs across the E-COSMOS field. The dashed lines indicate linear fits to the CLFs of each LAB sample. 
The flatter slopes and higher amplitudes of the proto-cluster–associated LAB CLFs suggest that the \lya luminosities and number densities of LABs are environment-dependent.
}
\label{fig:CLF}
\end{figure}
%----------------------------------------------------

%------------------------------------------------
\begin{figure*}
\includegraphics[width=\textwidth]{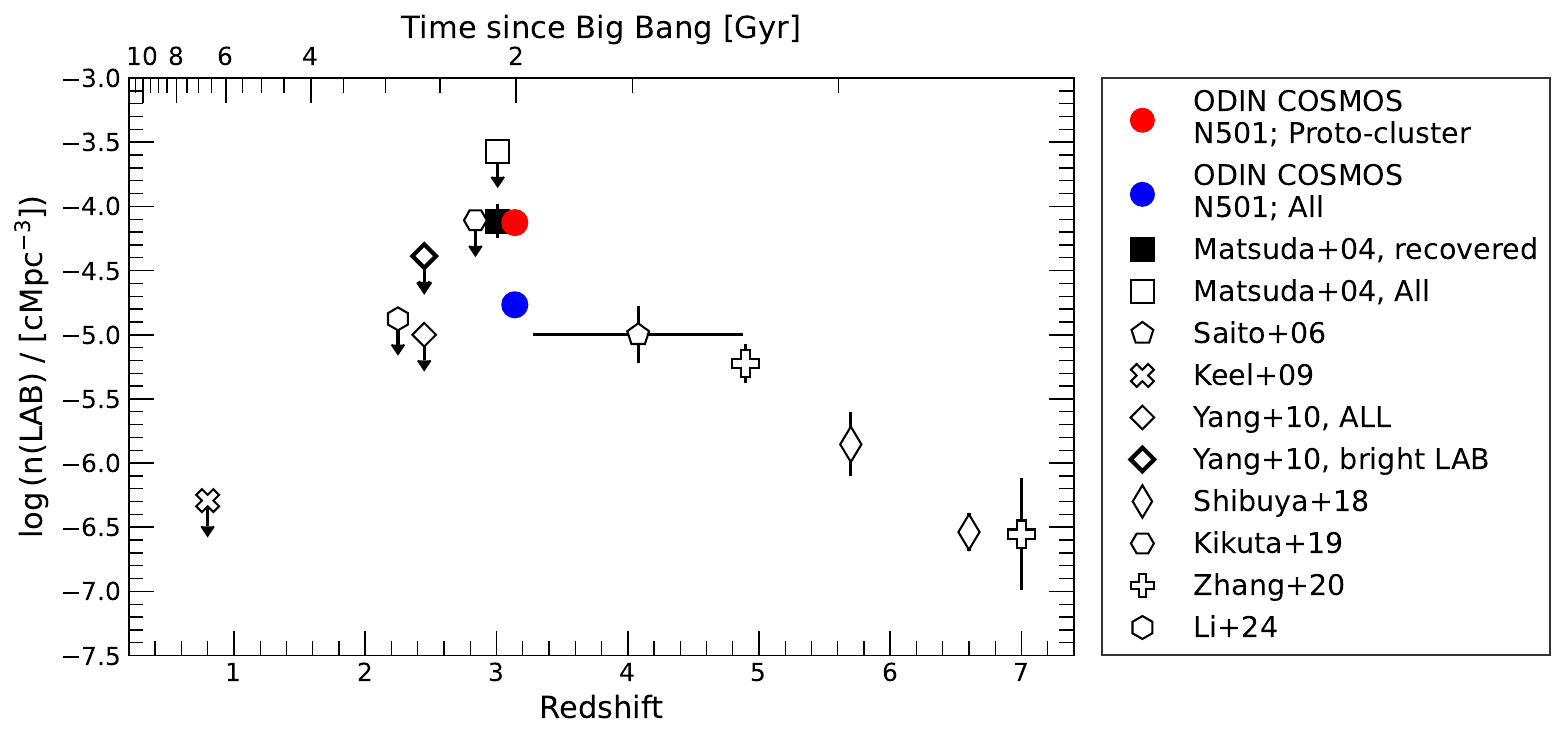}
\caption{
Number densities of LABs as a function of redshift (lower axis) and cosmic time (upper axis). Filled points indicate LAB samples whose number densities are directly comparable, while open points correspond to samples selected with different criteria and/or surface brightness depths, and are therefore treated as upper limits. 
The red circle represents the average number density at Complexes A and C.
Some symbomls are slightly shifted along the $x$-axis for clarity. The LAB number density appears to peak around Cosmic Noon ($z$ = 2--3). 
}
\label{fig:number_density}
\end{figure*}
%------------------------------------------------

Figure~\ref{fig:CLF} presents the cumulative luminosity function (CLF) of ODIN LABs across the entire E-COSMOS field, the CLF measured in the two proto-cluster complexes, and the CLFs of the SSA22 LABs \citep{Matsuda04}, located in the SSA22 proto-cluster (both original and recovered).
We fit the CLFs with a linear function in logarithmic space, $\log{n}=A \log{L(\mathrm{Ly\alpha})}+B$. The slope of the CLF for the full E-COSMOS LAB sample is $-1.4\pm0.1$, steeper than those of the proto-cluster-associated LABs ($-0.9\pm0.1$ and $-1.0\pm0.1$ for the recovered SSA22 LABs and for the ODIN LABs in Complex A and C, respectively).

Our initial results with still limited sample size (29 LABs within three proto-clusters) suggest the possibility of environmental effects on number density and luminosity of \lya blobs \citep[e.g.,][]{Matsuda11, Badescu17, Ramakrishnan23}.
To robustly characterize the luminosity function of LABs as a function of environment, an even larger sample is required. ODIN is poised to deliver such a dataset, with an expected sample of $\sim$1,000 LABs across multiple fields. This will enable detailed investigations into the environmental dependence of LAB luminosity functions and help disentangle the physical mechanisms that power their extended \lya emission.

\subsection{Number Density Over Cosmic Time}
\label{sec:number_density_evol}

We compare the number density of the ODIN LAB sample with those from previous surveys across a range of redshifts (Figure~\ref{fig:number_density}). 
For this comparison, we adopt the published values from each survey, taking only surface brightness dimming into account. We do {\it not} correct for differences in survey depth or selection method. This plot is intended to illustrate the current status of LAB surveys, not the accurate redshift evolution.
We compile narrow- and intermediate-band surveys with survey areas larger than 0.2 \sqdeg, comparable to that of \cite{Matsuda04}. We exclude shallow but wide surveys \citep[e.g.,][]{smith07, Yang09} that focused on large and bright LABs, as they tend to miss smaller LABs, making it difficult to compare number densities consistently.
A detailed analysis of larger sample of ODIN LABs, accounting for survey depth, selection methods, and completeness corrections, will be presented in a forthcoming paper (B.~Moon, in preparation).

We note that only the $z = 3.1$ samples (ODIN E-COSMOS and SSA22) are suitable for direct comparison as described in Section \ref{sec:f2f}. Therefore, all other number density measurements are treated either as upper limits or shown as their original values, for the reasons outlined below.
First, we adopt the non-detections from the low-redshift LAB survey by \cite{Keel09} as an upper limit.
Second, the samples at $z \sim 2.3$–2.4 \citep{Yang10, Mingyu24} are also treated as upper limits, since they would suffer from surface brightness dimming and drop out of the selection if placed at $z = 3.1$. This effect would lead to smaller apparent sizes and lower recovery fractions, thereby decreasing the inferred number densities. Note that the number densities reported by \cite{Mingyu24} are combined measurements from multiple survey fields covering $\sim$12 \sqdeg, without accounting for variations in survey depth or differences in size cuts.
Third, although the LAB sample at $z \sim 2.8$ reported by \cite{Kikuta19} is close in redshift to this work ($z \sim 3.1$), their surface brightness depth is twice as deep as ODIN. We therefore also treat their measurement as an upper limit.
Lastly, comparing higher-redshift LAB samples \citep{Saito06, Shibuya18, Zhang20} with our measurement is complex because the effects of flux boosting, shrinking sizes, and surface brightness depth must be considered simultaneously at the ODIN redshift ($z \sim 3.1$). Thus, we show their observed number densities as they are.

The LAB number density appears to peak around the Cosmic Noon ($z = 2$–3), mirroring the evolution of the cosmic star formation rate density \citep{Madau14}. 
\citet{Shibuya18} interpreted a similar redshift trend as evidence that the origin of LABs is related to star formation. However, we note that it could also be associated with AGN activity, given the similar shape of the black hole accretion rate density evolution \citep[e.g.,][]{Vito18}. A careful analysis of the redshift evolution of LABs, star formation, and AGN populations is therefore required to draw a firm conclusion.

The number density of ODIN LABs in proto-cluster regions (red dot) is consistent with that of the SSA22 LABs (black square), as well as with the upper limits set by \cite{Kikuta19}.
The overall ODIN LAB sample (blue dot) is also consistent with the measurements at $z \sim 2.3$, while all cluster LAB number densities at $z \sim 3$ (COSMOS and SSA22) are higher than those at other redshifts, demonstrating that environmental effects must be carefully accounted for when constructing the number density evolution.
To robustly trace the cosmic evolution of LABs, large-area surveys ($\gtrsim$10~deg$^2$), such as ODIN, with consistent sample selection across epochs, are essential.

\section{Summary} \label{sec:summary}

Using ODIN survey data in the E-COSMOS field, we identify a sample of 112 $z\sim 3.1$ Ly$\alpha$ blobs (LABs) with two complementary detection algorithms: one designed to find extended Ly$\alpha$ emission around LAEs (the extended LAE method), and the other newly developed to detect diffuse LABs using forced photometry with \texttt{Tractor}. These algorithms are being applied across the full $\sim 100$~deg$^2$ ODIN survey in three different redshift windows.

Our principle findings in this paper are:
\begin{itemize}    
    \item We spectroscopically confirm five LAB candidates through our new Gemini/GMOS observations, and 47 additional LABs through the DESI DR1 spectra and the literature. This includes ten LABs identified by the new \tractormethod pipeline, further supporting the robustness and reliability of our LAB sample.

    \item We present measurements of the number density and the luminosity function of ODIN LABs. The number density of proto-cluster-associated LABs is higher than the ODIN LABs across the full E-COSMOS field. The luminosity function of the entire ODIN LABs is 1.5 times steeper than the proto-cluster-associated LABs. These result suggest the \lya luminosities and number densities of LABs are environment-dependent. To robustly characterize the effect of environment, an even larger sample is required.

    \item The number density of LABs peaks around the Cosmic Noon ($z$ = 2--3), mirroring the cosmic star formation and the black hole accretion rate histories. To robustly trace the cosmic evolution of LABs, ODIN-like surveys with large areas and consistent sample selection across epochs, are essential.

\end{itemize}

\begin{acknowledgments}
Based on observations at Cerro Tololo Inter-American Observatory, NSF’s NOIRLab (Prop. ID 2020B-0201; PI: K.-S. Lee), which is managed by the Association of Universities for Research in Astronomy  under a cooperative agreement with the National Science Foundation.
We thank Yuichi Matsuda for providing the narrowband images of his \lya blobs.
BM and YY are supported by the Basic Science Research Program through the National Research Foundation of Korea funded by the Ministry of Science, ICT \& Future Planning (2019R1A2C4069803), and UST Young Scientist+ Research Program 2022 through the University of Science and Technology (2022YS45).
KSL acknowledges financial support from the National Science Foundation under grant Nos. AST-2206705, AST-2408359, and from the Ross-Lynn Purdue Research Foundations.
EG and NF acknowledge support from NSF grant AST-2206222. 
NF acknowledges support from NASA Astrophysics Data Analysis Program grant 80NSSC22K0487. 
This material is based upon work supported by the NSF Graduate Research Fellowship Program under Grant No. DGE-2233066 to NF. 
This work is supported by K-GMT Science Program (PID: GS-2021A-FT-211, GS-2021B-Q-110, GS-2023A-Q-110, GS-2025A-Q-110) of Korea Astronomy and Space Science Institute.
The Institute for Gravitation and the Cosmos is supported by the Eberly College of Science and the Office of the Senior Vice President for Research at the Pennsylvania State University.
LG also gratefully acknowledges financial support from ANID-MILENIO-NCN2024\_112, ANID BASAL project FB210003, and FONDECYT regular project number 1230591.
HSH acknowledges the support of the National Research Foundation of Korea (NRF) grant funded by the Korea government (MSIT), NRF-2021R1A2C1094577, and Hyunsong Educational \& Cultural Foundation.
SL acknowledges support from the National Research Foundation of Korea(NRF) grant (Nos. RS-2025-00573214 and 2021M3F7A1084525) funded by the Korea government(MSIT).
\end{acknowledgments}

\facilities{Blanco (DECam), Gemini:South (GMOS)}

\software{Source Extractor \citep{Bertin96},
          PSFEx \citep{PSFEx},
          Tractor \citep{Lang16},
          Astropy \citep{astropy13,astropy18,astropy22},
          Photutils \citep{photutils}
          }


\bibliography{master.LAB}{}

@Article{	  Aihara19,
  author	= {{Aihara}, Hiroaki and {AlSayyad}, Yusra and {Ando}, Makoto
		  and {Armstrong}, Robert and {Bosch}, James and {Egami},
		  Eiichi and {Furusawa}, Hisanori and {Furusawa}, Junko and
		  {Goulding}, Andy and {Harikane}, Yuichi and {Hikage},
		  Chiaki and {Ho}, Paul T.~P. and {Hsieh}, Bau-Ching and
		  {Huang}, Song and {Ikeda}, Hiroyuki and {Imanishi},
		  Masatoshi and {Ito}, Kei and {Iwata}, Ikuru and {Jaelani},
		  Anton T. and {Kakuma}, Ryota and {Kawana}, Kojiro and
		  {Kikuta}, Satoshi and {Kobayashi}, Umi and {Koike},
		  Michitaro and {Komiyama}, Yutaka and {Li}, Xiangchong and
		  {Liang}, Yongming and {Lin}, Yen-Ting and {Luo}, Wentao and
		  {Lupton}, Robert and {Lust}, Nate B. and {MacArthur},
		  Lauren A. and {Matsuoka}, Yoshiki and {Mineo}, Sogo and
		  {Miyatake}, Hironao and {Miyazaki}, Satoshi and {More},
		  Surhud and {Murata}, Ryoma and {Namiki}, Shigeru V. and
		  {Nishizawa}, Atsushi J. and {Oguri}, Masamune and {Okabe},
		  Nobuhiro and {Okamoto}, Sakurako and {Okura}, Yuki and
		  {Ono}, Yoshiaki and {Onodera}, Masato and {Onoue}, Masafusa
		  and {Osato}, Ken and {Ouchi}, Masami and {Shibuya},
		  Takatoshi and {Strauss}, Michael A. and {Sugiyama}, Naoshi
		  and {Suto}, Yasushi and {Takada}, Masahiro and {Takagi},
		  Yuhei and {Takata}, Tadafumi and {Takita}, Satoshi and
		  {Tanaka}, Masayuki and {Terai}, Tsuyoshi and {Toba},
		  Yoshiki and {Uchiyama}, Hisakazu and {Utsumi}, Yousuke and
		  {Wang}, Shiang-Yu and {Wang}, Wenting and {Yamada},
		  Yoshihiko},
  title		= "{Second data release of the Hyper Suprime-Cam Subaru
		  Strategic Program}",
  journal	= {\pasj},
  year		= 2019,
  month		= dec,
  volume	= {71},
  number	= {6},
  eid		= {114},
  pages		= {114},
  doi		= {10.1093/pasj/psz103},
  archiveprefix	= {arXiv},
  eprint	= {1905.12221},
  primaryclass	= {astro-ph.IM},
  adsurl	= {https://ui.adsabs.harvard.edu/abs/2019PASJ...71..114A}
}

@Article{	  Ao20,
  author	= {{Ao}, Yiping and {Zheng}, Zheng and {Henkel}, Christian
		  and {Nie}, Shiyu and {Beelen}, Alexandre and {Cen}, Renyue
		  and {Dijkstra}, Mark and {Francis}, Paul J. and {Geach},
		  James E. and {Kohno}, Kotaro and {Lehnert}, Matthew D. and
		  {Menten}, Karl M. and {Wang}, Junzhi and {Weiss}, Axel},
  title		= "{Infalling gas in a Lyman-{\ensuremath{\alpha}} blob}",
  journal	= {Nature Astronomy},
  year		= 2020,
  month		= mar,
  volume	= {4},
  pages		= {670-674},
  doi		= {10.1038/s41550-020-1033-3},
  archiveprefix	= {arXiv},
  eprint	= {2003.06099},
  primaryclass	= {astro-ph.GA},
  adsurl	= {https://ui.adsabs.harvard.edu/abs/2020NatAs...4..670A}
}

@Article{	  Badescu17,
  author	= {{B{\u{a}}descu}, Toma and {Yang}, Yujin and {Bertoldi},
		  Frank and {Zabludoff}, Ann and {Karim}, Alexander and
		  {Magnelli}, Benjamin},
  title		= "{Discovery of a Protocluster Associated with a
		  Ly{\ensuremath{\alpha}} Blob Pair at z = 2.3}",
  journal	= {\apj},
  year		= 2017,
  month		= aug,
  volume	= {845},
  number	= {2},
  eid		= {172},
  pages		= {172},
  doi		= {10.3847/1538-4357/aa8220},
  archiveprefix	= {arXiv},
  eprint	= {1708.00447},
  primaryclass	= {astro-ph.GA},
  adsurl	= {https://ui.adsabs.harvard.edu/abs/2017ApJ...845..172B}
}

@Article{	  Cai17,
  author	= {{Cai}, Zheng and {Fan}, Xiaohui and {Yang}, Yujin and
		  {Bian}, Fuyan and {Prochaska}, J. Xavier and {Zabludoff},
		  Ann and {McGreer}, Ian and {Zheng}, Zhen-Ya and {Green},
		  Richard and {Cantalupo}, Sebastiano and {Frye}, Brenda and
		  {Hamden}, Erika and {Jiang}, Linhua and {Kashikawa},
		  Nobunari and {Wang}, Ran},
  title		= "{Discovery of an Enormous Ly{\ensuremath{\alpha}} Nebula
		  in a Massive Galaxy Overdensity at z = 2.3}",
  journal	= {\apj},
  year		= 2017,
  month		= mar,
  volume	= {837},
  number	= {1},
  eid		= {71},
  pages		= {71},
  doi		= {10.3847/1538-4357/aa5d14},
  archiveprefix	= {arXiv},
  eprint	= {1609.04021},
  primaryclass	= {astro-ph.GA},
  adsurl	= {https://ui.adsabs.harvard.edu/abs/2017ApJ...837...71C}
}

@Article{	  Cantalupo14,
  author	= {{Cantalupo}, Sebastiano and {Arrigoni-Battaia}, Fabrizio
		  and {Prochaska}, J. Xavier and {Hennawi}, Joseph F. and
		  {Madau}, Piero},
  title		= "{A cosmic web filament revealed in
		  Lyman-{\ensuremath{\alpha}} emission around a luminous
		  high-redshift quasar}",
  journal	= {\nat},
  year		= 2014,
  month		= feb,
  volume	= {506},
  number	= {7486},
  pages		= {63-66},
  doi		= {10.1038/nature12898},
  archiveprefix	= {arXiv},
  eprint	= {1401.4469},
  primaryclass	= {astro-ph.CO},
  adsurl	= {https://ui.adsabs.harvard.edu/abs/2014Natur.506...63C}
}

@Article{	  Chang22,
  author	= {{Chang}, Seok-Jun and {Yang}, Yujin and {Seon}, Kwang-Il
		  and {Zabludoff}, Ann and {Lee}, Hee-Won},
  title		= "{Radiative Transfer in Ly{\ensuremath{\alpha}} Nebulae. I.
		  Modeling a Continuous or Clumpy Spherical Halo with a
		  Central Source}",
  journal	= {\apj},
  year		= 2023,
  month		= mar,
  volume	= {945},
  number	= {2},
  eid		= {100},
  pages		= {100},
  doi		= {10.3847/1538-4357/acac98},
  archiveprefix	= {arXiv},
  eprint	= {2212.09630},
  primaryclass	= {astro-ph.GA},
  adsurl	= {https://ui.adsabs.harvard.edu/abs/2023ApJ...945..100C}
}

@Article{	  Cool13,
  author	= {{Cool}, Richard J. and {Moustakas}, John and {Blanton},
		  Michael R. and {Burles}, Scott M. and {Coil}, Alison L. and
		  {Eisenstein}, Daniel J. and {Wong}, Kenneth C. and {Zhu},
		  Guangtun and {Aird}, James and {Bernstein}, Rebecca A. and
		  {Bolton}, Adam S. and {Hogg}, David W. and {Mendez},
		  Alexander J.},
  title		= "{The PRIsm MUlti-object Survey (PRIMUS). II. Data
		  Reduction and Redshift Fitting}",
  journal	= {\apj},
  year		= 2013,
  month		= apr,
  volume	= {767},
  number	= {2},
  eid		= {118},
  pages		= {118},
  doi		= {10.1088/0004-637X/767/2/118},
  archiveprefix	= {arXiv},
  eprint	= {1303.2672},
  primaryclass	= {astro-ph.CO},
  adsurl	= {https://ui.adsabs.harvard.edu/abs/2013ApJ...767..118C}
}

@Article{	  Daddi20,
  author	= {{Daddi}, E. and {Valentino}, F. and {Rich}, R.~M. and
		  {Neill}, J.~D. and {Gronke}, M. and {O'Sullivan}, D. and
		  {Elbaz}, D. and {Bournaud}, F. and {Finoguenov}, A. and
		  {Marchal}, A. and {Delvecchio}, I. and {Jin}, S. and {Liu},
		  D. and {Strazzullo}, V. and {Calabro}, A. and {Coogan}, R.
		  and {D'Eugenio}, C. and {Gobat}, R. and {Kalita}, B.~S. and
		  {Laursen}, P. and {Martin}, D.~C. and {Puglisi}, A. and
		  {Schinnerer}, E. and {Wang}, T.},
  title		= "{Three Lyman-{\ensuremath{\alpha}}-emitting filaments
		  converging to a massive galaxy group at z = 2.91:
		  discussing the case for cold gas infall}",
  journal	= {\aap},
  year		= 2021,
  month		= may,
  volume	= {649},
  eid		= {A78},
  pages		= {A78},
  doi		= {10.1051/0004-6361/202038700},
  archiveprefix	= {arXiv},
  eprint	= {2006.11089},
  primaryclass	= {astro-ph.GA},
  adsurl	= {https://ui.adsabs.harvard.edu/abs/2021A&A...649A..78D}
}

@Article{	  Daddi22,
  author	= {{Daddi}, E. and {Rich}, R.~M. and {Valentino}, F. and
		  {Jin}, S. and {Delvecchio}, I. and {Liu}, D. and
		  {Strazzullo}, V. and {Neill}, J. and {Gobat}, R. and
		  {Finoguenov}, A. and {Bournaud}, F. and {Elbaz}, D. and
		  {Kalita}, B.~S. and {O'Sullivan}, D. and {Wang}, T.},
  title		= "{Evidence for Cold-stream to Hot-accretion Transition as
		  Traced by Ly{\ensuremath{\alpha}} Emission from Groups and
		  Clusters at 2 < z < 3.3}",
  journal	= {\apjl},
  year		= 2022,
  month		= feb,
  volume	= {926},
  number	= {2},
  eid		= {L21},
  pages		= {L21},
  doi		= {10.3847/2041-8213/ac531f},
  archiveprefix	= {arXiv},
  eprint	= {2202.03715},
  primaryclass	= {astro-ph.CO},
  adsurl	= {https://ui.adsabs.harvard.edu/abs/2022ApJ...926L..21D}
}

@Article{	  Dey05,
  author	= {{Dey}, Arjun and {Bian}, Chao and {Soifer}, Baruch T. and
		  {Brand}, Kate and {Brown}, Michael J.~I. and {Chaffee},
		  Frederic H. and {Le Floc'h}, Emeric and {Hill}, Grant and
		  {Houck}, James R. and {Jannuzi}, Buell T. and {Rieke},
		  Marcia and {Weedman}, Daniel and {Brodwin}, Mark and
		  {Eisenhardt}, Peter},
  title		= "{Discovery of a Large \raisebox{-0.5ex}\textasciitilde200
		  kpc Gaseous Nebula at z \raisebox{-0.5ex}\textasciitilde
		  2.7 with the Spitzer Space Telescope}",
  journal	= {\apj},
  year		= 2005,
  month		= aug,
  volume	= {629},
  number	= {2},
  pages		= {654-666},
  doi		= {10.1086/430775},
  archiveprefix	= {arXiv},
  eprint	= {astro-ph/0503632},
  primaryclass	= {astro-ph},
  adsurl	= {https://ui.adsabs.harvard.edu/abs/2005ApJ...629..654D}
}

@Article{	  Dey19,
  author	= {{Dey}, Arjun and {Schlegel}, David J. and {Lang}, Dustin
		  and {Blum}, Robert and {Burleigh}, Kaylan and {Fan},
		  Xiaohui and {Findlay}, Joseph R. and {Finkbeiner}, Doug and
		  {Herrera}, David and {Juneau}, St{\'e}phanie and
		  {Landriau}, Martin and {Levi}, Michael and {McGreer}, Ian
		  and {Meisner}, Aaron and {Myers}, Adam D. and {Moustakas},
		  John and {Nugent}, Peter and {Patej}, Anna and {Schlafly},
		  Edward F. and {Walker}, Alistair R. and {Valdes}, Francisco
		  and {Weaver}, Benjamin A. and {Y{\`e}che}, Christophe and
		  {Zou}, Hu and {Zhou}, Xu and {Abareshi}, Behzad and
		  {Abbott}, T.~M.~C. and {Abolfathi}, Bela and {Aguilera}, C.
		  and {Alam}, Shadab and {Allen}, Lori and {Alvarez}, A. and
		  {Annis}, James and {Ansarinejad}, Behzad and {Aubert},
		  Marie and {Beechert}, Jacqueline and {Bell}, Eric F. and
		  {BenZvi}, Segev Y. and {Beutler}, Florian and {Bielby},
		  Richard M. and {Bolton}, Adam S. and {Brice{\~n}o},
		  C{\'e}sar and {Buckley-Geer}, Elizabeth J. and {Butler},
		  Karen and {Calamida}, Annalisa and {Carlberg}, Raymond G.
		  and {Carter}, Paul and {Casas}, Ricard and {Castander},
		  Francisco J. and {Choi}, Yumi and {Comparat}, Johan and
		  {Cukanovaite}, Elena and {Delubac}, Timoth{\'e}e and
		  {DeVries}, Kaitlin and {Dey}, Sharmila and {Dhungana},
		  Govinda and {Dickinson}, Mark and {Ding}, Zhejie and
		  {Donaldson}, John B. and {Duan}, Yutong and {Duckworth},
		  Christopher J. and {Eftekharzadeh}, Sarah and {Eisenstein},
		  Daniel J. and {Etourneau}, Thomas and {Fagrelius}, Parker
		  A. and {Farihi}, Jay and {Fitzpatrick}, Mike and
		  {Font-Ribera}, Andreu and {Fulmer}, Leah and
		  {G{\"a}nsicke}, Boris T. and {Gaztanaga}, Enrique and
		  {George}, Koshy and {Gerdes}, David W. and {Gontcho}, Satya
		  Gontcho A. and {Gorgoni}, Claudio and {Green}, Gregory and
		  {Guy}, Julien and {Harmer}, Diane and {Hernandez}, M. and
		  {Honscheid}, Klaus and {Huang}, Lijuan Wendy and {James},
		  David J. and {Jannuzi}, Buell T. and {Jiang}, Linhua and
		  {Joyce}, Richard and {Karcher}, Armin and {Karkar}, Sonia
		  and {Kehoe}, Robert and {Kneib}, Jean-Paul and
		  {Kueter-Young}, Andrea and {Lan}, Ting-Wen and {Lauer}, Tod
		  R. and {Le Guillou}, Laurent and {Le Van Suu}, Auguste and
		  {Lee}, Jae Hyeon and {Lesser}, Michael and {Perreault
		  Levasseur}, Laurence and {Li}, Ting S. and {Mann}, Justin
		  L. and {Marshall}, Robert and {Mart{\'\i}nez-V{\'a}zquez},
		  C.~E. and {Martini}, Paul and {du Mas des Bourboux},
		  H{\'e}lion and {McManus}, Sean and {Meier}, Tobias Gabriel
		  and {M{\'e}nard}, Brice and {Metcalfe}, Nigel and
		  {Mu{\~n}oz-Guti{\'e}rrez}, Andrea and {Najita}, Joan and
		  {Napier}, Kevin and {Narayan}, Gautham and {Newman},
		  Jeffrey A. and {Nie}, Jundan and {Nord}, Brian and
		  {Norman}, Dara J. and {Olsen}, Knut A.~G. and {Paat},
		  Anthony and {Palanque-Delabrouille}, Nathalie and {Peng},
		  Xiyan and {Poppett}, Claire L. and {Poremba}, Megan R. and
		  {Prakash}, Abhishek and {Rabinowitz}, David and {Raichoor},
		  Anand and {Rezaie}, Mehdi and {Robertson}, A.~N. and {Roe},
		  Natalie A. and {Ross}, Ashley J. and {Ross}, Nicholas P.
		  and {Rudnick}, Gregory and {Safonova}, Sasha and {Saha},
		  Abhijit and {S{\'a}nchez}, F. Javier and {Savary}, Elodie
		  and {Schweiker}, Heidi and {Scott}, Adam and {Seo},
		  Hee-Jong and {Shan}, Huanyuan and {Silva}, David R. and
		  {Slepian}, Zachary and {Soto}, Christian and {Sprayberry},
		  David and {Staten}, Ryan and {Stillman}, Coley M. and
		  {Stupak}, Robert J. and {Summers}, David L. and {Sien Tie},
		  Suk and {Tirado}, H. and {Vargas-Maga{\~n}a}, Mariana and
		  {Vivas}, A. Katherina and {Wechsler}, Risa H. and
		  {Williams}, Doug and {Yang}, Jinyi and {Yang}, Qian and
		  {Yapici}, Tolga and {Zaritsky}, Dennis and {Zenteno}, A.
		  and {Zhang}, Kai and {Zhang}, Tianmeng and {Zhou}, Rongpu
		  and {Zhou}, Zhimin},
  title		= "{Overview of the DESI Legacy Imaging Surveys}",
  journal	= {\aj},
  year		= 2019,
  month		= may,
  volume	= {157},
  number	= {5},
  eid		= {168},
  pages		= {168},
  doi		= {10.3847/1538-3881/ab089d},
  archiveprefix	= {arXiv},
  eprint	= {1804.08657},
  primaryclass	= {astro-ph.IM},
  adsurl	= {https://ui.adsabs.harvard.edu/abs/2019AJ....157..168D}
}

@Article{	  Erb11,
  author	= {{Erb}, Dawn K. and {Bogosavljevi{\'c}}, Milan and
		  {Steidel}, Charles C.},
  title		= "{Filamentary Large-scale Structure Traced by Six
		  Ly{\ensuremath{\alpha}} Blobs at z = 2.3}",
  journal	= {\apjl},
  year		= 2011,
  month		= oct,
  volume	= {740},
  number	= {1},
  eid		= {L31},
  pages		= {L31},
  doi		= {10.1088/2041-8205/740/1/L31},
  archiveprefix	= {arXiv},
  eprint	= {1109.2167},
  primaryclass	= {astro-ph.CO},
  adsurl	= {https://ui.adsabs.harvard.edu/abs/2011ApJ...740L..31E}
}

@Article{	  Eunchong20,
  author	= {{Kim}, Eunchong and {Yang}, Yujin and {Zabludoff}, Ann and
		  {Smith}, Paul and {Jannuzi}, Buell and {Lee}, Myung Gyoon
		  and {Hwang}, Narae and {Park}, Byeong-Gon},
  title		= "{What Makes Ly{\ensuremath{\alpha}} Nebulae Glow? Mapping
		  the Polarization of LABd05}",
  journal	= {\apj},
  year		= 2020,
  month		= may,
  volume	= {894},
  number	= {1},
  eid		= {33},
  pages		= {33},
  doi		= {10.3847/1538-4357/ab837f},
  archiveprefix	= {arXiv},
  eprint	= {2003.13915},
  primaryclass	= {astro-ph.GA},
  adsurl	= {https://ui.adsabs.harvard.edu/abs/2020ApJ...894...33K}
}

@Article{	  Fabrizio18a,
  author	= {{Arrigoni Battaia}, Fabrizio and {Prochaska}, J. Xavier
		  and {Hennawi}, Joseph F. and {Obreja}, Aura and {Buck},
		  Tobias and {Cantalupo}, Sebastiano and {Dutton}, Aaron A.
		  and {Macci{\`o}}, Andrea V.},
  title		= "{Inspiraling halo accretion mapped in Ly
		  {\ensuremath{\alpha}} emission around a z
		  {\ensuremath{\sim}} 3 quasar}",
  journal	= {\mnras},
  year		= 2018,
  month		= jan,
  volume	= {473},
  number	= {3},
  pages		= {3907-3940},
  doi		= {10.1093/mnras/stx2465},
  archiveprefix	= {arXiv},
  eprint	= {1709.08228},
  primaryclass	= {astro-ph.GA},
  adsurl	= {https://ui.adsabs.harvard.edu/abs/2018MNRAS.473.3907A}
}

@Article{	  Fabrizio22,
  author	= {{Arrigoni Battaia}, Fabrizio and {Chen}, Chian-Chou and
		  {Liu}, Hau-Yu Baobab and {De Breuck}, Carlos and
		  {Galametz}, Maud and {Fumagalli}, Michele and {Yang}, Yujin
		  and {Zanella}, Anita and {Man}, Allison and {Obreja}, Aura
		  and {Prochaska}, J. Xavier and {Ba{\~n}ados}, Eduardo and
		  {Hennawi}, Joseph F. and {Farina}, Emanuele P. and {Zwaan},
		  Martin A. and {Decarli}, Roberto and {Lusso}, Elisabeta},
  title		= "{A Multiwavelength Study of ELAN Environments
		  (AMUSE$^{2}$). Mass Budget, Satellites Spin Alignment, and
		  Gas Infall in a Massive z 3 Quasar Host Halo}",
  journal	= {\apj},
  year		= 2022,
  month		= may,
  volume	= {930},
  number	= {1},
  eid		= {72},
  pages		= {72},
  doi		= {10.3847/1538-4357/ac5a4d},
  archiveprefix	= {arXiv},
  eprint	= {2111.15392},
  primaryclass	= {astro-ph.GA},
  adsurl	= {https://ui.adsabs.harvard.edu/abs/2022ApJ...930...72A}
}

@Article{	  Francis01,
  author	= {{Francis}, Paul J. and {Williger}, Gerard M. and
		  {Collins}, Nicholas R. and {Palunas}, Povilas and
		  {Malumuth}, Eliot M. and {Woodgate}, Bruce E. and
		  {Teplitz}, Harry I. and {Smette}, Alain and {Sutherland},
		  Ralph S. and {Danks}, Anthony C. and {Hill}, Robert S. and
		  {Lindler}, Donald and {Kimble}, Randy A. and {Heap}, Sara
		  R. and {Hutchings}, John B.},
  title		= "{A Pair of Compact Red Galaxies at Redshift 2.38, Immersed
		  in a 100 Kiloparsec Scale Ly{\ensuremath{\alpha}} Nebula}",
  journal	= {\apj},
  year		= 2001,
  month		= jun,
  volume	= {554},
  number	= {2},
  pages		= {1001-1011},
  doi		= {10.1086/321417},
  archiveprefix	= {arXiv},
  eprint	= {astro-ph/0102263},
  primaryclass	= {astro-ph},
  adsurl	= {https://ui.adsabs.harvard.edu/abs/2001ApJ...554.1001F}
}

@Article{	  Geach09,
  author	= {{Geach}, J.~E. and {Alexander}, D.~M. and {Lehmer}, B.~D.
		  and {Smail}, Ian and {Matsuda}, Y. and {Chapman}, S.~C. and
		  {Scharf}, C.~A. and {Ivison}, R.~J. and {Volonteri}, M. and
		  {Yamada}, T. and {Blain}, A.~W. and {Bower}, R.~G. and
		  {Bauer}, F.~E. and {Basu-Zych}, A.},
  title		= "{The Chandra Deep Protocluster Survey:
		  Ly{\ensuremath{\alpha}} Blobs are Powered by Heating, Not
		  Cooling}",
  journal	= {\apj},
  year		= 2009,
  month		= jul,
  volume	= {700},
  number	= {1},
  pages		= {1-9},
  doi		= {10.1088/0004-637X/700/1/1},
  archiveprefix	= {arXiv},
  eprint	= {0904.0452},
  primaryclass	= {astro-ph.CO},
  adsurl	= {https://ui.adsabs.harvard.edu/abs/2009ApJ...700....1G}
}

@Article{	  Geach16,
  author	= {{Geach}, J.~E. and {Narayanan}, D. and {Matsuda}, Y. and
		  {Hayes}, M. and {Mas-Ribas}, Ll. and {Dijkstra}, M. and
		  {Steidel}, C.~C. and {Chapman}, S.~C. and {Feldmann}, R.
		  and {Avison}, A. and {Agertz}, O. and {Ao}, Y. and
		  {Birkinshaw}, M. and {Bremer}, M.~N. and {Clements}, D.~L.
		  and {Dannerbauer}, H. and {Farrah}, D. and {Harrison},
		  C.~M. and {Kubo}, M. and {Micha{\l}owski}, M.~J. and
		  {Scott}, Douglas and {Smith}, D.~J.~B. and {Spaans}, M. and
		  {Simpson}, J.~M. and {Swinbank}, A.~M. and {Taniguchi}, Y.
		  and {van der Werf}, P. and {Verma}, A. and {Yamada}, T.},
  title		= "{ALMA Observations of Ly{\ensuremath{\alpha}} Blob 1: Halo
		  Substructure Illuminated from Within}",
  journal	= {\apj},
  year		= 2016,
  month		= nov,
  volume	= {832},
  number	= {1},
  eid		= {37},
  pages		= {37},
  doi		= {10.3847/0004-637X/832/1/37},
  archiveprefix	= {arXiv},
  eprint	= {1608.02941},
  primaryclass	= {astro-ph.GA},
  adsurl	= {https://ui.adsabs.harvard.edu/abs/2016ApJ...832...37G}
}

@Article{	  Gronke17b,
  author	= {{Gronke}, Max},
  title		= "{Modeling 237 Lyman-{\ensuremath{\alpha}} spectra of the
		  MUSE-Wide survey}",
  journal	= {\aap},
  year		= 2017,
  month		= dec,
  volume	= {608},
  eid		= {A139},
  pages		= {A139},
  doi		= {10.1051/0004-6361/201731791},
  archiveprefix	= {arXiv},
  eprint	= {1709.07008},
  primaryclass	= {astro-ph.GA},
  adsurl	= {https://ui.adsabs.harvard.edu/abs/2017A&A...608A.139G}
}

@Article{	  Hasinger18,
  author	= {{Hasinger}, G. and {Capak}, P. and {Salvato}, M. and
		  {Barger}, A.~J. and {Cowie}, L.~L. and {Faisst}, A. and
		  {Hemmati}, S. and {Kakazu}, Y. and {Kartaltepe}, J. and
		  {Masters}, D. and {Mobasher}, B. and {Nayyeri}, H. and
		  {Sanders}, D. and {Scoville}, N.~Z. and {Suh}, H. and
		  {Steinhardt}, C. and {Yang}, Fengwei},
  title		= "{The DEIMOS 10K Spectroscopic Survey Catalog of the COSMOS
		  Field}",
  journal	= {\apj},
  year		= 2018,
  month		= may,
  volume	= {858},
  number	= {2},
  eid		= {77},
  pages		= {77},
  doi		= {10.3847/1538-4357/aabacf},
  archiveprefix	= {arXiv},
  eprint	= {1803.09251},
  primaryclass	= {astro-ph.GA},
  adsurl	= {https://ui.adsabs.harvard.edu/abs/2018ApJ...858...77H}
}

@Article{	  Hayes11,
  author	= {{Hayes}, Matthew and {Scarlata}, Claudia and {Siana},
		  Brian},
  title		= "{Central powering of the largest
		  Lyman-{\ensuremath{\alpha}} nebula is revealed by polarized
		  radiation}",
  journal	= {\nat},
  year		= 2011,
  month		= aug,
  volume	= {476},
  number	= {7360},
  pages		= {304-307},
  doi		= {10.1038/nature10320},
  archiveprefix	= {arXiv},
  eprint	= {1108.3332},
  primaryclass	= {astro-ph.CO},
  adsurl	= {https://ui.adsabs.harvard.edu/abs/2011Natur.476..304H}
}

@Article{	  Hennawi15,
  author	= {{Hennawi}, Joseph F. and {Prochaska}, J. Xavier and
		  {Cantalupo}, Sebastiano and {Arrigoni-Battaia}, Fabrizio},
  title		= "{Quasar quartet embedded in giant nebula reveals rare
		  massive structure in distant universe}",
  journal	= {Science},
  year		= 2015,
  month		= may,
  volume	= {348},
  number	= {6236},
  pages		= {779-783},
  doi		= {10.1126/science.aaa5397},
  archiveprefix	= {arXiv},
  eprint	= {1505.03786},
  primaryclass	= {astro-ph.GA},
  adsurl	= {https://ui.adsabs.harvard.edu/abs/2015Sci...348..779H}
}

@Article{	  Hogg08,
  author	= {{Hogg}, David W.},
  title		= "{Data analysis recipes: Choosing the binning for a
		  histogram}",
  journal	= {arXiv e-prints},
  year		= 2008,
  month		= jul,
  eid		= {arXiv:0807.4820},
  pages		= {arXiv:0807.4820},
  doi		= {10.48550/arXiv.0807.4820},
  archiveprefix	= {arXiv},
  eprint	= {0807.4820},
  primaryclass	= {physics.data-an},
  adsurl	= {https://ui.adsabs.harvard.edu/abs/2008arXiv0807.4820H}
}

@Article{	  Hong19,
  author	= {{Hong}, Sungryong and {Dey}, Arjun and {Lee}, Kyoung-Soo
		  and {Orsi}, {\'A}lvaro A. and {Gebhardt}, Karl and
		  {Vogelsberger}, Mark and {Hernquist}, Lars and {Xue}, Rui
		  and {Jung}, Intae and {Finklestein}, Steven L. and
		  {Tuttle}, Sarah and {Boylan-Kolchin}, Michael},
  title		= "{Statistics of two-point correlation and network topology
		  for Ly {\ensuremath{\alpha}} emitters at z
		  {\ensuremath{\approx}} 2.67}",
  journal	= {\mnras},
  year		= 2019,
  month		= mar,
  volume	= {483},
  number	= {3},
  pages		= {3950-3970},
  doi		= {10.1093/mnras/sty3219},
  archiveprefix	= {arXiv},
  eprint	= {1811.10631},
  primaryclass	= {astro-ph.CO},
  adsurl	= {https://ui.adsabs.harvard.edu/abs/2019MNRAS.483.3950H}
}

@Article{	  Kato18,
  author	= {{Kato}, Yuta and {Matsuda}, Yuichi and {Iono}, Daisuke and
		  {Hatsukade}, Bunyo and {Umehata}, Hideki and {Kohno},
		  Kotaro and {Alexander}, David M. and {Ao}, Yiping and
		  {Chapman}, Scott C. and {Hayes}, Matthew and {Kubo}, Mariko
		  and {Lehmer}, Bret D. and {Malkan}, Matthew A. and
		  {Michiyama}, Tomonari and {Nagao}, Tohru and {Saito},
		  Tomoki and {Tanaka}, Ichi and {Taniguchi}, Yoshiaki},
  title		= "{A high dust emissivity index {\ensuremath{\beta}} for a
		  CO-faint galaxy in a filamentary Ly{\ensuremath{\alpha}}
		  nebula at z = 3.1}",
  journal	= {\pasj},
  year		= 2018,
  month		= oct,
  volume	= {70},
  number	= {5},
  eid		= {L6},
  pages		= {L6},
  doi		= {10.1093/pasj/psy087},
  archiveprefix	= {arXiv},
  eprint	= {1807.07933},
  primaryclass	= {astro-ph.GA},
  adsurl	= {https://ui.adsabs.harvard.edu/abs/2018PASJ...70L...6K}
}

@Article{	  Keel09,
  author	= {{Keel}, William C. and {White}, Raymond E., III and
		  {Chapman}, Scott and {Windhorst}, Rogier A.},
  title		= "{The Disappearance of Ly{\ensuremath{\alpha}} Blobs: A
		  Galex Search at z = 0.8}",
  journal	= {\aj},
  year		= 2009,
  month		= sep,
  volume	= {138},
  number	= {3},
  pages		= {986-990},
  doi		= {10.1088/0004-6256/138/3/986},
  archiveprefix	= {arXiv},
  eprint	= {0907.2201},
  primaryclass	= {astro-ph.GA},
  adsurl	= {https://ui.adsabs.harvard.edu/abs/2009AJ....138..986K}
}

@Article{	  Keel99,
  author	= {{Keel}, William C. and {Cohen}, Seth H. and {Windhorst},
		  Rogier A. and {Waddington}, Ian},
  title		= "{Evidence for Large-Scale Structure at z
		  \raisebox{-0.5ex}\textasciitilde 2.4 from
		  Ly{\ensuremath{\alpha}} Imaging}",
  journal	= {\aj},
  year		= 1999,
  month		= dec,
  volume	= {118},
  number	= {6},
  pages		= {2547-2560},
  doi		= {10.1086/301139},
  archiveprefix	= {arXiv},
  eprint	= {astro-ph/9908183},
  primaryclass	= {astro-ph},
  adsurl	= {https://ui.adsabs.harvard.edu/abs/1999AJ....118.2547K}
}

@Article{	  Kikuta19,
  author	= {{Kikuta}, Satoshi and {Matsuda}, Yuichi and {Cen}, Renyue
		  and {Steidel}, Charles C. and {Yagi}, Masafumi and
		  {Hayashino}, Tomoki and {Imanishi}, Masatoshi and
		  {Komiyama}, Yutaka and {Momose}, Rieko and {Saito},
		  Tomoki},
  title		= "{Ly{\ensuremath{\alpha}} view around a z = 2.84
		  hyperluminous QSO at a node of the cosmic
		  web$^{{\textdagger}}$}",
  journal	= {\pasj},
  year		= 2019,
  month		= jun,
  volume	= {71},
  number	= {3},
  eid		= {L2},
  pages		= {L2},
  doi		= {10.1093/pasj/psz055},
  archiveprefix	= {arXiv},
  eprint	= {1904.07747},
  primaryclass	= {astro-ph.GA},
  adsurl	= {https://ui.adsabs.harvard.edu/abs/2019PASJ...71L...2K}
}

@Article{	  Kriek07,
  author	= {{Kriek}, Mariska and {van Dokkum}, Pieter G. and {Franx},
		  Marijn and {Illingworth}, Garth D. and {Coppi}, Paolo and
		  {F{\"o}rster Schreiber}, Natascha M. and {Gawiser}, Eric
		  and {Labb{\'e}}, Ivo and {Lira}, Paulina and {Marchesini},
		  Danilo and {Quadri}, Ryan and {Rudnick}, Gregory and
		  {Taylor}, Edward N. and {Urry}, C. Megan and {van der
		  Werf}, Paul P.},
  title		= "{The Origin of Line Emission in Massive z
		  \raisebox{-0.5ex}\textasciitilde 2.3 Galaxies: Evidence for
		  Cosmic Downsizing of AGN Host Galaxies}",
  journal	= {\apj},
  year		= 2007,
  month		= nov,
  volume	= {669},
  number	= {2},
  pages		= {776-790},
  doi		= {10.1086/520789},
  archiveprefix	= {arXiv},
  eprint	= {astro-ph/0611724},
  primaryclass	= {astro-ph},
  adsurl	= {https://ui.adsabs.harvard.edu/abs/2007ApJ...669..776K}
}

@Article{	  Kusakabe22,
  author	= {{Kusakabe}, Haruka and {Verhamme}, Anne and {Blaizot},
		  J{\'e}r{\'e}my and {Garel}, Thibault and {Wisotzki}, Lutz
		  and {Leclercq}, Floriane and {Bacon}, Roland and {Schaye},
		  Joop and {Gallego}, Sofia G. and {Kerutt}, Josephine and
		  {Matthee}, Jorryt and {Maseda}, Michael and {Nanayakkara},
		  Themiya and {Pell{\'o}}, Roser and {Richard}, Johan and
		  {Tresse}, Laurence and {Urrutia}, Tanya and {Vitte},
		  Elo{\"\i}se},
  title		= "{The MUSE eXtremely Deep Field: Individual detections of
		  Ly{\ensuremath{\alpha}} haloes around rest-frame
		  UV-selected galaxies at z ≃ 2.9-4.4}",
  journal	= {\aap},
  year		= 2022,
  month		= apr,
  volume	= {660},
  eid		= {A44},
  pages		= {A44},
  doi		= {10.1051/0004-6361/202142302},
  archiveprefix	= {arXiv},
  eprint	= {2201.07257},
  primaryclass	= {astro-ph.GA},
  adsurl	= {https://ui.adsabs.harvard.edu/abs/2022A&A...660A..44K}
}

@Misc{		  Lang16,
  author	= {{Lang}, Dustin and {Hogg}, David W. and {Mykytyn}, David},
  title		= "{The Tractor: Probabilistic astronomical source detection
		  and measurement}",
  howpublished	= {Astrophysics Source Code Library, record ascl:1604.008},
  year		= 2016,
  month		= apr,
  eid		= {ascl:1604.008},
  pages		= {ascl:1604.008},
  archiveprefix	= {ascl},
  eprint	= {1604.008},
  adsurl	= {https://ui.adsabs.harvard.edu/abs/2016ascl.soft04008L}
}

@Article{	  Langen23,
  author	= {{Langen}, Vivienne and {Cantalupo}, Sebastiano and
		  {Steidel}, Charles C. and {Chen}, Yuguang and {Pezzulli},
		  Gabriele and {Gallego}, Sofia G.},
  title		= "{Characterizing the circumgalactic medium of quasars at z
		  2.2 through H {\ensuremath{\alpha}} and Ly
		  {\ensuremath{\alpha}} emission}",
  journal	= {\mnras},
  year		= 2023,
  month		= mar,
  volume	= {519},
  number	= {4},
  pages		= {5099-5113},
  doi		= {10.1093/mnras/stac3205},
  archiveprefix	= {arXiv},
  eprint	= {2303.05531},
  primaryclass	= {astro-ph.GA},
  adsurl	= {https://ui.adsabs.harvard.edu/abs/2023MNRAS.519.5099L}
}

@Article{	  Leclercq17,
  author	= {{Leclercq}, Floriane and {Bacon}, Roland and {Wisotzki},
		  Lutz and {Mitchell}, Peter and {Garel}, Thibault and
		  {Verhamme}, Anne and {Blaizot}, J{\'e}r{\'e}my and
		  {Hashimoto}, Takuya and {Herenz}, Edmund Christian and
		  {Conseil}, Simon and {Cantalupo}, Sebastiano and {Inami},
		  Hanae and {Contini}, Thierry and {Richard}, Johan and
		  {Maseda}, Michael and {Schaye}, Joop and {Marino},
		  Raffaella Anna and {Akhlaghi}, Mohammad and {Brinchmann},
		  Jarle and {Carollo}, Marcella},
  title		= "{The MUSE Hubble Ultra Deep Field Survey. VIII. Extended
		  Lyman-{\ensuremath{\alpha}} haloes around high-z
		  star-forming galaxies}",
  journal	= {\aap},
  year		= 2017,
  month		= dec,
  volume	= {608},
  eid		= {A8},
  pages		= {A8},
  doi		= {10.1051/0004-6361/201731480},
  archiveprefix	= {arXiv},
  eprint	= {1710.10271},
  primaryclass	= {astro-ph.GA},
  adsurl	= {https://ui.adsabs.harvard.edu/abs/2017A&A...608A...8L}
}

@Article{	  Lilly09,
  author	= {{Lilly}, Simon J. and {Le Brun}, Vincent and {Maier},
		  Christian and {Mainieri}, Vincenzo and {Mignoli}, Marco and
		  {Scodeggio}, Marco and {Zamorani}, Gianni and {Carollo},
		  Marcella and {Contini}, Thierry and {Kneib}, Jean-Paul and
		  {Le F{\`e}vre}, Olivier and {Renzini}, Alvio and
		  {Bardelli}, Sandro and {Bolzonella}, Micol and {Bongiorno},
		  Angela and {Caputi}, Karina and {Coppa}, Graziano and
		  {Cucciati}, Olga and {de la Torre}, Sylvain and {de Ravel},
		  Loic and {Franzetti}, Paolo and {Garilli}, Bianca and
		  {Iovino}, Angela and {Kampczyk}, Pawel and {Kovac},
		  Katarina and {Knobel}, Christian and {Lamareille}, Fabrice
		  and {Le Borgne}, Jean-Francois and {Pello}, Roser and
		  {Peng}, Yingjie and {P{\'e}rez-Montero}, Enrique and
		  {Ricciardelli}, Elena and {Silverman}, John D. and
		  {Tanaka}, Masayuki and {Tasca}, Lidia and {Tresse},
		  Laurence and {Vergani}, Daniela and {Zucca}, Elena and
		  {Ilbert}, Olivier and {Salvato}, Mara and {Oesch}, Pascal
		  and {Abbas}, Umi and {Bottini}, Dario and {Capak}, Peter
		  and {Cappi}, Alberto and {Cassata}, Paolo and {Cimatti},
		  Andrea and {Elvis}, Martin and {Fumana}, Marco and {Guzzo},
		  Luigi and {Hasinger}, Gunther and {Koekemoer}, Anton and
		  {Leauthaud}, Alexei and {Maccagni}, Dario and {Marinoni},
		  Christian and {McCracken}, Henry and {Memeo}, Pierdomenico
		  and {Meneux}, Baptiste and {Porciani}, Cristiano and
		  {Pozzetti}, Lucia and {Sanders}, David and {Scaramella},
		  Roberto and {Scarlata}, Claudia and {Scoville}, Nick and
		  {Shopbell}, Patrick and {Taniguchi}, Yoshiaki},
  title		= "{The zCOSMOS 10k-Bright Spectroscopic Sample}",
  journal	= {\apjs},
  year		= 2009,
  month		= oct,
  volume	= {184},
  number	= {2},
  pages		= {218-229},
  doi		= {10.1088/0067-0049/184/2/218},
  adsurl	= {https://ui.adsabs.harvard.edu/abs/2009ApJS..184..218L}
}

@Article{	  Madau14,
  author	= {{Madau}, Piero and {Dickinson}, Mark},
  title		= "{Cosmic Star-Formation History}",
  journal	= {\araa},
  year		= 2014,
  month		= aug,
  volume	= {52},
  pages		= {415-486},
  doi		= {10.1146/annurev-astro-081811-125615},
  archiveprefix	= {arXiv},
  eprint	= {1403.0007},
  primaryclass	= {astro-ph.CO},
  adsurl	= {https://ui.adsabs.harvard.edu/abs/2014ARA&A..52..415M}
}

@Article{	  Matsuda04,
  author	= {{Matsuda}, Yuichi and {Yamada}, Toru and {Hayashino},
		  Tomoki and {Tamura}, Hajime and {Yamauchi}, Ryosuke and
		  {Ajiki}, Masaru and {Fujita}, Shinobu S. and {Murayama},
		  Takashi and {Nagao}, Tohru and {Ohta}, Kouji and {Okamura},
		  Sadanori and {Ouchi}, Masami and {Shimasaku}, Kazuhiro and
		  {Shioya}, Yasuhiro and {Taniguchi}, Yoshiaki},
  title		= "{A Subaru Search for Ly{\ensuremath{\alpha}} Blobs in and
		  around the Protocluster Region At Redshift z = 3.1}",
  journal	= {\aj},
  year		= 2004,
  month		= aug,
  volume	= {128},
  number	= {2},
  pages		= {569-584},
  doi		= {10.1086/422020},
  archiveprefix	= {arXiv},
  eprint	= {astro-ph/0405221},
  primaryclass	= {astro-ph},
  adsurl	= {https://ui.adsabs.harvard.edu/abs/2004AJ....128..569M}
}

@Article{	  Matsuda11,
  author	= {{Matsuda}, Y. and {Yamada}, T. and {Hayashino}, T. and
		  {Yamauchi}, R. and {Nakamura}, Y. and {Morimoto}, N. and
		  {Ouchi}, M. and {Ono}, Y. and {Kousai}, K. and {Nakamura},
		  E. and {Horie}, M. and {Fujii}, T. and {Umemura}, M. and
		  {Mori}, M.},
  title		= "{The Subaru Ly{\ensuremath{\alpha}} blob survey: a sample
		  of 100-kpc Ly{\ensuremath{\alpha}} blobs at z = 3}",
  journal	= {\mnras},
  year		= 2011,
  month		= jan,
  volume	= {410},
  number	= {1},
  pages		= {L13-L17},
  doi		= {10.1111/j.1745-3933.2010.00969.x},
  archiveprefix	= {arXiv},
  eprint	= {1010.2877},
  primaryclass	= {astro-ph.CO},
  adsurl	= {https://ui.adsabs.harvard.edu/abs/2011MNRAS.410L..13M}
}

@ARTICLE{Mingyu24,
       author = {{Li}, Mingyu and {Zhang}, Haibin and {Cai}, Zheng and {Liang}, Yongming and {Kashikawa}, Nobunari and {Ma}, Ke and {Fan}, Xiaohui and {Prochaska}, J. Xavier and {Emonts}, Bjorn H.~C. and {Wang}, Xin and {Wu}, Yunjing and {Zhang}, Shiwu and {Li}, Qiong and {Johnson}, Sean D. and {Yue}, Minghao and {Arrigoni Battaia}, Fabrizio and {Cantalupo}, Sebastiano and {Hennawi}, Joseph F. and {Kikuta}, Satoshi and {Ning}, Yuanhang and {Ouchi}, Masami and {Shimakawa}, Rhythm and {Wang}, Ben and {Wang}, Weichen and {Zheng}, Zheng and {Zheng}, Zhen-Ya},
        title = "{MAMMOTH-Subaru. II. Diverse Populations of Circumgalactic Ly{\ensuremath{\alpha}} Nebulae at Cosmic Noon}",
      journal = {\apjs},
         year = 2024,
        month = dec,
       volume = {275},
       number = {2},
          eid = {27},
        pages = {27},
          doi = {10.3847/1538-4365/ad812c},
archivePrefix = {arXiv},
       eprint = {2405.13113},
 primaryClass = {astro-ph.GA},
       adsurl = {https://ui.adsabs.harvard.edu/abs/2024ApJS..275...27L}
}

@Article{	  Momcheva16,
  author	= {{Momcheva}, Ivelina G. and {Brammer}, Gabriel B. and {van
		  Dokkum}, Pieter G. and {Skelton}, Rosalind E. and
		  {Whitaker}, Katherine E. and {Nelson}, Erica J. and
		  {Fumagalli}, Mattia and {Maseda}, Michael V. and {Leja},
		  Joel and {Franx}, Marijn and {Rix}, Hans-Walter and
		  {Bezanson}, Rachel and {Da Cunha}, Elisabete and {Dickey},
		  Claire and {F{\"o}rster Schreiber}, Natascha M. and
		  {Illingworth}, Garth and {Kriek}, Mariska and {Labb{\'e}},
		  Ivo and {Ulf Lange}, Johannes and {Lundgren}, Britt F. and
		  {Magee}, Daniel and {Marchesini}, Danilo and {Oesch},
		  Pascal and {Pacifici}, Camilla and {Patel}, Shannon G. and
		  {Price}, Sedona and {Tal}, Tomer and {Wake}, David A. and
		  {van der Wel}, Arjen and {Wuyts}, Stijn},
  title		= "{The 3D-HST Survey: Hubble Space Telescope WFC3/G141 Grism
		  Spectra, Redshifts, and Emission Line Measurements for
		  \raisebox{-0.5ex}\textasciitilde 100,000 Galaxies}",
  journal	= {\apjs},
  year		= 2016,
  month		= aug,
  volume	= {225},
  number	= {2},
  eid		= {27},
  pages		= {27},
  doi		= {10.3847/0067-0049/225/2/27},
  archiveprefix	= {arXiv},
  eprint	= {1510.02106},
  primaryclass	= {astro-ph.GA},
  adsurl	= {https://ui.adsabs.harvard.edu/abs/2016ApJS..225...27M}
}

@Article{	  Momose16,
  author	= {{Momose}, Rieko and {Ouchi}, Masami and {Nakajima},
		  Kimihiko and {Ono}, Yoshiaki and {Shibuya}, Takatoshi and
		  {Shimasaku}, Kazuhiro and {Yuma}, Suraphong and {Mori},
		  Masao and {Umemura}, Masayuki},
  title		= "{Statistical properties of diffuse Ly{\ensuremath{\alpha}}
		  haloes around star-forming galaxies at z
		  {\ensuremath{\sim}} 2}",
  journal	= {\mnras},
  year		= 2016,
  month		= apr,
  volume	= {457},
  number	= {3},
  pages		= {2318-2330},
  doi		= {10.1093/mnras/stw021},
  archiveprefix	= {arXiv},
  eprint	= {1509.09001},
  primaryclass	= {astro-ph.GA},
  adsurl	= {https://ui.adsabs.harvard.edu/abs/2016MNRAS.457.2318M}
}

@ARTICLE{Firestone24,
       author = {{Firestone}, Nicole M. and {Gawiser}, Eric and {Ramakrishnan}, Vandana and {Lee}, Kyoung-Soo and {Valdes}, Francisco and {Park}, Changbom and {Yang}, Yujin and {Ciardullo}, Robin and {Artale}, Mar{\'\i}a Celeste and {Benda}, Barbara and {Broussard}, Adam and {Eid}, Lana and {Farooq}, Rameen and {Gronwall}, Caryl and {Guaita}, Lucia and {Gwyn}, Stephen and {Hwang}, Ho Seong and {Im}, Sang Hyeok and {Jeong}, Woong-Seob and {Karthikeyan}, Shreya and {Lang}, Dustin and {Moon}, Byeongha and {Padilla}, Nelson and {Sawicki}, Marcin and {Seo}, Eunsuk and {Singh}, Akriti and {Song}, Hyunmi and {Troncoso Iribarren}, Paulina},
        title = "{ODIN: Improved Narrowband Ly{\ensuremath{\alpha}} Emitter Selection Techniques for z = 2.4, 3.1, and 4.5}",
      journal = {\apj},
         year = 2024,
        month = oct,
       volume = {974},
       number = {2},
          eid = {217},
        pages = {217},
          doi = {10.3847/1538-4357/ad71c9},
archivePrefix = {arXiv},
       eprint = {2312.16075},
 primaryClass = {astro-ph.GA},
       adsurl = {https://ui.adsabs.harvard.edu/abs/2024ApJ...974..217F}
}

@Article{	  Ouchi09,
  author	= {{Ouchi}, Masami and {Ono}, Yoshiaki and {Egami}, Eiichi
		  and {Saito}, Tomoki and {Oguri}, Masamune and {McCarthy},
		  Patrick J. and {Farrah}, Duncan and {Kashikawa}, Nobunari
		  and {Momcheva}, Ivelina and {Shimasaku}, Kazuhiro and
		  {Nakanishi}, Kouichiro and {Furusawa}, Hisanori and
		  {Akiyama}, Masayuki and {Dunlop}, James S. and {Mortier},
		  Angela M.~J. and {Okamura}, Sadanori and {Hayashi}, Masao
		  and {Cirasuolo}, Michele and {Dressler}, Alan and {Iye},
		  Masanori and {Jarvis}, Matt J. and {Kodama}, Tadayuki and
		  {Martin}, Crystal L. and {McLure}, Ross J. and {Ohta},
		  Kouji and {Yamada}, Toru and {Yoshida}, Michitoshi},
  title		= "{Discovery of a Giant Ly{\ensuremath{\alpha}} Emitter Near
		  the Reionization Epoch}",
  journal	= {\apj},
  year		= 2009,
  month		= may,
  volume	= {696},
  number	= {2},
  pages		= {1164-1175},
  doi		= {10.1088/0004-637X/696/2/1164},
  archiveprefix	= {arXiv},
  eprint	= {0807.4174},
  primaryclass	= {astro-ph},
  adsurl	= {https://ui.adsabs.harvard.edu/abs/2009ApJ...696.1164O}
}

@Article{	  Ouchi20,
  author	= {{Ouchi}, Masami and {Ono}, Yoshiaki and {Shibuya},
		  Takatoshi},
  title		= "{Observations of the Lyman-{\ensuremath{\alpha}}
		  Universe}",
  journal	= {\araa},
  year		= 2020,
  month		= aug,
  volume	= {58},
  pages		= {617-659},
  doi		= {10.1146/annurev-astro-032620-021859},
  archiveprefix	= {arXiv},
  eprint	= {2012.07960},
  primaryclass	= {astro-ph.GA},
  adsurl	= {https://ui.adsabs.harvard.edu/abs/2020ARA&A..58..617O}
}

@Article{	  Paris14,
  author	= {{P{\^a}ris}, Isabelle and {Petitjean}, Patrick and
		  {Aubourg}, {\'E}ric and {Ross}, Nicholas P. and {Myers},
		  Adam D. and {Streblyanska}, Alina and {Bailey}, Stephen and
		  {Hall}, Patrick B. and {Strauss}, Michael A. and
		  {Anderson}, Scott F. and {Bizyaev}, Dmitry and {Borde},
		  Arnaud and {Brinkmann}, J. and {Bovy}, Jo and {Brandt},
		  William N. and {Brewington}, Howard and {Brownstein}, Joel
		  R. and {Cook}, Benjamin A. and {Ebelke}, Garrett and {Fan},
		  Xiaohui and {Filiz Ak}, Nurten and {Finley}, Hayley and
		  {Font-Ribera}, Andreu and {Ge}, Jian and {Hamann}, Fred and
		  {Ho}, Shirley and {Jiang}, Linhua and {Kinemuchi}, Karen
		  and {Malanushenko}, Elena and {Malanushenko}, Viktor and
		  {Marchante}, Moses and {McGreer}, Ian D. and {McMahon},
		  Richard G. and {Miralda-Escud{\'e}}, Jordi and {Muna},
		  Demitri and {Noterdaeme}, Pasquier and {Oravetz}, Daniel
		  and {Palanque-Delabrouille}, Nathalie and {Pan}, Kaike and
		  {Perez-Fournon}, Isma{\"e}l and {Pieri}, Matthew and
		  {Riffel}, Rog{\'e}rio and {Schlegel}, David J. and
		  {Schneider}, Donald P. and {Simmons}, Audrey and {Viel},
		  Matteo and {Weaver}, Benjamin A. and {Wood-Vasey}, W.
		  Michael and {Y{\`e}che}, Christophe and {York}, Donald G.},
  title		= "{The Sloan Digital Sky Survey quasar catalog: tenth data
		  release}",
  journal	= {\aap},
  year		= 2014,
  month		= mar,
  volume	= {563},
  eid		= {A54},
  pages		= {A54},
  doi		= {10.1051/0004-6361/201322691},
  archiveprefix	= {arXiv},
  eprint	= {1311.4870},
  primaryclass	= {astro-ph.CO},
  adsurl	= {https://ui.adsabs.harvard.edu/abs/2014A&A...563A..54P}
}

@Article{	  Prescott08,
  author	= {{Prescott}, Moire K.~M. and {Kashikawa}, Nobunari and
		  {Dey}, Arjun and {Matsuda}, Yuichi},
  title		= "{The Overdense Environment of a Large
		  Ly{\ensuremath{\alpha}} Nebula at z {\ensuremath{\approx}}
		  2.7}",
  journal	= {\apjl},
  year		= 2008,
  month		= may,
  volume	= {678},
  number	= {2},
  pages		= {L77},
  doi		= {10.1086/588606},
  archiveprefix	= {arXiv},
  eprint	= {0803.4230},
  primaryclass	= {astro-ph},
  adsurl	= {https://ui.adsabs.harvard.edu/abs/2008ApJ...678L..77P}
}

@Article{	  Prescott12a,
  author	= {{Prescott}, Moire K.~M. and {Dey}, Arjun and {Brodwin},
		  Mark and {Chaffee}, Frederic H. and {Desai}, Vandana and
		  {Eisenhardt}, Peter and {Le Floc'h}, Emeric and {Jannuzi},
		  Buell T. and {Kashikawa}, Nobunari and {Matsuda}, Yuichi
		  and {Soifer}, B.~T.},
  title		= "{Resolving the Galaxies within a Giant
		  Ly{\ensuremath{\alpha}} Nebula: Witnessing the Formation of
		  a Galaxy Group?}",
  journal	= {\apj},
  year		= 2012,
  month		= jun,
  volume	= {752},
  number	= {2},
  eid		= {86},
  pages		= {86},
  doi		= {10.1088/0004-637X/752/2/86},
  archiveprefix	= {arXiv},
  eprint	= {1111.0630},
  primaryclass	= {astro-ph.CO},
  adsurl	= {https://ui.adsabs.harvard.edu/abs/2012ApJ...752...86P}
}

@Article{	  Prescott12b,
  author	= {{Prescott}, Moire K.~M. and {Dey}, Arjun and {Jannuzi},
		  Buell T.},
  title		= "{A Successful Broadband Survey for Giant
		  Ly{\ensuremath{\alpha}} Nebulae. I. Survey Design and
		  Candidate Selection}",
  journal	= {\apj},
  year		= 2012,
  month		= apr,
  volume	= {748},
  number	= {2},
  eid		= {125},
  pages		= {125},
  doi		= {10.1088/0004-637X/748/2/125},
  archiveprefix	= {arXiv},
  eprint	= {1111.2603},
  primaryclass	= {astro-ph.CO},
  adsurl	= {https://ui.adsabs.harvard.edu/abs/2012ApJ...748..125P}
}

@Article{	  Ramakrishnan23,
  author	= {{Ramakrishnan}, Vandana and {Moon}, Byeongha and {Im},
		  Sang Hyeok and {Farooq}, Rameen and {Lee}, Kyoung-Soo and
		  {Gawiser}, Eric and {Yang}, Yujin and {Park}, Changbom and
		  {Hwang}, Ho Seong and {Valdes}, Francisco and {Artale},
		  Maria Celeste and {Ciardullo}, Robin and {Dey}, Arjun and
		  {Gronwall}, Caryl and {Guaita}, Lucia and {Jeong},
		  Woong-Seob and {Padilla}, Nelson and {Singh}, Akriti and
		  {Zabludoff}, Ann},
  title		= "{ODIN: Where Do Ly{\ensuremath{\alpha}} Blobs Live?
		  Contextualizing Blob Environments within Large-scale
		  Structure}",
  journal	= {\apj},
  year		= 2023,
  month		= jul,
  volume	= {951},
  number	= {2},
  eid		= {119},
  pages		= {119},
  doi		= {10.3847/1538-4357/acd341},
  archiveprefix	= {arXiv},
  eprint	= {2302.07860},
  primaryclass	= {astro-ph.GA},
  adsurl	= {https://ui.adsabs.harvard.edu/abs/2023ApJ...951..119R}
}

@ARTICLE{Ramakrishnan24,
       author = {{Ramakrishnan}, Vandana and {Lee}, Kyoung-Soo and {Artale}, Maria Celeste and {Gawiser}, Eric and {Yang}, Yujin and {Park}, Changbom and {Chiang}, Yi-Kuan and {Ciardullo}, Robin and {Dey}, Arjun and {Gronwall}, Caryl and {Guaita}, Lucia and {Hwang}, Ho Seong and {Im}, Sang Hyeok and {Jeong}, Woong-Seob and {Kim}, Seongjae and {Kumar}, Ankit and {Lee}, Jaehyun and {Lee}, Seong-Kook and {Moon}, Byeongha and {Padilla}, Nelson and {Pope}, Alexandra and {Popescu}, Roxana and {Singh}, Akriti and {Song}, Hyunmi and {Troncoso}, Paulina and {Valdes}, Francisco and {Zabludoff}, Ann},
        title = "{ODIN: Identifying Protoclusters and Cosmic Filaments Traced by Ly{\ensuremath{\alpha}}-emitting Galaxies}",
      journal = {\apj},
         year = 2024,
        month = dec,
       volume = {977},
       number = {1},
          eid = {119},
        pages = {119},
          doi = {10.3847/1538-4357/ad83cb},
archivePrefix = {arXiv},
       eprint = {2406.08645},
 primaryClass = {astro-ph.GA},
       adsurl = {https://ui.adsabs.harvard.edu/abs/2024ApJ...977..119R}
}

@ARTICLE{Ramakrishnan25,
       author = {{Ramakrishnan}, Vandana and {Lee}, Kyoung-Soo and {Firestone}, Nicole and {Gawiser}, Eric and {Artale}, Maria Celeste and {Gronwall}, Caryl and {Guaita}, Lucia and {Hwang}, Ho Seong and {Im}, Sang Hyeok and {Jeong}, Woong-Seob and {Kim}, Seongjae and {Kumar}, Ankit and {Lee}, Jaehyun and {Moon}, Byeongha and {Padilla}, Nelson and {Park}, Changbom and {Singh}, Akriti and {Song}, Hyunmi and {Troncoso Iribarren}, Paulina and {Yang}, Yujin},
        title = "{ODIN: High Clustering Strength of Protoclusters at Cosmic Noon}",
      journal = {\apj},
         year = 2025,
        month = apr,
       volume = {982},
       number = {2},
          eid = {74},
        pages = {74},
          doi = {10.3847/1538-4357/adb624},
archivePrefix = {arXiv},
       eprint = {2410.18341},
 primaryClass = {astro-ph.GA},
       adsurl = {https://ui.adsabs.harvard.edu/abs/2025ApJ...982...74R}
}

@Article{	  Rosdahl12,
  author	= {{Rosdahl}, J. and {Blaizot}, J.},
  title		= "{Extended Ly{\ensuremath{\alpha}} emission from cold
		  accretion streams}",
  journal	= {\mnras},
  year		= 2012,
  month		= jun,
  volume	= {423},
  number	= {1},
  pages		= {344-366},
  doi		= {10.1111/j.1365-2966.2012.20883.x},
  archiveprefix	= {arXiv},
  eprint	= {1112.4408},
  primaryclass	= {astro-ph.CO},
  adsurl	= {https://ui.adsabs.harvard.edu/abs/2012MNRAS.423..344R}
}

@Article{	  Saito06,
  author	= {{Saito}, Tomoki and {Shimasaku}, Kazuhiro and {Okamura},
		  Sadanori and {Ouchi}, Masami and {Akiyama}, Masayuki and
		  {Yoshida}, Michitoshi},
  title		= "{Systematic Survey of Extended Ly{\ensuremath{\alpha}}
		  Sources over z \raisebox{-0.5ex}\textasciitilde 3-5}",
  journal	= {\apj},
  year		= 2006,
  month		= sep,
  volume	= {648},
  number	= {1},
  pages		= {54-66},
  doi		= {10.1086/505678},
  archiveprefix	= {arXiv},
  eprint	= {astro-ph/0605360},
  primaryclass	= {astro-ph},
  adsurl	= {https://ui.adsabs.harvard.edu/abs/2006ApJ...648...54S}
}

@Article{	  Shibuya18,
  author	= {{Shibuya}, Takatoshi and {Ouchi}, Masami and {Konno},
		  Akira and {Higuchi}, Ryo and {Harikane}, Yuichi and {Ono},
		  Yoshiaki and {Shimasaku}, Kazuhiro and {Taniguchi},
		  Yoshiaki and {Kobayashi}, Masakazu A.~R. and {Kajisawa},
		  Masaru and {Nagao}, Tohru and {Furusawa}, Hisanori and
		  {Goto}, Tomotsugu and {Kashikawa}, Nobunari and {Komiyama},
		  Yutaka and {Kusakabe}, Haruka and {Lee}, Chien-Hsiu and
		  {Momose}, Rieko and {Nakajima}, Kimihiko and {Tanaka},
		  Masayuki and {Wang}, Shiang-Yu and {Yuma}, Suraphong},
  title		= "{SILVERRUSH. II. First catalogs and properties of
		  {\ensuremath{\sim}}2000 Ly{\ensuremath{\alpha}} emitters
		  and blobs at z {\ensuremath{\sim}} 6-7 identified over the
		  14-21 deg$^{2}$ sky$^{*}$}",
  journal	= {\pasj},
  year		= 2018,
  month		= jan,
  volume	= {70},
  eid		= {S14},
  pages		= {S14},
  doi		= {10.1093/pasj/psx122},
  archiveprefix	= {arXiv},
  eprint	= {1704.08140},
  primaryclass	= {astro-ph.GA},
  adsurl	= {https://ui.adsabs.harvard.edu/abs/2018PASJ...70S..14S}
}

@Article{	  Silverman15,
  author	= {{Silverman}, J.~D. and {Kashino}, D. and {Sanders}, D. and
		  {Kartaltepe}, J.~S. and {Arimoto}, N. and {Renzini}, A. and
		  {Rodighiero}, G. and {Daddi}, E. and {Zahid}, J. and
		  {Nagao}, T. and {Kewley}, L.~J. and {Lilly}, S.~J. and
		  {Sugiyama}, N. and {Baronchelli}, I. and {Capak}, P. and
		  {Carollo}, C.~M. and {Chu}, J. and {Hasinger}, G. and
		  {Ilbert}, O. and {Juneau}, S. and {Kajisawa}, M. and
		  {Koekemoer}, A.~M. and {Kovac}, K. and {Le F{\`e}vre}, O.
		  and {Masters}, D. and {McCracken}, H.~J. and {Onodera}, M.
		  and {Schulze}, A. and {Scoville}, N. and {Strazzullo}, V.
		  and {Taniguchi}, Y.},
  title		= "{The FMOS-COSMOS Survey of Star-forming Galaxies at
		  z\raisebox{-0.5ex}\textasciitilde1.6. III. Survey Design,
		  Performance, and Sample Characteristics}",
  journal	= {\apjs},
  year		= 2015,
  month		= sep,
  volume	= {220},
  number	= {1},
  eid		= {12},
  pages		= {12},
  doi		= {10.1088/0067-0049/220/1/12},
  archiveprefix	= {arXiv},
  eprint	= {1409.0447},
  primaryclass	= {astro-ph.GA},
  adsurl	= {https://ui.adsabs.harvard.edu/abs/2015ApJS..220...12S}
}

@Article{	  Soo23,
  author	= {{Lee}, Kyoung-Soo and {Gawiser}, Eric and {Park}, Changbom
		  and {Yang}, Yujin and {Valdes}, Francisco and {Lang},
		  Dustin and {Ramakrishnan}, Vandana and {Moon}, Byeongha and
		  {Firestone}, Nicole and {Appleby}, Stephen and {Artale},
		  Maria Celeste and {Andrews}, Moira and {Bauer}, Franz and
		  {Benda}, Barbara and {Broussard}, Adam and {Chiang},
		  Yi-Kuan and {Ciardullo}, Robin and {Dey}, Arjun and
		  {Farooq}, Rameen and {Gronwall}, Caryl and {Guaita}, Lucia
		  and {Huang}, Yun and {Hwang}, Ho Seong and {Im}, Sang Hyeok
		  and {Jeong}, Woong-Seob and {Karthikeyan}, Shreya and
		  {Kim}, Hwihyun and {Kim}, Seongjae and {Kumar}, Ankit and
		  {Nagaraj}, Gautam R. and {Nantais}, Julie and {Padilla},
		  Nelson and {Park}, Jaehong and {Pope}, Alexandra and
		  {Popescu}, Roxana and {Schlegel}, David and {Seo}, Eunsuk
		  and {Singh}, Akriti and {Song}, Hyunmi and {Troncoso},
		  Paulina and {Vivas}, A. Katherina and {Zabludoff}, Ann and
		  {Zenteno}, Alfredo},
  title		= "{The One-hundred-deg$^{2}$ DECam Imaging in Narrowbands
		  (ODIN): Survey Design and Science Goals}",
  journal	= {\apj},
  year		= 2024,
  month		= feb,
  volume	= {962},
  number	= {1},
  eid		= {36},
  pages		= {36},
  doi		= {10.3847/1538-4357/ad165e},
  archiveprefix	= {arXiv},
  eprint	= {2309.10191},
  primaryclass	= {astro-ph.GA},
  adsurl	= {https://ui.adsabs.harvard.edu/abs/2024ApJ...962...36L}
}

@Article{	  Steidel00,
  author	= {{Steidel}, Charles C. and {Adelberger}, Kurt L. and
		  {Shapley}, Alice E. and {Pettini}, Max and {Dickinson},
		  Mark and {Giavalisco}, Mauro},
  title		= "{Ly{\ensuremath{\alpha}} Imaging of a Proto-Cluster Region
		  at <z>=3.09}",
  journal	= {\apj},
  year		= 2000,
  month		= mar,
  volume	= {532},
  number	= {1},
  pages		= {170-182},
  doi		= {10.1086/308568},
  archiveprefix	= {arXiv},
  eprint	= {astro-ph/9910144},
  primaryclass	= {astro-ph},
  adsurl	= {https://ui.adsabs.harvard.edu/abs/2000ApJ...532..170S}
}

@Article{	  Vito18,
  author	= {{Vito}, F. and {Brandt}, W.~N. and {Yang}, G. and {Gilli},
		  R. and {Luo}, B. and {Vignali}, C. and {Xue}, Y.~Q. and
		  {Comastri}, A. and {Koekemoer}, A.~M. and {Lehmer}, B.~D.
		  and {Liu}, T. and {Paolillo}, M. and {Ranalli}, P. and
		  {Schneider}, D.~P. and {Shemmer}, O. and {Volonteri}, M.
		  and {Wang}, J.},
  title		= "{High-redshift AGN in the Chandra Deep Fields: the
		  obscured fraction and space density of the sub-L$_{*}$
		  population}",
  journal	= {\mnras},
  year		= 2018,
  month		= jan,
  volume	= {473},
  number	= {2},
  pages		= {2378-2406},
  doi		= {10.1093/mnras/stx2486},
  archiveprefix	= {arXiv},
  eprint	= {1709.07892},
  primaryclass	= {astro-ph.GA},
  adsurl	= {https://ui.adsabs.harvard.edu/abs/2018MNRAS.473.2378V}
}

@Article{	  Yamada12,
  author	= {{Yamada}, T. and {Matsuda}, Y. and {Kousai}, K. and
		  {Hayashino}, T. and {Morimoto}, N. and {Umemura}, M.},
  title		= "{Profiles of Ly{\ensuremath{\alpha}} Emission Lines of the
		  Emitters at z = 3.1}",
  journal	= {\apj},
  year		= 2012,
  month		= may,
  volume	= {751},
  number	= {1},
  eid		= {29},
  pages		= {29},
  doi		= {10.1088/0004-637X/751/1/29},
  archiveprefix	= {arXiv},
  eprint	= {1203.3633},
  primaryclass	= {astro-ph.CO},
  adsurl	= {https://ui.adsabs.harvard.edu/abs/2012ApJ...751...29Y}
}

@Article{	  Yang09,
  author	= {{Yang}, Yujin and {Zabludoff}, Ann and {Tremonti}, Christy
		  and {Eisenstein}, Daniel and {Dav{\'e}}, Romeel},
  title		= "{Extended Ly{\ensuremath{\alpha}} Nebulae at z ≃ 2.3: An
		  Extremely Rare and Strongly Clustered Population?}",
  journal	= {\apj},
  year		= 2009,
  month		= mar,
  volume	= {693},
  number	= {2},
  pages		= {1579-1587},
  doi		= {10.1088/0004-637X/693/2/1579},
  archiveprefix	= {arXiv},
  eprint	= {0811.3446},
  primaryclass	= {astro-ph},
  adsurl	= {https://ui.adsabs.harvard.edu/abs/2009ApJ...693.1579Y}
}

@Article{	  Yang10,
  author	= {{Yang}, Yujin and {Zabludoff}, Ann and {Eisenstein},
		  Daniel and {Dav{\'e}}, Romeel},
  title		= "{Strong Field-to-field Variation of
		  Ly{\ensuremath{\alpha}} Nebulae Populations at z
		  \raisebox{-0.5ex}\textasciitilde= 2.3}",
  journal	= {\apj},
  year		= 2010,
  month		= aug,
  volume	= {719},
  number	= {2},
  pages		= {1654-1671},
  doi		= {10.1088/0004-637X/719/2/1654},
  archiveprefix	= {arXiv},
  eprint	= {1008.2776},
  primaryclass	= {astro-ph.CO},
  adsurl	= {https://ui.adsabs.harvard.edu/abs/2010ApJ...719.1654Y}
}

@Article{	  Yang11,
  author	= {{Yang}, Yujin and {Zabludoff}, Ann and {Jahnke}, Knud and
		  {Eisenstein}, Daniel and {Dav{\'e}}, Romeel and {Shectman},
		  Stephen A. and {Kelson}, Daniel D.},
  title		= "{Gas Kinematics in Ly{\ensuremath{\alpha}} Nebulae}",
  journal	= {\apj},
  year		= 2011,
  month		= jul,
  volume	= {735},
  number	= {2},
  eid		= {87},
  pages		= {87},
  doi		= {10.1088/0004-637X/735/2/87},
  archiveprefix	= {arXiv},
  eprint	= {1104.3597},
  primaryclass	= {astro-ph.CO},
  adsurl	= {https://ui.adsabs.harvard.edu/abs/2011ApJ...735...87Y}
}

@Article{	  Yang14b,
  author	= {{Yang}, Yujin and {Zabludoff}, Ann and {Jahnke}, Knud and
		  {Dav{\'e}}, Romeel},
  title		= "{The Properties of Ly{\ensuremath{\alpha}} Nebulae: Gas
		  Kinematics from Nonresonant Lines}",
  journal	= {\apj},
  year		= 2014,
  month		= oct,
  volume	= {793},
  number	= {2},
  eid		= {114},
  pages		= {114},
  doi		= {10.1088/0004-637X/793/2/114},
  archiveprefix	= {arXiv},
  eprint	= {1407.6801},
  primaryclass	= {astro-ph.GA},
  adsurl	= {https://ui.adsabs.harvard.edu/abs/2014ApJ...793..114Y}
}

@Article{	  You2017,
  author	= {{You}, Chang and {Zabludoff}, Ann and {Smith}, Paul and
		  {Yang}, Yujin and {Kim}, Eunchong and {Jannuzi}, Buell and
		  {Prescott}, Moire K.~M. and {Matsuda}, Yuichi and {Lee},
		  Myung Gyoon},
  title		= "{Mapping the Polarization of the Radio-Loud
		  Ly{\ensuremath{\alpha}} Nebula B3 J2330+3927}",
  journal	= {\apj},
  year		= 2017,
  month		= jan,
  volume	= {834},
  number	= {2},
  eid		= {182},
  pages		= {182},
  doi		= {10.3847/1538-4357/834/2/182},
  archiveprefix	= {arXiv},
  eprint	= {1611.05506},
  primaryclass	= {astro-ph.GA},
  adsurl	= {https://ui.adsabs.harvard.edu/abs/2017ApJ...834..182Y}
}

@ARTICLE{Zhang25,
       author = {{Zhang}, Haibin and {Cai}, Zheng and {Li}, Mingyu and {Liang}, Yongming and {Kashikawa}, Nobunari and {Ma}, Ke and {Wu}, Yunjing and {Li}, Qiong and {Johnson}, Sean D. and {Kikuta}, Satoshi and {Ouchi}, Masami and {Fan}, Xiaohui and {Ning}, Yuanhang},
        title = "{MAMMOTH-Subaru. IV. Large Scale Structure and Clustering Analysis of Ly{\ensuremath{\alpha}} Emitters and Ly{\ensuremath{\alpha}} Blobs at z = 2.2{\textendash}2.3}",
      journal = {\apj},
         year = 2025,
        month = mar,
       volume = {981},
       number = {1},
          eid = {70},
        pages = {70},
          doi = {10.3847/1538-4357/adb41b},
archivePrefix = {arXiv},
       eprint = {2301.07359},
 primaryClass = {astro-ph.GA},
       adsurl = {https://ui.adsabs.harvard.edu/abs/2025ApJ...981...70Z}
}

@ARTICLE{Umeda25,
       author = {{Umeda}, Hiroya and {Ouchi}, Masami and {Kikuta}, Satoshi and {Harikane}, Yuichi and {Ono}, Yoshiaki and {Shibuya}, Takatoshi and {Inoue}, Akio K. and {Shimasaku}, Kazuhiro and {Liang}, Yongming and {Matsumoto}, Akinori and {Saito}, Shun and {Kusakabe}, Haruka and {Kageura}, Yuta and {Nakane}, Minami},
        title = "{SILVERRUSH. XIV. Ly{\ensuremath{\alpha}} Luminosity Functions and Angular Correlation Functions from 20,000 Ly{\ensuremath{\alpha}} Emitters at z {\ensuremath{\sim}} 2.2{\textendash}7.3 from up to 24 deg$^{2}$ HSC-SSP and CHORUS Surveys: Linking the Postreionization Epoch to the Heart of Reionization}",
      journal = {\apjs},
         year = 2025,
        month = apr,
       volume = {277},
       number = {2},
          eid = {37},
        pages = {37},
          doi = {10.3847/1538-4365/adb1c0},
archivePrefix = {arXiv},
       eprint = {2411.15495},
 primaryClass = {astro-ph.GA},
       adsurl = {https://ui.adsabs.harvard.edu/abs/2025ApJS..277...37U}
}

@ARTICLE{Umehata19,
       author = {{Umehata}, H. and {Fumagalli}, M. and {Smail}, I. and {Matsuda}, Y. and {Swinbank}, A.~M. and {Cantalupo}, S. and {Sykes}, C. and {Ivison}, R.~J. and {Steidel}, C.~C. and {Shapley}, A.~E. and {Vernet}, J. and {Yamada}, T. and {Tamura}, Y. and {Kubo}, M. and {Nakanishi}, K. and {Kajisawa}, M. and {Hatsukade}, B. and {Kohno}, K.},
        title = "{Gas filaments of the cosmic web located around active galaxies in a protocluster}",
      journal = {Science},
         year = 2019,
        month = oct,
       volume = {366},
       number = {6461},
        pages = {97-100},
          doi = {10.1126/science.aaw5949},
archivePrefix = {arXiv},
       eprint = {1910.01324},
 primaryClass = {astro-ph.GA},
       adsurl = {https://ui.adsabs.harvard.edu/abs/2019Sci...366...97U}
}

@ARTICLE{Herrera25,
       author = {{Herrera}, Danisbel and {Gawiser}, Eric and {Benda}, Barbara and {Firestone}, Nicole M. and {Ramakrishnan}, Vandana and {Moon}, Byeongha and {Lee}, Kyoung-Soo and {Park}, Changbom and {Valdes}, Francisco and {Yang}, Yujin and {Artale}, Mar{\'\i}a Celeste and {Ciardullo}, Robin and {Gronwall}, Caryl and {Guaita}, Lucia and {Hwang}, Ho Seong and {Kennedy}, Jacob and {Kumar}, Ankit and {Zabludoff}, Ann},
        title = "{ODIN: Clustering Analysis of 14,000 Ly{\ensuremath{\alpha}}-emitting Galaxies at z = 2.4, 3.1, and 4.5}",
      journal = {\apjl},
         year = 2025,
        month = aug,
       volume = {988},
       number = {2},
          eid = {L57},
        pages = {L57},
          doi = {10.3847/2041-8213/adec82},
archivePrefix = {arXiv},
       eprint = {2503.17824},
 primaryClass = {astro-ph.GA},
       adsurl = {https://ui.adsabs.harvard.edu/abs/2025ApJ...988L..57H}
}

@ARTICLE{Palunas04,
       author = {{Palunas}, Povilas and {Teplitz}, Harry I. and {Francis}, Paul J. and {Williger}, Gerard M. and {Woodgate}, Bruce E.},
        title = "{The Distribution of Ly{\ensuremath{\alpha}}-Emitting Galaxies at z=2.38}",
      journal = {\apj},
         year = 2004,
        month = feb,
       volume = {602},
       number = {2},
        pages = {545-554},
          doi = {10.1086/381145},
archivePrefix = {arXiv},
       eprint = {astro-ph/0311279},
 primaryClass = {astro-ph},
       adsurl = {https://ui.adsabs.harvard.edu/abs/2004ApJ...602..545P}
}

@ARTICLE{GMOS,
       author = {{Hook}, I.~M. and {J{\o}rgensen}, Inger and {Allington-Smith}, J.~R. and {Davies}, R.~L. and {Metcalfe}, N. and {Murowinski}, R.~G. and {Crampton}, D.},
        title = "{The Gemini-North Multi-Object Spectrograph: Performance in Imaging, Long-Slit, and Multi-Object Spectroscopic Modes}",
      journal = {\pasp},
         year = 2004,
        month = may,
       volume = {116},
       number = {819},
        pages = {425-440},
          doi = {10.1086/383624},
       adsurl = {https://ui.adsabs.harvard.edu/abs/2004PASP..116..425H}
}

@ARTICLE{Zhang20,
       author = {{Zhang}, Haibin and {Ouchi}, Masami and {Itoh}, Ryohei and {Shibuya}, Takatoshi and {Ono}, Yoshiaki and {Harikane}, Yuichi and {Inoue}, Akio K. and {Rauch}, Michael and {Kikuchihara}, Shotaro and {Nakajima}, Kimihiko and {Yajima}, Hidenobu and {Arata}, Shohei and {Abe}, Makito and {Iwata}, Ikuru and {Kashikawa}, Nobunari and {Kawanomoto}, Satoshi and {Kikuta}, Satoshi and {Kobayashi}, Masakazu A.~R. and {Kusakabe}, Haruka and {Mawatari}, Ken and {Nagao}, Tohru and {Shimasaku}, Kazuhiro and {Taniguchi}, Yoshiaki},
        title = "{CHORUS. III. Photometric and Spectroscopic Properties of Ly{\ensuremath{\alpha}} Blobs at z = 4.9-7.0}",
      journal = {\apj},
         year = 2020,
        month = mar,
       volume = {891},
       number = {2},
          eid = {177},
        pages = {177},
          doi = {10.3847/1538-4357/ab7917},
archivePrefix = {arXiv},
       eprint = {1905.09841},
 primaryClass = {astro-ph.GA},
       adsurl = {https://ui.adsabs.harvard.edu/abs/2020ApJ...891..177Z}
}

@ARTICLE{Bertin96,
       author = {{Bertin}, E. and {Arnouts}, S.},
        title = "{SExtractor: Software for source extraction.}",
      journal = {\aaps},
         year = 1996,
        month = jun,
       volume = {117},
        pages = {393-404},
          doi = {10.1051/aas:1996164},
       adsurl = {https://ui.adsabs.harvard.edu/abs/1996A&AS..117..393B}
}

@ARTICLE{Smith07,
       author = {{Smith}, Daniel J.~B. and {Jarvis}, Matt J.},
        title = "{Evidence for cold accretion onto a massive galaxy at high redshift?}",
      journal = {\mnras},
         year = 2007,
        month = jun,
       volume = {378},
       number = {1},
        pages = {L49-L53},
          doi = {10.1111/j.1745-3933.2007.00318.x},
archivePrefix = {arXiv},
       eprint = {astro-ph/0703522},
 primaryClass = {astro-ph},
       adsurl = {https://ui.adsabs.harvard.edu/abs/2007MNRAS.378L..49S}
}

@ARTICLE{Nagaraj25,
       author = {{Nagaraj}, Gautam and {Ciardullo}, Robin and {Gronwall}, Caryl and {Ramakrishnan}, Vandana and {Lee}, Kyoung-Soo and {Gawiser}, Eric and {Firestone}, Nicole M. and {Ramgopal}, Govind and {Aguilar}, J. and {Ahlen}, Steven and {Bianchi}, Davide and {Brooks}, David and {Castander}, Francisco Javier and {Claybaugh}, Todd and {Cuceu}, Andrei and {de la Macorra}, Axel and {Dey}, Arjun and {Dey}, Biprateep and {Doel}, Peter and {Forero-Romero}, Jaime and {Gaztanaga}, Enrique and {Gontcho}, Satya Gontcho A and {Gutierrez}, Gaston and {Herrera-Alcantar}, Hiram K. and {Honscheid}, Klaus and {Ishak}, Mustapha and {Kehoe}, Robert and {Kirkby}, David and {Kisner}, T. and {Kremin}, Anthony and {Landriau}, Martin and {Le Guillou}, Laurent and {Levi}, Michael and {Magneville}, Christophe and {Manera}, Marc and {Martini}, Paul and {Meisner}, Aaron M. and {Miquel}, Ramon and {Moustakas}, John and {Palanque-Delabrouille}, Nathalie and {Prada}, Francisco and {Perez-Rafols}, Ignasi and {Rossi}, Graziano and {Samushia}, Lado and {Sanchez}, Eusebio and {Schlegel}, David J. and {Schubnell}, Michael F. and {Seo}, Hee-Jong and {Silber}, Joseph H. and {Sprayberry}, David and {Tarle}, Gregory and {Valdes}, Francisco and {Weaver}, Benjamin A. and {White}, Martin and {Zhou}, Rongpu and {Zou}, Hu},
        title = "{ODIN: The LAE Ly{\ensuremath{\alpha}} Luminosity Function over Cosmic Time and Environmental Density}",
      journal = {arXiv e-prints},
         year = 2025,
        month = jun,
          eid = {arXiv:2506.14510},
        pages = {arXiv:2506.14510},
          doi = {10.48550/arXiv.2506.14510},
archivePrefix = {arXiv},
       eprint = {2506.14510},
 primaryClass = {astro-ph.GA},
       adsurl = {https://ui.adsabs.harvard.edu/abs/2025arXiv250614510N}
}

@ARTICLE{Lusso19,
       author = {{Lusso}, E. and {Fumagalli}, M. and {Fossati}, M. and {Mackenzie}, R. and {Bielby}, R.~M. and {Arrigoni Battaia}, F. and {Cantalupo}, S. and {Cooke}, R. and {Cristiani}, S. and {Dayal}, P. and {D'Odorico}, V. and {Haardt}, F. and {Lofthouse}, E. and {Morris}, S. and {Peroux}, C. and {Prichard}, L. and {Rafelski}, M. and {Simcoe}, R. and {Swinbank}, A.~M. and {Theuns}, T.},
        title = "{The MUSE Ultra Deep Field (MUDF) - I. Discovery of a group of Ly{\ensuremath{\alpha}} nebulae associated with a bright z {\ensuremath{\approx}} 3.23 quasar pair}",
      journal = {\mnras},
         year = 2019,
        month = may,
       volume = {485},
       number = {1},
        pages = {L62-L67},
          doi = {10.1093/mnrasl/slz032},
archivePrefix = {arXiv},
       eprint = {1903.00483},
 primaryClass = {astro-ph.GA},
       adsurl = {https://ui.adsabs.harvard.edu/abs/2019MNRAS.485L..62L}
}

@ARTICLE{Tornotti25,
       author = {{Tornotti}, Davide and {Fumagalli}, Michele and {Fossati}, Matteo and {Benitez-Llambay}, Alejandro and {Izquierdo-Villalba}, David and {Travascio}, Andrea and {Arrigoni Battaia}, Fabrizio and {Cantalupo}, Sebastiano and {Beckett}, Alexander and {Bonoli}, Silvia and {Dayal}, Pratika and {D'Odorico}, Valentina and {Dutta}, Rajeshwari and {Lusso}, Elisabeta and {Peroux}, Celine and {Rafelski}, Marc and {Revalski}, Mitchell and {Spinoso}, Daniele and {Swinbank}, Mark},
        title = "{High-definition imaging of a filamentary connection between a close quasar pair at z = 3}",
      journal = {Nature Astronomy},
         year = 2025,
        month = apr,
       volume = {9},
        pages = {577-588},
          doi = {10.1038/s41550-024-02463-w},
archivePrefix = {arXiv},
       eprint = {2406.17035},
 primaryClass = {astro-ph.CO},
       adsurl = {https://ui.adsabs.harvard.edu/abs/2025NatAs...9..577T}
}

@ARTICLE{Lang16b,
       author = {{Lang}, Dustin and {Hogg}, David W. and {Schlegel}, David J.},
        title = "{WISE Photometry for 400 Million SDSS Sources}",
      journal = {\aj},
         year = 2016,
        month = feb,
       volume = {151},
       number = {2},
          eid = {36},
        pages = {36},
          doi = {10.3847/0004-6256/151/2/36},
archivePrefix = {arXiv},
       eprint = {1410.7397},
 primaryClass = {astro-ph.IM},
       adsurl = {https://ui.adsabs.harvard.edu/abs/2016AJ....151...36L}
}

@ARTICLE{DESI_DR1,
        author = {{DESI Collaboration} and {Abdul-Karim}, M. and {Adame}, A.~G. and {Aguado}, D. and {Aguilar}, J. and {Ahlen}, S. and {Alam}, S. and {Aldering}, G. and {Alexander}, D.~M. and {Alfarsy}, R. and {Allen}, L. and {Allende Prieto}, C. and {Alves}, O. and {Anand}, A. and {Andrade}, U. and {Armengaud}, E. and {Avila}, S. and {Aviles}, A. and {Awan}, H. and {Bailey}, S. and {Baleato Lizancos}, A. and {Ballester}, O. and {Bault}, A. and {Bautista}, J. and {BenZvi}, S. and {Beraldo e Silva}, L. and {Bermejo-Climent}, J.~R. and {Beutler}, F. and {Bianchi}, D. and {Blake}, C. and {Blum}, R. and {Bolton}, A.~S. and {Bonici}, M. and {Brieden}, S. and {Brodzeller}, A. and {Brooks}, D. and {Buckley-Geer}, E. and {Burtin}, E. and {Canning}, R. and {Carnero Rosell}, A. and {Carr}, A. and {Carrilho}, P. and {Casas}, L. and {Castander}, F.~J. and {Cereskaite}, R. and {Cervantes-Cota}, J.~L. and {Chaussidon}, E. and {Chaves-Montero}, J. and {Chen}, S. and {Chen}, X. and {Claybaugh}, T. and {Cole}, S. and {Cooper}, A.~P. and {Cousinou}, M.-C. and {Cuceu}, A. and {Davis}, T.~M. and {Dawson}, K.~S. and {de Belsunce}, R. and {de la Cruz}, R. and {de la Macorra}, A. and {de Mattia}, A. and {Deiosso}, N. and {Della Costa}, J. and {Demina}, R. and {Demirbozan}, U. and {DeRose}, J. and {Dey}, A. and {Dey}, B. and {Ding}, J. and {Ding}, Z. and {Doel}, P. and {Douglass}, K. and {Dowicz}, M. and {Ebina}, H. and {Edelstein}, J. and {Eisenstein}, D.~J. and {Elbers}, W. and {Emas}, N. and {Escoffier}, S. and {Fagrelius}, P. and {Fan}, X. and {Fanning}, K. and {Fawcett}, V.~A. and {Fern'{a}ndez-Garc'{i}a}, E. and {Ferraro}, S. and {Findlay}, N. and {Font-Ribera}, A. and {Forero-Romero}, J.~E. and {Forero-S'{a}nchez}, D. and {Frenk}, C.~S. and {G"{a}nsicke}, B.~T. and {Galbany}, L. and {Garc'{i}a-Bellido}, J. and {Garcia-Quintero}, C. and {Garrison}, L.~H. and {Gazta~{n}aga}, E. and {Gil-Mar'{i}n}, H. and {Gnedin}, O.~Y. and {Gontcho}, S. Gontcho A and {Gonzalez-Morales}, A.~X. and {Gonzalez-Perez}, V. and {Gordon}, C. and {Graur}, O. and {Green}, D. and {Gruen}, D. and {Gsponer}, R. and {Guandalin}, C. and {Gutierrez}, G. and {Guy}, J. and {Hahn}, C. and {Han}, J.~J. and {Han}, J. and {He}, S. and {Herrera-Alcantar}, H.~K. and {Honscheid}, K. and {Hou}, J. and {Howlett}, C. and {Huterer}, D. and {Ir\v{s}i\v{c}}, V. and {Ishak}, M. and {Jacques}, A. and {Jimenez}, J. and {Jing}, Y.~P. and {Joachimi}, B. and {Joudaki}, S. and {Joyce}, R. and {Jullo}, E. and {Juneau}, S. and {Kara\c{c}ayl{\i}}, N.~G. and {Karim}, T. and {Kehoe}, R. and {Kent}, S. and {Khederlarian}, A. and {Kirkby}, D. and {Kisner}, T. and {Kitaura}, F.-S. and {Kizhuprakkat}, N. and {Kong}, H. and {Koposov}, S.~E. and {Kremin}, A. and {Krolewski}, A. and {Lahav}, O. and {Lai}, Y. and {Lamman}, C. and {Lan}, T.-W. and {Landriau}, M. and {Lang}, D. and {Lange}, J.~U. and {Lasker}, J. and {Le Goff}, J.~M. and {Le Guillou}, L. and {Leauthaud}, A. and {Levi}, M.~E. and {Li}, S. and {Li}, T.~S. and {Lodha}, K. and {Lokken}, M. and {Luo}, Y. and {Magneville}, C. and {Manera}, M. and {Manser}, C.~J. and {Margala}, D. and {Martini}, P. and {Maus}, M. and {McCullough}, J. and {McDonald}, P. and {Medina}, G.~E. and {Medina-Varela}, L. and {Meisner}, A. and {Mena-Fern'{a}ndez}, J. and {Menegas}, A. and {Mezcua}, M. and {Miquel}, R. and {Montero-Camacho}, P. and {Moon}, J. and {Moustakas}, J. and {Mu~{n}oz-Guti'{e}rrez}, A. and {Mu~{n}oz-Santos}, D. and {Myers}, A.~D. and {Myles}, J. and {Nadathur}, S. and {Najita}, J. and {Napolitano}, L. and {Newman}, J.~A. and {Nikakhtar}, F. and {Nikutta}, R. and {Niz}, G. and {Noriega}, H.~E. and {Padmanabhan}, N. and {Paillas}, E. and {Palanque-Delabrouille}, N. and {Palmese}, A. and {Pan}, J. and {Pan}, Z. and {Parkinson}, D. and {Peacock}, J. and {Percival}, W.~J. and {P'{e}rez-Fern'{a}ndez}, A. and {P'{e}rez-R`afols}, I. and {Peterson}, P.},
        title = "{Data Release 1 of the Dark Energy Spectroscopic Instrument}",
      journal = {arXiv e-prints},
         year = 2025,
        month = mar,
          eid = {arXiv:2503.14745},
        pages = {arXiv:2503.14745},
          doi = {10.48550/arXiv.2503.14745},
archivePrefix = {arXiv},
       eprint = {2503.14745},
 primaryClass = {astro-ph.CO},
       adsurl = {https://ui.adsabs.harvard.edu/abs/2025arXiv250314745D}
}

@ARTICLE{Hill21_hetdex,
       author = {{Hill}, Gary J. and {Lee}, Hanshin and {MacQueen}, Phillip J. and {Kelz}, Andreas and {Drory}, Niv and {Vattiat}, Brian L. and {Good}, John M. and {Ramsey}, Jason and {Kriel}, Herman and {Peterson}, Trent and {DePoy}, D.~L. and {Gebhardt}, Karl and {Marshall}, J.~L. and {Tuttle}, Sarah E. and {Bauer}, Svend M. and {Chonis}, Taylor S. and {Fabricius}, Maximilian H. and {Froning}, Cynthia and {H{\"a}user}, Marco and {Indahl}, Briana L. and {Jahn}, Thomas and {Landriau}, Martin and {Leck}, Ron and {Montesano}, Francesco and {Prochaska}, Travis and {Snigula}, Jan M. and {Zeimann}, Greg and {Bryant}, Randy and {Damm}, George and {Fowler}, J.~R. and {Janowiecki}, Steven and {Martin}, Jerry and {Mrozinski}, Emily and {Odewahn}, Stephen and {Rostopchin}, Sergey and {Shetrone}, Matthew and {Spencer}, Renny and {Mentuch Cooper}, Erin and {Armandroff}, Taft and {Bender}, Ralf and {Dalton}, Gavin and {Hopp}, Ulrich and {Komatsu}, Eiichiro and {Nicklas}, Harald and {Ramsey}, Lawrence W. and {Roth}, Martin M. and {Schneider}, Donald P. and {Sneden}, Chris and {Steinmetz}, Matthias},
        title = "{The HETDEX Instrumentation: Hobby-Eberly Telescope Wide-field Upgrade and VIRUS}",
      journal = {\aj},
         year = 2021,
        month = dec,
       volume = {162},
       number = {6},
          eid = {298},
        pages = {298},
          doi = {10.3847/1538-3881/ac2c02},
archivePrefix = {arXiv},
       eprint = {2110.03843},
 primaryClass = {astro-ph.IM},
       adsurl = {https://ui.adsabs.harvard.edu/abs/2021AJ....162..298H}
}

@ARTICLE{astropy13,
       author = {{Astropy Collaboration} and {Robitaille}, Thomas P. and {Tollerud}, Erik J. and {Greenfield}, Perry and {Droettboom}, Michael and {Bray}, Erik and {Aldcroft}, Tom and {Davis}, Matt and {Ginsburg}, Adam and {Price-Whelan}, Adrian M. and {Kerzendorf}, Wolfgang E. and {Conley}, Alexander and {Crighton}, Neil and {Barbary}, Kyle and {Muna}, Demitri and {Ferguson}, Henry and {Grollier}, Fr{\'e}d{\'e}ric and {Parikh}, Madhura M. and {Nair}, Prasanth H. and {Unther}, Hans M. and {Deil}, Christoph and {Woillez}, Julien and {Conseil}, Simon and {Kramer}, Roban and {Turner}, James E.~H. and {Singer}, Leo and {Fox}, Ryan and {Weaver}, Benjamin A. and {Zabalza}, Victor and {Edwards}, Zachary I. and {Azalee Bostroem}, K. and {Burke}, D.~J. and {Casey}, Andrew R. and {Crawford}, Steven M. and {Dencheva}, Nadia and {Ely}, Justin and {Jenness}, Tim and {Labrie}, Kathleen and {Lim}, Pey Lian and {Pierfederici}, Francesco and {Pontzen}, Andrew and {Ptak}, Andy and {Refsdal}, Brian and {Servillat}, Mathieu and {Streicher}, Ole},
        title = "{Astropy: A community Python package for astronomy}",
      journal = {\aap},
         year = 2013,
        month = oct,
       volume = {558},
          eid = {A33},
        pages = {A33},
          doi = {10.1051/0004-6361/201322068},
archivePrefix = {arXiv},
       eprint = {1307.6212},
 primaryClass = {astro-ph.IM},
       adsurl = {https://ui.adsabs.harvard.edu/abs/2013A&A...558A..33A}
}

@ARTICLE{astropy18,
       author = {{Astropy Collaboration} and {Price-Whelan}, A.~M. and {Sip{\H{o}}cz}, B.~M. and {G{\"u}nther}, H.~M. and {Lim}, P.~L. and {Crawford}, S.~M. and {Conseil}, S. and {Shupe}, D.~L. and {Craig}, M.~W. and {Dencheva}, N. and {Ginsburg}, A. and {VanderPlas}, J.~T. and {Bradley}, L.~D. and {P{\'e}rez-Su{\'a}rez}, D. and {de Val-Borro}, M. and {Aldcroft}, T.~L. and {Cruz}, K.~L. and {Robitaille}, T.~P. and {Tollerud}, E.~J. and {Ardelean}, C. and {Babej}, T. and {Bach}, Y.~P. and {Bachetti}, M. and {Bakanov}, A.~V. and {Bamford}, S.~P. and {Barentsen}, G. and {Barmby}, P. and {Baumbach}, A. and {Berry}, K.~L. and {Biscani}, F. and {Boquien}, M. and {Bostroem}, K.~A. and {Bouma}, L.~G. and {Brammer}, G.~B. and {Bray}, E.~M. and {Breytenbach}, H. and {Buddelmeijer}, H. and {Burke}, D.~J. and {Calderone}, G. and {Cano Rodr{\'\i}guez}, J.~L. and {Cara}, M. and {Cardoso}, J.~V.~M. and {Cheedella}, S. and {Copin}, Y. and {Corrales}, L. and {Crichton}, D. and {D'Avella}, D. and {Deil}, C. and {Depagne}, {\'E}. and {Dietrich}, J.~P. and {Donath}, A. and {Droettboom}, M. and {Earl}, N. and {Erben}, T. and {Fabbro}, S. and {Ferreira}, L.~A. and {Finethy}, T. and {Fox}, R.~T. and {Garrison}, L.~H. and {Gibbons}, S.~L.~J. and {Goldstein}, D.~A. and {Gommers}, R. and {Greco}, J.~P. and {Greenfield}, P. and {Groener}, A.~M. and {Grollier}, F. and {Hagen}, A. and {Hirst}, P. and {Homeier}, D. and {Horton}, A.~J. and {Hosseinzadeh}, G. and {Hu}, L. and {Hunkeler}, J.~S. and {Ivezi{\'c}}, {\v{Z}}. and {Jain}, A. and {Jenness}, T. and {Kanarek}, G. and {Kendrew}, S. and {Kern}, N.~S. and {Kerzendorf}, W.~E. and {Khvalko}, A. and {King}, J. and {Kirkby}, D. and {Kulkarni}, A.~M. and {Kumar}, A. and {Lee}, A. and {Lenz}, D. and {Littlefair}, S.~P. and {Ma}, Z. and {Macleod}, D.~M. and {Mastropietro}, M. and {McCully}, C. and {Montagnac}, S. and {Morris}, B.~M. and {Mueller}, M. and {Mumford}, S.~J. and {Muna}, D. and {Murphy}, N.~A. and {Nelson}, S. and {Nguyen}, G.~H. and {Ninan}, J.~P. and {N{\"o}the}, M. and {Ogaz}, S. and {Oh}, S. and {Parejko}, J.~K. and {Parley}, N. and {Pascual}, S. and {Patil}, R. and {Patil}, A.~A. and {Plunkett}, A.~L. and {Prochaska}, J.~X. and {Rastogi}, T. and {Reddy Janga}, V. and {Sabater}, J. and {Sakurikar}, P. and {Seifert}, M. and {Sherbert}, L.~E. and {Sherwood-Taylor}, H. and {Shih}, A.~Y. and {Sick}, J. and {Silbiger}, M.~T. and {Singanamalla}, S. and {Singer}, L.~P. and {Sladen}, P.~H. and {Sooley}, K.~A. and {Sornarajah}, S. and {Streicher}, O. and {Teuben}, P. and {Thomas}, S.~W. and {Tremblay}, G.~R. and {Turner}, J.~E.~H. and {Terr{\'o}n}, V. and {van Kerkwijk}, M.~H. and {de la Vega}, A. and {Watkins}, L.~L. and {Weaver}, B.~A. and {Whitmore}, J.~B. and {Woillez}, J. and {Zabalza}, V. and {Astropy Contributors}},
        title = "{The Astropy Project: Building an Open-science Project and Status of the v2.0 Core Package}",
      journal = {\aj},
         year = 2018,
        month = sep,
       volume = {156},
       number = {3},
          eid = {123},
        pages = {123},
          doi = {10.3847/1538-3881/aabc4f},
archivePrefix = {arXiv},
       eprint = {1801.02634},
 primaryClass = {astro-ph.IM},
       adsurl = {https://ui.adsabs.harvard.edu/abs/2018AJ....156..123A}
}

@ARTICLE{astropy22,
       author = {{Astropy Collaboration} and {Price-Whelan}, Adrian M. and {Lim}, Pey Lian and {Earl}, Nicholas and {Starkman}, Nathaniel and {Bradley}, Larry and {Shupe}, David L. and {Patil}, Aarya A. and {Corrales}, Lia and {Brasseur}, C.~E. and {N{\"o}the}, Maximilian and {Donath}, Axel and {Tollerud}, Erik and {Morris}, Brett M. and {Ginsburg}, Adam and {Vaher}, Eero and {Weaver}, Benjamin A. and {Tocknell}, James and {Jamieson}, William and {van Kerkwijk}, Marten H. and {Robitaille}, Thomas P. and {Merry}, Bruce and {Bachetti}, Matteo and {G{\"u}nther}, H. Moritz and {Aldcroft}, Thomas L. and {Alvarado-Montes}, Jaime A. and {Archibald}, Anne M. and {B{\'o}di}, Attila and {Bapat}, Shreyas and {Barentsen}, Geert and {Baz{\'a}n}, Juanjo and {Biswas}, Manish and {Boquien}, M{\'e}d{\'e}ric and {Burke}, D.~J. and {Cara}, Daria and {Cara}, Mihai and {Conroy}, Kyle E. and {Conseil}, Simon and {Craig}, Matthew W. and {Cross}, Robert M. and {Cruz}, Kelle L. and {D'Eugenio}, Francesco and {Dencheva}, Nadia and {Devillepoix}, Hadrien A.~R. and {Dietrich}, J{\"o}rg P. and {Eigenbrot}, Arthur Davis and {Erben}, Thomas and {Ferreira}, Leonardo and {Foreman-Mackey}, Daniel and {Fox}, Ryan and {Freij}, Nabil and {Garg}, Suyog and {Geda}, Robel and {Glattly}, Lauren and {Gondhalekar}, Yash and {Gordon}, Karl D. and {Grant}, David and {Greenfield}, Perry and {Groener}, Austen M. and {Guest}, Steve and {Gurovich}, Sebastian and {Handberg}, Rasmus and {Hart}, Akeem and {Hatfield-Dodds}, Zac and {Homeier}, Derek and {Hosseinzadeh}, Griffin and {Jenness}, Tim and {Jones}, Craig K. and {Joseph}, Prajwel and {Kalmbach}, J. Bryce and {Karamehmetoglu}, Emir and {Ka{\l}uszy{\'n}ski}, Miko{\l}aj and {Kelley}, Michael S.~P. and {Kern}, Nicholas and {Kerzendorf}, Wolfgang E. and {Koch}, Eric W. and {Kulumani}, Shankar and {Lee}, Antony and {Ly}, Chun and {Ma}, Zhiyuan and {MacBride}, Conor and {Maljaars}, Jakob M. and {Muna}, Demitri and {Murphy}, N.~A. and {Norman}, Henrik and {O'Steen}, Richard and {Oman}, Kyle A. and {Pacifici}, Camilla and {Pascual}, Sergio and {Pascual-Granado}, J. and {Patil}, Rohit R. and {Perren}, Gabriel I. and {Pickering}, Timothy E. and {Rastogi}, Tanuj and {Roulston}, Benjamin R. and {Ryan}, Daniel F. and {Rykoff}, Eli S. and {Sabater}, Jose and {Sakurikar}, Parikshit and {Salgado}, Jes{\'u}s and {Sanghi}, Aniket and {Saunders}, Nicholas and {Savchenko}, Volodymyr and {Schwardt}, Ludwig and {Seifert-Eckert}, Michael and {Shih}, Albert Y. and {Jain}, Anany Shrey and {Shukla}, Gyanendra and {Sick}, Jonathan and {Simpson}, Chris and {Singanamalla}, Sudheesh and {Singer}, Leo P. and {Singhal}, Jaladh and {Sinha}, Manodeep and {Sip{\H{o}}cz}, Brigitta M. and {Spitler}, Lee R. and {Stansby}, David and {Streicher}, Ole and {{\v{S}}umak}, Jani and {Swinbank}, John D. and {Taranu}, Dan S. and {Tewary}, Nikita and {Tremblay}, Grant R. and {de Val-Borro}, Miguel and {Van Kooten}, Samuel J. and {Vasovi{\'c}}, Zlatan and {Verma}, Shresth and {de Miranda Cardoso}, Jos{\'e} Vin{\'\i}cius and {Williams}, Peter K.~G. and {Wilson}, Tom J. and {Winkel}, Benjamin and {Wood-Vasey}, W.~M. and {Xue}, Rui and {Yoachim}, Peter and {Zhang}, Chen and {Zonca}, Andrea and {Astropy Project Contributors}},
        title = "{The Astropy Project: Sustaining and Growing a Community-oriented Open-source Project and the Latest Major Release (v5.0) of the Core Package}",
      journal = {\apj},
         year = 2022,
        month = aug,
       volume = {935},
       number = {2},
          eid = {167},
        pages = {167},
          doi = {10.3847/1538-4357/ac7c74},
archivePrefix = {arXiv},
       eprint = {2206.14220},
 primaryClass = {astro-ph.IM},
       adsurl = {https://ui.adsabs.harvard.edu/abs/2022ApJ...935..167A}
}

@software{PSFEx,
       author = {{Bertin}, Emmanuel},
        title = "{PSFEx: Point Spread Function Extractor}",
 howpublished = {Astrophysics Source Code Library, record ascl:1301.001},
         year = 2013,
        month = jan,
          eid = {ascl:1301.001},
archivePrefix = {ascl},
       eprint = {1301.001},
       adsurl = {https://ui.adsabs.harvard.edu/abs/2013ascl.soft01001B}
}

@ARTICLE{Flaugher15,
       author = {{Flaugher}, B. and {Diehl}, H.~T. and {Honscheid}, K. and {Abbott}, T.~M.~C. and {Alvarez}, O. and {Angstadt}, R. and {Annis}, J.~T. and {Antonik}, M. and {Ballester}, O. and {Beaufore}, L. and {Bernstein}, G.~M. and {Bernstein}, R.~A. and {Bigelow}, B. and {Bonati}, M. and {Boprie}, D. and {Brooks}, D. and {Buckley-Geer}, E.~J. and {Campa}, J. and {Cardiel-Sas}, L. and {Castander}, F.~J. and {Castilla}, J. and {Cease}, H. and {Cela-Ruiz}, J.~M. and {Chappa}, S. and {Chi}, E. and {Cooper}, C. and {da Costa}, L.~N. and {Dede}, E. and {Derylo}, G. and {DePoy}, D.~L. and {de Vicente}, J. and {Doel}, P. and {Drlica-Wagner}, A. and {Eiting}, J. and {Elliott}, A.~E. and {Emes}, J. and {Estrada}, J. and {Fausti Neto}, A. and {Finley}, D.~A. and {Flores}, R. and {Frieman}, J. and {Gerdes}, D. and {Gladders}, M.~D. and {Gregory}, B. and {Gutierrez}, G.~R. and {Hao}, J. and {Holland}, S.~E. and {Holm}, S. and {Huffman}, D. and {Jackson}, C. and {James}, D.~J. and {Jonas}, M. and {Karcher}, A. and {Karliner}, I. and {Kent}, S. and {Kessler}, R. and {Kozlovsky}, M. and {Kron}, R.~G. and {Kubik}, D. and {Kuehn}, K. and {Kuhlmann}, S. and {Kuk}, K. and {Lahav}, O. and {Lathrop}, A. and {Lee}, J. and {Levi}, M.~E. and {Lewis}, P. and {Li}, T.~S. and {Mandrichenko}, I. and {Marshall}, J.~L. and {Martinez}, G. and {Merritt}, K.~W. and {Miquel}, R. and {Mu{\~n}oz}, F. and {Neilsen}, E.~H. and {Nichol}, R.~C. and {Nord}, B. and {Ogando}, R. and {Olsen}, J. and {Palaio}, N. and {Patton}, K. and {Peoples}, J. and {Plazas}, A.~A. and {Rauch}, J. and {Reil}, K. and {Rheault}, J.-P. and {Roe}, N.~A. and {Rogers}, H. and {Roodman}, A. and {Sanchez}, E. and {Scarpine}, V. and {Schindler}, R.~H. and {Schmidt}, R. and {Schmitt}, R. and {Schubnell}, M. and {Schultz}, K. and {Schurter}, P. and {Scott}, L. and {Serrano}, S. and {Shaw}, T.~M. and {Smith}, R.~C. and {Soares-Santos}, M. and {Stefanik}, A. and {Stuermer}, W. and {Suchyta}, E. and {Sypniewski}, A. and {Tarle}, G. and {Thaler}, J. and {Tighe}, R. and {Tran}, C. and {Tucker}, D. and {Walker}, A.~R. and {Wang}, G. and {Watson}, M. and {Weaverdyck}, C. and {Wester}, W. and {Woods}, R. and {Yanny}, B. and {DES Collaboration}},
        title = "{The Dark Energy Camera}",
      journal = {\aj},
         year = 2015,
        month = nov,
       volume = {150},
       number = {5},
          eid = {150},
        pages = {150},
          doi = {10.1088/0004-6256/150/5/150},
archivePrefix = {arXiv},
       eprint = {1504.02900},
 primaryClass = {astro-ph.IM},
       adsurl = {https://ui.adsabs.harvard.edu/abs/2015AJ....150..150F}
}

@ARTICLE{Sersic,
       author = {{S{\'e}rsic}, J.~L.},
        title = "{Influence of the atmospheric and instrumental dispersion on the brightness distribution in a galaxy}",
      journal = {Boletin de la Asociacion Argentina de Astronomia La Plata Argentina},
         year = 1963,
        month = feb,
       volume = {6},
        pages = {41-43},
       adsurl = {https://ui.adsabs.harvard.edu/abs/1963BAAA....6...41S}
}

@ARTICLE{deVaucouleurs,
       author = {{de Vaucouleurs}, Gerard},
        title = "{Recherches sur les Nebuleuses Extragalactiques}",
      journal = {Annales d'Astrophysique},
         year = 1948,
        month = jan,
       volume = {11},
        pages = {247},
       adsurl = {https://ui.adsabs.harvard.edu/abs/1948AnAp...11..247D}
}

@ARTICLE{Ramakrishnan25b,
       author = {{Ramakrishnan}, Vandana and {Ortiz}, Ashley and {Moon}, Byeongha and {Jun}, Eunsoo and {Schlegel}, David and {Lee}, Kyoung-Soo and {Aguilar}, Jessica Nicole and {Artale}, Maria Celeste and {Brooks}, David and {Candela Cerdosino}, Maria and {Ciardullo}, Robin and {Claybaugh}, Todd and {Cuceu}, Andrei and {de la Macorra}, Axel and {Dey}, Arjun and {Firestone}, Nicole M. and {Font-Ribera}, Andreu and {Forero-Romero}, Jaime E. and {Gawiser}, Eric and {Gazta{\~n}aga}, Enrique and {Gronwall}, Caryl and {Guaita}, Lucia and {Gutierrez}, Gaston and {Hong}, Sungryong and {Hwang}, Ho Seong and {Im}, Sang Hyeok and {Troncoso Iribarren}, Paulina and {Jeong}, Woong-Seob and {Joyce}, Dick and {Kumar}, Ankit and {Lamman}, Claire and {Landriau}, Martin and {Lee}, Seong-Kook and {Lee}, Jaehyun and {Meisner}, Aaron and {Miquel}, Ramon and {Moustakas}, John and {Nadathur}, Seshadri and {Nagaraj}, Gautam and {Nantais}, Julie and {Padilla}, Nelson and {Park}, Changbom and {Percival}, Will and {Prada}, Francisco and {P{\'e}rez-R{\`a}fols}, Ignasi and {Rossi}, Graziano and {Sanchez}, Eusebio and {Silber}, Joseph Harry and {Song}, Hyunmi and {Sprayberry}, David and {Tarl{\'e}}, Gregory and {Valdes}, Francisco and {Yang}, Yujin and {Zabludoff}, Ann and {Zou}, Hu},
        title = "{ODIN: Characterizing the Three-dimensional Structure of Two Protocluster Complexes at $z = 3.1$}",
      journal = {arXiv e-prints},
         year = 2025,
        month = nov,
          eid = {arXiv:2511.11826},
        pages = {arXiv:2511.11826},
          doi = {10.48550/arXiv.2511.11826},
archivePrefix = {arXiv},
       eprint = {2511.11826},
 primaryClass = {astro-ph.GA},
       adsurl = {https://ui.adsabs.harvard.edu/abs/2025arXiv251111826R}
}

@software{photutils,
  author       = {Larry Bradley and
                  Brigitta Sip{\H o}cz and
                  Thomas Robitaille and
                  Erik Tollerud and
                  Z\`e Vin{\'{\i}}cius and
                  Christoph Deil and
                  Kyle Barbary and
                  Tom J Wilson and
                  Ivo Busko and
                  Axel Donath and
                  Hans Moritz G{\"u}nther and
                  Mihai Cara and
                  P. L. Lim and
                  Sebastian Me{\ss}linger and
                  Zach Burnett and
                  Simon Conseil and
                  Michael Droettboom and
                  Azalee Bostroem and
                  E. M. Bray and
                  Lars Andersen Bratholm and
                  William Jamieson and
                  Adam Ginsburg and
                  Geert Barentsen and
                  Matt Craig and
                  Sergio Pascual and
                  Shivangee Rathi and
                  Marshall Perrin and
                  Brett M. Morris},
  title        = {astropy/photutils: 2.2.0},
  month        = feb,
  year         = 2025,
  publisher    = {Zenodo},
  version      = {2.2.0},
  doi          = {10.5281/zenodo.14889440},
  url          = {https://doi.org/10.5281/zenodo.14889440},
  swhid        = {swh:1:dir:11159107f27a28985192ed1118b1f2055709d093
                   ;origin=https://doi.org/10.5281/zenodo.596036;visi
                   t=swh:1:snp:ae8c4a55d349d43e53cfe9ce92e678fcfe840f
                   3b;anchor=swh:1:rel:0117f67e8888adcdfc85308287dd9c
                   854b466389;path=astropy-photutils-ffb96c5
                  },
}
\bibliographystyle{aasjournalv7}

%----------------------------------------------------------------------
\appendix

\section{Surface Brightness Profile of LABs and LAEs}
\label{app:SB_profile}

\restartappendixnumbering

\begin{figure}[h]
\includegraphics[width=0.47\textwidth]{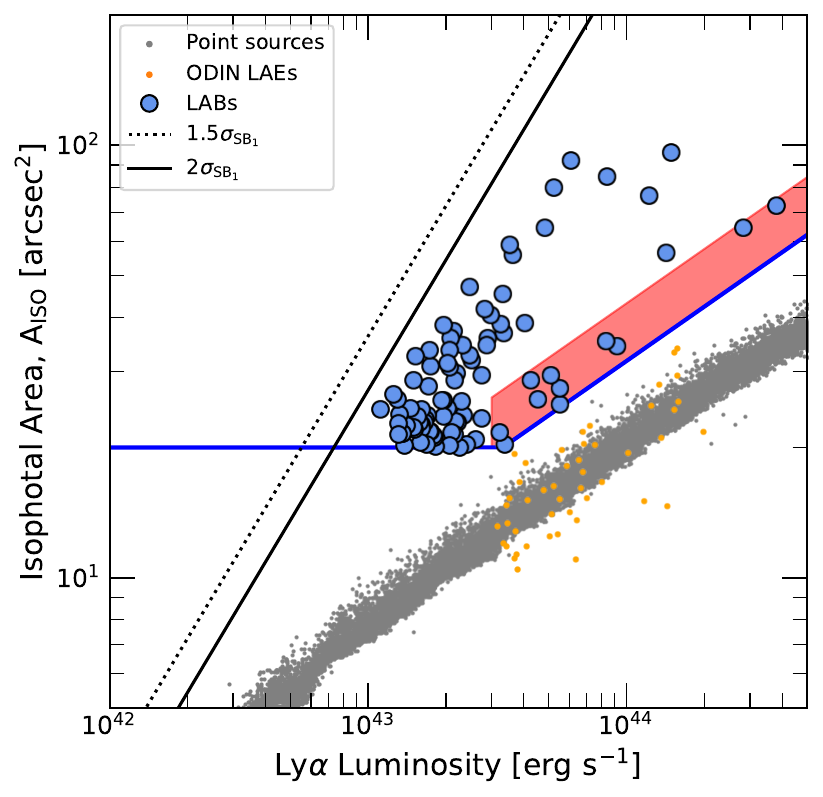}
\includegraphics[width=0.48\textwidth]{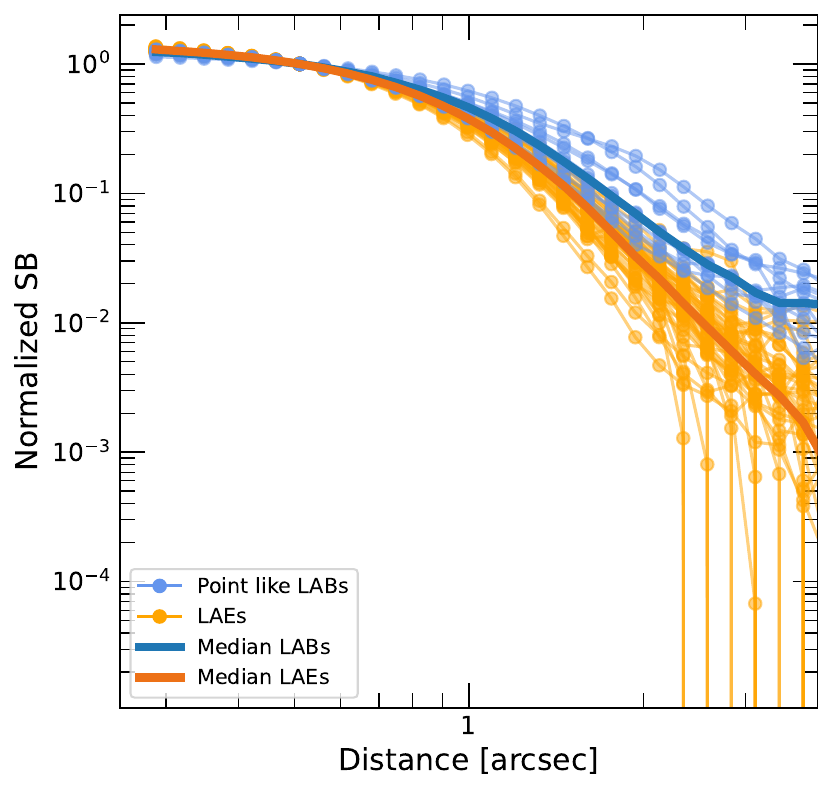}
\caption{
{\bf Left}: 
The blue circles represent the 89 LAB candidates selected using the \traditional method; the red shaded region indicates LABs near the point-source size–luminosity relation. 
The LABs closest to the point-source luminosity–size relation (defined as those within the $3\sigma$–$5\sigma$ scatter estimated for point sources with \lya luminosity $3 \times 10^{43}$ \unitcgslum) are highlighted in the red-shaded region.
The ODIN LAEs used in the comparison are shown as orange dots and also have \lya luminosities above $3 \times 10^{43}$\unitcgslum.
{\bf Right}: Surface brightness profiles of LABs and LAEs at 0.5\arcsec, corresponding to the half of the seeing in the $N501$ band. The bordered lines indicates median of the 1D profiles. LABs (blue solid line and circles) exhibit more spatially extended profiles than LAEs (orange solid line and circles), indicating that LABs near the point-source branch are indeed more extended than LAEs with comparable \lya luminosities.
\label{fig:app_size_lum}}
\end{figure}

To confirm the nature of LAB candidates near the size-luminosity relation of simulated point sources, we compare their surface brightness profiles with those of LAEs. Specifically, we select LABs that lie within 5$\sigma$ of the size–luminosity relation of the simulated point sources and have \lya\ luminosities above $3 \times 10^{43}$ \unitcgslum (Figure~\ref{fig:app_size_lum}, left). For comparison, we use the independently identified LAEs within the same luminosity range. These selections yield 11 LAB candidates above the point-source relation and 47 LAEs lying along the relation.

We extract surface brightness profiles from both LABs and LAEs using {\it Ellipse} in Python {\tt Photutils} package. Each profile is normalized to the surface brightness at 0.5\arcsec, which corresponds to the half of seeing in the $N501$ band. Figure~\ref{fig:app_size_lum} (right) shows that the profiles of LAB candidates are more spatially extended than LAEs, indicating that LABs near the point-source branch are indeed more extended than LAEs with similar luminosities. These different profiles support the classification of these LABs as genuinely extended sources, despite their proximity to the point-source relation in the size–luminosity space.

\section{LAB Catalog} \label{app:cat1}

%----------------------------------------------------------------------
% Table 1
\startlongtable
\movetabledown=60mm
\begin{rotatetable}
\begin{deluxetable*}{crrcccccccc}
\tablecaption{ODIN LAB Catalog \label{tab:lab_cat}}
\tablewidth{0pt}
\tablehead{
\colhead{} & 
\colhead{} & 
\colhead{} & 
\colhead{} & 
\colhead{} & 
\colhead{} & 
\colhead{} &
\multicolumn{2}{c}{LAB Selection} &
\colhead{} & 
\colhead{} \\
\cline{8-9}
\colhead{Name} & 
\colhead{R.A.} & 
\colhead{Decl.} & 
\colhead{$L$(\lya)} & 
\colhead{$A_{\rm iso}$\tablenotemark{1}} & 
\colhead{\lya Size} & 
\colhead{\frecv} &
\colhead{Extended-LAE} & 
\colhead{\tractormethod} & 
\colhead{Redshift} & 
\colhead{Note} \\
\colhead{} & 
\colhead{deg} & 
\colhead{deg} & 
\colhead{10$^{43}$\unitcgslum} & 
\colhead{\sqarcsec} & 
\colhead{pkpc} & 
\colhead{} &
\colhead{} &
\colhead{} &
\colhead{} &
\colhead{} \\
\colhead{(1)} & 
\colhead{(2)} & 
\colhead{(3)} & 
\colhead{(4)} & 
\colhead{(5)} & 
\colhead{(6)} &
\colhead{(7)} & 
\colhead{(8)} & 
\colhead{(9)} & 
\colhead{(10)} & 
\colhead{(11)} 
}
\startdata
ODIN-COSMOS-z3p1-LAB001 & 151.791042 & 2.873806 & 37.41$\pm$0.24 & 72.2$\pm$18.2 & 185 & 1.000 &       Y & \nodata &                 \nodata &                           \nodata \\
ODIN-COSMOS-z3p1-LAB002 & 151.073375 & 2.225167 & 27.90$\pm$0.14 &  64.5$\pm$6.9 & 107 & 1.000 &       Y &       Y &                 \nodata &                              DESI \\
ODIN-COSMOS-z3p1-LAB003 & 151.041375 & 2.513778 & 14.73$\pm$0.17 &  95.7$\pm$7.9 & 110 & 1.000 &       Y &       Y &                  3.1324 &                               R25 \\
ODIN-COSMOS-z3p1-LAB004 & 148.726667 & 2.786306 & 14.07$\pm$0.15 &  56.3$\pm$7.8 & 102 & 1.000 &       Y &       Y &                  3.0825 &         This work,SDSS-DR14Q,DESI \\
ODIN-COSMOS-z3p1-LAB005 & 150.474375 & 3.068111 & 12.06$\pm$0.16 &  76.1$\pm$8.6 &  99 & 1.000 &       Y &       Y &                 \nodata &                           \nodata \\
ODIN-COSMOS-z3p1-LAB006 & 149.489417 & 3.582750 &  9.11$\pm$0.12 &  34.4$\pm$5.9 &  76 & 0.979 &       Y &       Y &                 \nodata &                           \nodata \\
ODIN-COSMOS-z3p1-LAB007 & 150.619958 & 2.671500 &  8.31$\pm$0.16 & 84.6$\pm$12.4 & 108 & 1.000 &       Y &       Y &                  3.1493 &                PRIMUS,ZCOSMOS,R25 \\
ODIN-COSMOS-z3p1-LAB008 & 149.085250 & 2.535417 &  8.28$\pm$0.11 &  35.4$\pm$6.0 &  81 & 0.993 &       Y &       Y &                 \nodata &                              DESI \\
ODIN-COSMOS-z3p1-LAB009 & 148.972500 & 1.593694 &  6.01$\pm$0.17 & 92.0$\pm$21.6 & 152 & 1.000 &       Y &       Y &                 \nodata &                           \nodata \\
ODIN-COSMOS-z3p1-LAB010 & 150.764458 & 3.091083 &  5.49$\pm$0.09 &  25.2$\pm$3.8 &  62 & 0.858 &       Y &       Y &                 \nodata &                              DESI \\
ODIN-COSMOS-z3p1-LAB011 & 149.274458 & 1.165694 &  5.45$\pm$0.09 &  27.4$\pm$8.5 &  63 & 0.930 &       Y & \nodata &                  3.1066 &                   SDSS-DR14Q,DESI \\
ODIN-COSMOS-z3p1-LAB012 & 148.848917 & 2.638417 &  5.21$\pm$0.16 & 79.8$\pm$10.4 & 103 & 1.000 &       Y &       Y &                  3.1189 &                         This work \\
ODIN-COSMOS-z3p1-LAB013 & 149.497875 & 0.735222 &  5.04$\pm$0.12 &  29.4$\pm$3.9 &  73 & 0.995 &       Y &       Y &                 \nodata &                              DESI \\
ODIN-COSMOS-z3p1-LAB014 & 149.452250 & 1.667333 &  4.77$\pm$0.15 & 64.5$\pm$15.7 & 108 & 1.000 &       Y & \nodata & 3.1326\tablenotemark{2} &                         This work \\
ODIN-COSMOS-z3p1-LAB015 & 148.857625 & 2.136722 &  4.50$\pm$0.12 &  25.9$\pm$7.4 &  64 & 0.921 &       Y & \nodata &                 \nodata &                              DESI \\
ODIN-COSMOS-z3p1-LAB016 & 150.489833 & 1.148167 &  4.23$\pm$0.09 &  28.7$\pm$3.5 &  59 & 1.000 &       Y &       Y &                 \nodata &                           \nodata \\
ODIN-COSMOS-z3p1-LAB017 & 149.969250 & 2.304861 &  4.00$\pm$0.12 &  38.9$\pm$5.6 &  69 & 1.000 &       Y &       Y &                    3.15 &                        DEIMOS 10k \\
ODIN-COSMOS-z3p1-LAB018 & 149.704083 & 2.657806 &  3.61$\pm$0.14 &  55.9$\pm$8.5 & 100 & 1.000 &       Y &       Y &                  3.1335 &                    This work,DESI \\
ODIN-COSMOS-z3p1-LAB019 & 149.240000 & 2.479167 &  3.51$\pm$0.14 &  58.8$\pm$7.7 &  92 & 1.000 &       Y &       Y &                 \nodata &                           \nodata \\
ODIN-COSMOS-z3p1-LAB020 & 151.733208 & 3.074139 &  3.35$\pm$0.12 &  20.4$\pm$4.2 &  61 & 0.855 &       Y &       Y &                 \nodata &                           \nodata 
\enddata
\tablecomments{
Column (1) LAB ID, (2)--(3) coordinates, (4) \lya luminosity, (5) isophotal size, (6) projected end-to-end \lya extent (7) recovery fraction from Section~\ref{sec:recovery}, (8)--(9) whether each LAB is identified by \traditional and/or \tractormethod methods, (10) spectroscopic redshift, (11) notes and source of the spectroscopic confirmation. `R25' indicates Keck/DEIMOS observations from \citet{Ramakrishnan25b}. The `DESI' tag indicates DESI DR1 spectra \citep{DESI_DR1}.
Only the first 20 sources are listed. 
}
\tablenotetext{1}{The uncertainty of \lya size is defined as the standard deviation of the recovered \lya sizes from the recovery test (Section~\ref{sec:recovery}).}
\tablenotetext{2}{The peak of \lya emission coincided with a bad column (Figure~\ref{fig:GMOS}).}
\end{deluxetable*}
\end{rotatetable}
%----------------------------------------------------------------------



\end{document}